\documentclass[aps,pra,twocolumn,showpacs,notitlepage,superscriptaddress,letterpaper]{revtex4-2}
\usepackage{graphicx, amsmath, amssymb, amsfonts, pifont, bm, cancel}
\usepackage[usenames]{color}
\usepackage{subfigure}
\usepackage[normalem]{ulem} 
\usepackage{circuitikz}
\usepackage{hyperref}
\usepackage{upgreek}

\definecolor{MyDarkBlue}{rgb}{0,0,1}
\definecolor{darkgreen}{rgb}{0.0, 0.5, 0.0}

\renewcommand\Re{\operatorname{Re}}

\newcommand{\beq}{\begin{equation}}
\newcommand{\eeq}{\end{equation}}
\newcommand{\bel}{\begin{align*}}
\newcommand{\tamam}{\end{align*}}

\newcommand{\ket}[1]{|#1\rangle}

\newcommand{\beqa}{\begin{eqnarray}}             
\newcommand{\eeqa}{\end{eqnarray}}               
\newcommand{\bra}[1]{\langle#1\vert}                 

\newcommand{\taud}{\tau_{\text{d}}}
\newcommand{\omegamod}{\omega_{\text{mod}}}
\newcommand{\omegage}{\omega_{ge}}

\newcommand{\omegaef}{\omega_{ef}}

\newcommand{\ketg}{\ket{g}}
\newcommand{\kete}{\ket{e}}
\newcommand{\ketf}{\ket{f}}

\newcommand{\gvacMMuc}{g_{\text{uc}}}

\newcommand{\GammaoneD}{\Gamma_{\text{1D}}}

\newcommand{\fiftyOhm}{$50$-$\Omega$}

\newcommand{\aout}{ \left \langle a_{\text{out}} \right \rangle }
\newcommand{\fluxout}{\left \langle {a^\dagger}_{\text{out}} a_{\text{out}} \right \rangle }
\usepackage{physics}

\newcommand{\aoutt}{ \left \langle a_{\text{out}}(t) \right \rangle }
\newcommand{\fluxoutt}{\left \langle {a^\dagger}_{\text{out}}(t) a_{\text{out}}(t) \right \rangle }
\usepackage{physics}

\begin{document}

\title{Deterministic Generation of Multidimensional Photonic Cluster States with a Single Quantum Emitter}

\author{Vinicius~S.~Ferreira}
\thanks{These authors contributed equally}
\affiliation{Kavli Nanoscience Institute and Thomas J. Watson, Sr., Laboratory of Applied Physics, California Institute of Technology, Pasadena, California 91125, USA.}
\affiliation{Institute for Quantum Information and Matter, California Institute of Technology, Pasadena, California 91125, USA.}
\author{Gihwan Kim}
\thanks{These authors contributed equally}
\affiliation{Kavli Nanoscience Institute and Thomas J. Watson, Sr., Laboratory of Applied Physics, California Institute of Technology, Pasadena, California 91125, USA.}
\affiliation{Institute for Quantum Information and Matter, California Institute of Technology, Pasadena, California 91125, USA.}
\author{Andreas Butler}
\thanks{These authors contributed equally}
\affiliation{Kavli Nanoscience Institute and Thomas J. Watson, Sr., Laboratory of Applied Physics, California Institute of Technology, Pasadena, California 91125, USA.}
\affiliation{Institute for Quantum Information and Matter, California Institute of Technology, Pasadena, California 91125, USA.}
\author{Hannes Pichler}
\affiliation{Institute for Theoretical Physics, University of Innsbruck, Innsbruck A-6020, Austria}
\affiliation{Institute for Quantum Optics and Quantum Information, Austrian Academy of Sciences, Innsbruck A-6020, Austria}

\author{Oskar~Painter}
\email{opainter@caltech.edu}
\homepage{http://copilot.caltech.edu}
\affiliation{Kavli Nanoscience Institute and Thomas J. Watson, Sr., Laboratory of Applied Physics, California Institute of Technology, Pasadena, California 91125, USA.}
\affiliation{Institute for Quantum Information and Matter, California Institute of Technology, Pasadena, California 91125, USA.}

\date{\today}

\begin{abstract}

Multidimensional photonic graph states, such as cluster states, have prospective applications in quantum metrology, secure quantum communication, and measurement-based quantum computation. However, to date, generation of multidimensional cluster states of photonic qubits has relied on probabilistic methods that limit the scalability of typical generation schemes in optical systems. Here we present an experimental implementation in the microwave domain of a resource-efficient scheme for the deterministic generation of 2D photonic cluster states. By utilizing a coupled resonator array as a slow-light waveguide, a single flux-tunable transmon qubit as a quantum emitter, and a second auxiliary transmon as a switchable mirror, we achieve rapid, shaped emission of entangled photon wavepackets,
and selective time-delayed feedback of photon wavepackets to the emitter qubit. We leverage these capabilities to generate a 2D cluster state of four photons with 70\% fidelity, as verified by tomographic reconstruction of the quantum state. We discuss how our scheme could be straightforwardly extended to the generation of even larger cluster states, of even higher dimension, thereby expanding the scope and practical utility  of such states for quantum information processing tasks.

\end{abstract}
\maketitle

\clearpage

\section{Introduction}
\label{intro}
 
Quantum entanglement is generally regarded 
as a necessary resource for exceeding classical performance limits in tasks such as quantum computing, quantum communication, and quantum metrology \cite{wootters1998quantum, horodecki2009quantum, bennett1998quantum, kimble2008quantum, jozsa1997entanglement}. 
In the optical domain, where photons are the ubiquitous carriers of quantum information, multi-partite entangled states are key resources for various quantum computation and networking protocols \cite{gisin2007quantum, kempe1999multiparticle}. Of particular importance are multi-dimensional cluster states, a subset of the family of entangled graph states, which are highly flexible resource states with utility in measurement-based quantum computing \cite{raussendorf2001one, raussendorf2003measurement, raussendorf2007topological, briegel2009measurement}, quantum metrology~\cite{friis2017flexible, shettell2020graph}, and decoherence protected preservation and teleportation of quantum information~\cite{briegel2001persistent, muralidharan2008quantum, schlingemann2001quantum, bell2014experimental}.

Reliable generation of cluster states of photonic qubits by conventional optical means remains an outstanding challenge 
due to the reliance on probabilistic photon entanglement heralding schemes, a by-product of the complexity and inefficiency of optical set ups
~\cite{nielsen2004optical, browne2005resource, kok2007linear}. Thus, there has been significant interest in achieving generation of such multi-dimensional cluster states by deterministic, resource-efficient means. Notable among these are schemes that involve sequential emission of entangled photons via control of only one or a small number of quantum emitters \cite{lindner2009proposal, schwartz2016deterministic}. Note that while sequential emission from a single coherent emitter is sufficient to generate 1D cluster states, higher dimensional cluster states require more emitters or an additional memory element. A promising approach is based on delay lines generating a time-delayed feedback mechanism, expanding the class of cluster states that can be generated with a single emitter \cite{pichler2017universal, wan2021fault, shi2021deterministic, zhan2020deterministic, xu2018generate}. 

Superconducting circuit QED system are a natural fit to implement such protocols. In contrast to atomic-optical systems where the finite atom-photon cooperativity is often a limiting factor \cite{goban2014atom, corzo2019waveguide}, superconducting circuit QED systems enjoy a strong qubit-(microwave)photon coupling that far exceeds the strength of other dissipative channels due to the ease of creating microwave circuits at a deep subwavelength scale \cite{blais2004cavity}. Indeed, there has been significant progress over the last decade in leveraging superconducting qubits to generate, manipulate, and measure non-classical states of light, including 1D cluster states \cite{eichler2011experimental, hoi2012generation, lang2013correlations, eichler2015exploring, kannan2020generating, besse2020realizing}. However, to date, deterministic generation of higher dimensional photonic cluster states via a single quantum emitter has yet to be realized.

In this work we go beyond the previous state-of-the-art by using time-delayed quantum feedback, implemented using an integrated slow-light waveguide, for the generation of multipartite entangled photonic states, thereby achieving generation of a 2D cluster state of microwave photons. 
Our system consists of two superconducting flux-tunable transmon qubits coupled to the two ends of a slow-light waveguide that serves as a delay line. One qubit serves as our quantum emitter, generating shaped photon pulses with durations as short as 30 ns. The other qubit serves as a switchable mirror for selective reflection of emitted photons. In conjunction with the slow light waveguide this mirror allows us to introduce a time-delayed feedback mechanism, which is essential to our multidimensional entanglement generation scheme (following the proposal in Ref. \cite{pichler2017universal}). We characterize the time-delayed feedback between the emitter qubit and previously emitted photons through quantum process tomography, and certify via quantum state tomography the generation of a 2D cluster state of four photons with a  fidelity of 70\%. Finally, we comment on how straightforward hardware and design improvements could increase the size of generated cluster states by an order of magnitude, and allow for the generation of 3D cluster states. Thus, our demonstrated results pave the way for deterministic, resource-efficient synthesis of multi-dimensional photonic quantum resource states, and their use in quantum information science.

\begin{figure*}[tbp]
\centering
\includegraphics[width = \textwidth]{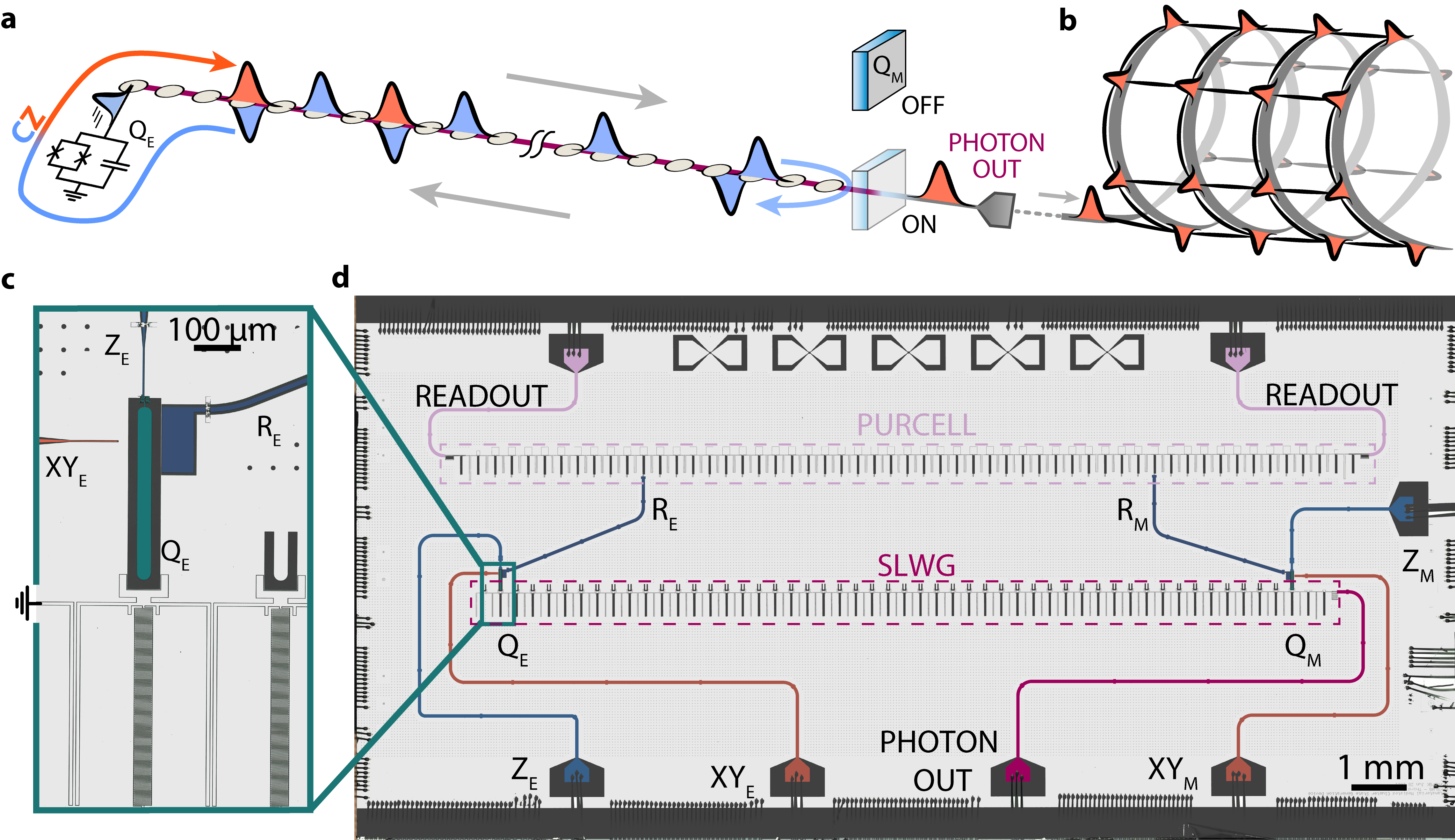}
\caption{\textbf{Deterministic Generation of 2D Cluster States with a Single Emitter Qubit.} \textbf{a,} General scheme for generation of 2D cluster states of photons via a single quantum emitter qubit and time-delayed feedback. The qubit is made to emit a pulse train of nearest-neighbor entangled photons (colored blue) into a delay line. Via control of a switchable mirror, each emitted photon pulse is reflected at the end of the delay line for re-scattering (blue to orange color change) by the emitter qubit. In the device used in this work, superconducting transmon qubits are used as both the quantum emitter ($Q_E)$ and the switchable mirror ($Q_M$), and an array of weakly coupled resonators comprise a slow-light waveguide, which serves as a single-ended delay line. \textbf{b,} Visualization of the resulting entanglement structure. \textbf{c,d,} False-color optical images of the device, comprising a slow-light waveguide (SLWG) and two transmon qubits (false color green), with each qubit coupled to a readout resonator (false color dark blue), a XY control-line (false color red), and a Z fast flux-line (false color light blue). The readout resonators are probed through a coupled resonator array Purcell filter connected to CPW feed-lines (false color lilac). The emitted photons exit the slow-light waveguide and device via a CPW feedline (false color dark purple).} 
\label{fig:Device}
\end{figure*}

\section{Results}
\label{results}

\textit{Cluster State Generation Protocol -} We first discuss the general scheme we use to generate multidimensional cluster states on a conceptual level. Our approach is based on the proposal in Ref.~\cite{pichler2017universal}, and is illustrated in Fig. \ref{fig:Device}a. In our  scheme we couple a quantum emitter to the terminated end of a single-ended, low-group velocity waveguide. This quantum emitter has two stable states $\ketg$ and $\kete$, as well as a radiative state $\ketf$ which is highly damped to the waveguide and decays to the $\kete$ state. 

Starting with the emitter in the $\ketg$ state, the protocol first involves the generation of entanglement between the emitter and a photon mode using two coherent pulses:  a first $\pi_{ge}/2$ pulse generates an equal superposition of the $\ketg$ and $\kete$ state,  then a $\pi_{ef}$ pulse transfers the amplitude from the $\kete$ state to the $\ketf$ state, which subsequently decays back to the $\kete$ state by emission of a photon into the waveguide. This leaves the emitter and the first emitted photon in the maximally entangled state $\ket{\psi} = \left( \ketg\ket{0}_1 + \kete\ket{1}_1 \right)/\sqrt{2}$. 
Repeating this 2-pulse control sequence leads to sequential emission of a train of entangled photonic time-bin qubits with a nearest-neighbor entanglement structure equivalent to the one of a 1D cluster state \cite{lindner2009proposal, schwartz2016deterministic} into the slow-light waveguide (illustrated in Fig. \ref{fig:Device}a as the blue colored pulses).

This train of sequentially emitted pulses is then reflected back towards the emitter by a switchable mirror at the other end of the delay line. After a full roundtrip the photons thus scatter from the quantum emitter and pick up a state-dependent scattering phase (illustrated in Fig. \ref{fig:Device}a as a color change in the pulses from blue to orange). Specifically, if the emitter qubit is in state $\kete$, the returning photon is resonantly coupled to the $\kete \rightarrow \ketf$ transition, and acquires a scattering phase of $\pi$. However, if the emitter qubit is in state $\ketg$, then the returning photon is not resonant with any transition, and no scattering phase is acquired. Thus, this scattering process effectively implements a controlled $CZ$ gate between the emitter qubit and the returning photonic qubit of the form $\ketg\bra{g} \bigotimes \mathbb{I} + \kete\bra{e} \otimes \sigma_z$

This combination of the sequential emission process and the state-dependent scattering process allows us to synthesize the 2D cluster state: through judicious control of the switchable mirror and emitter qubit, we ensure that all sequentially emitted photons scatter from the emitter qubit exactly once and thereafter are allowed to leave the waveguide. The resulting entanglement structure of the outgoing photon pulse train is that of a 2D cluster state with shifted periodic boundary conditions, as illustrated in Fig. \ref{fig:Device}b (see \cite{pichler2017universal} and Appendix \ref{App:tomo} for a quantum circuit representation of this protocol).
Nearest neighbor entanglement in this photonic pulse train is derived from the sequential emission of photons representing one of the two dimensions, whereas entanglement along the other dimension results from the time-delayed scattering process. Importantly, the extent of the second dimension is set by the number of photon pulses that can be generated during one round trip time $\taud$, highlighting the role of the time-delayed feedback. 

\textit{Device Description -} Inspired by this proposal, we fabricated the device shown in Fig. \ref{App:device}c,d in order to achieve a practical realization of this scheme. We implement the requisite delay line as a single-ended slow-light waveguide (SLWG), which is comprised of a periodic coupled resonator array of 52 resonators \cite{ferreira2021collapse}. The output port of the SLWG is connected to a coplanar waveguide (CPW) through which emitted itinerant photons leave the device for amplification and subsequent measurement at the digitizer (see Appendix \ref{App:Fab_Meas} for details on the measurement output chain of the device). The round-trip delay of the slow-light waveguide is $\taud = 240$ ns. The resonator array is terminated at one end via a capacitance between the leftmost unit cell and the ground plane. At the other end of the resonator array the last two boundary resonators are modified relative to the unit cells
in order to effectively match the Bloch impedance of the periodic structure to the characteristic impedance of the output CPW (for further details and design principles of this resonator array slow-light waveguide, see ref. \cite{ferreira2021collapse} and Appendix \ref{App:Fab_Meas}). The resulting transfer function of such a slow-light waveguide is that of a flat ``passband" of finite bandwidth for guided modes, 
and a sharp extinction of transmission outside of the passband due to the sharp decline in the photonic density of states (DOS) of the periodic structure occurring at the bandedges. The width of the passband is $4J$, where $J$ is the coupling between unit cells in the resonator array; in our device $J/2\pi = 34$ MHz (giving a passband width of 136 MHz) and the passband center frequency is $\omega_p/2\pi = 4.82$ GHz. 

On the terminated end of the slow-light waveguide we couple the emitter qubit $Q_E$, while at the other end of waveguide we couple another qubit $Q_M$. The mirror qubit is effectively side-coupled to the slow-light waveguide, allowing it to act as a high reflectivity mirror for single photons if the ratio between its decay rate into the waveguide and its decoherence rate into other channels, $\GammaoneD/\Gamma'$, is sufficiently high \cite{shen2005coherent}. 
Each qubit is coupled to its own XY control line for single-qubit control, a Z control line for rapid flux tuning of the qubit transition frequency, and a CPW readout resonator (R) coupled to a Purcell filter for dispersive readout of the qubit state (the Purcell filter in this work is also comprised of a coupled resonator array; for more details, see Appendix \ref{App:Fab_Meas}). 
At zero flux bias, the transition frequency between $Q_E$'s ground state (denoted $\ketg$) and first excited state (denoted $\kete$) is $\omegage^E/2\pi = 6.21$ GHz, and the transition frequency between the first excited state and second excited state (denoted $\ketf$) is $\omegaef^E/2\pi = 5.93$ GHz, with associated anharmonicity of $\eta^E/2\pi = (\omegaef^E -\omegage^E)/2\pi = -273$ MHz. The center frequency of the $Q_E$ readout resonator at this bias is $\omega_r^E/2\pi = 7.67$ GHz, and its induced dispersive shift $2\chi^E$ is given by $\chi^E/2\pi = 2.1$ MHz. The same quantities parameterizing the mirror qubit at its zero flux bias are given by $\omegage^M/2\pi = 6.44$ GHz, $\eta^M/2\pi = -280$ MHz, $\omega_r^M/2\pi = 7.47$ GHz, and $\chi^M/2\pi = 3.4$ MHz (see Appendix \ref{App:Fab_Meas} for further qubit characterization details).

\begin{figure}[tbp]
\centering
\includegraphics[width = \columnwidth]{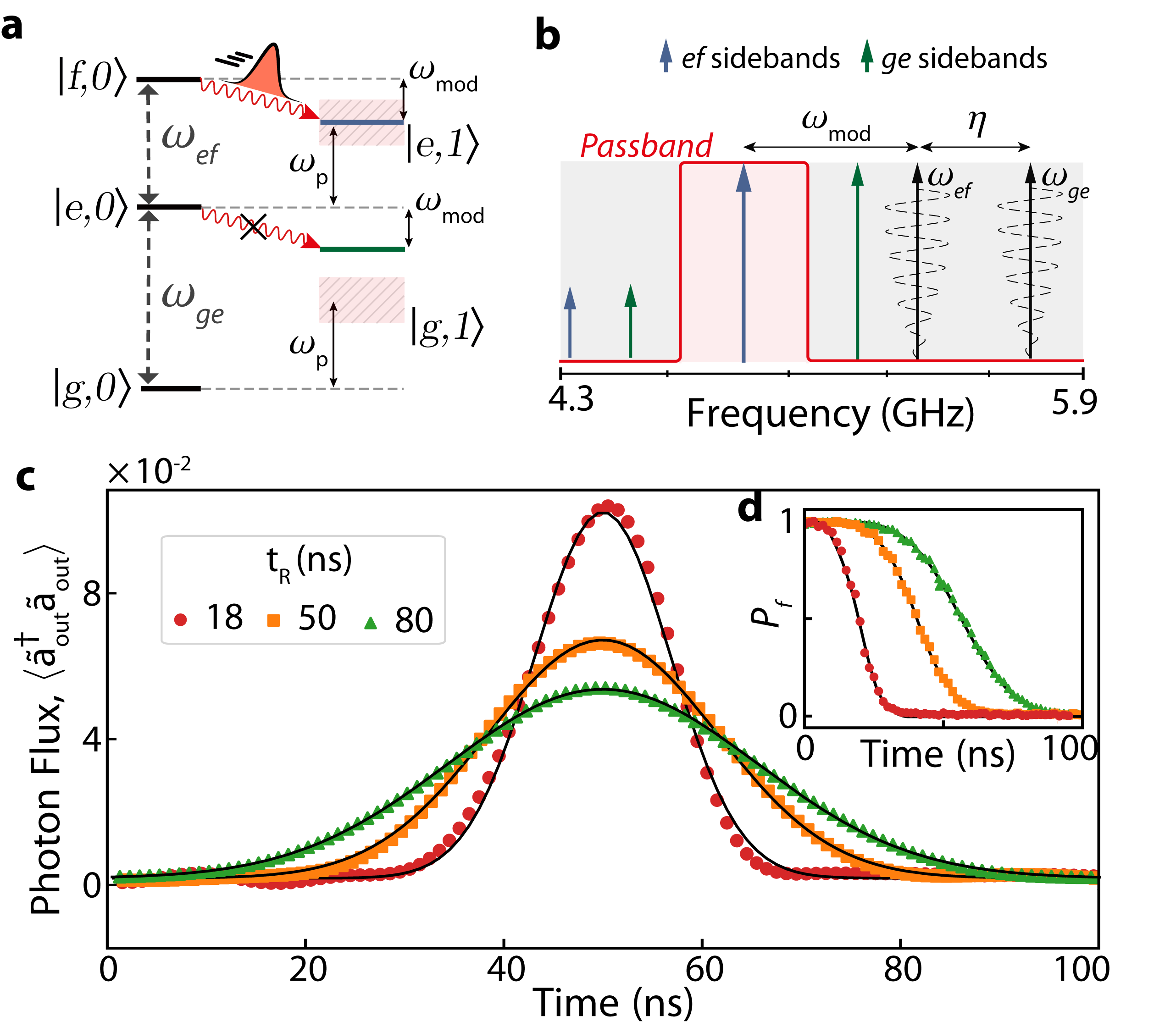}
\caption{\textbf{Emission of shaped photons pulses via flux-modulation.} \textbf{a} Effective level diagram of the qubit-waveguide system, showing the photon emission process. In the $\ket{i,n}$ notation, $i$ denotes the state of the transmon emitter qubit, and $n$ denotes the number of photons in the slow-light waveguide; additionally, $\omega_{m}$ is the flux-modulation frequency, and $\omega_p$ is the center frequency of the waveguide's passband. When $\omegaef - \omegamod = \omega_p$, the qubit's levels assume a three-state ladder system where only the $\ketf$ state is selectively damped to the slow-light waveguide.   \textbf{b} Simplified frequency spectrum of the emitter qubit under flux-modulation, where $\eta$ is the qubit anharmonicity. The flux-modulation waveform is depicted as a dashed black line; the waveform's modulated amplitude directly maps to a modulated emission rate into the waveguide that allows for shaped emission of photon pulses. \textbf{c} Measured photon flux $\langle {a^\dagger}_{\text{out}} a_{\text{out}} \rangle$ (dots) of shaped emitted pulses, in normalized units. Black lines are Gaussian fits. \textbf{d} Measured $\ketf$ population during shaped emission. Black lines are the scaled integral of the Gaussian fits of subfigure \textbf{c}.}
\label{fig:Emission}
\end{figure}

Crucially, due to the finite width of the slow-light waveguide passband and its sharp bandedges, it is possible to tune the $\kete \rightarrow \ketf$ transition frequency into resonance with $\omega_p$ and achieve large emission rates of the $\ketf$ state, while simultaneously protecting the $\kete$ state from decay if the $\ketg \rightarrow \kete$ transition frequency is situated outside the passband of the waveguide (where the DOS of the periodic structure is negligible). In our system we naturally achieve this configuration, where our anhamonicity $\eta^E/2 \pi$ of $\sim$280 MHz allows us to situate the $\kete \rightarrow \ketf$ transition frequency inside the passband of 136 MHz width, while maintaining the $\kete \rightarrow \ketg$ outside the passband. Thus, with these parameters, the first three levels of the transmon comprise the aforementioned necessary ladder level structure for cluster state generation. 
We stress that
the sharp bandedges of the waveguide allow us to engineer remarkably large emission rates of $\GammaoneD^{ef}/2\pi = 2\GammaoneD^{ge}/2\pi \approx 140$ MHz for $Q_E$, while strongly suppressing decay of the $\ketg \rightarrow \kete$ transition to single kHz rates, even though $\eta$ is comparable to $\GammaoneD^{ge}$.

\textit{Shaped Photon Emission -}  For generation of 2D cluster states as we have described, it is crucial to be able to control the shape of emitted photon pulses. This allows us to mitigate the effects of the waveguide's residual dispersion near $\omega_p$, and to improve the fidelity of the CZ gate after a photon round trip. For this it is necessary to control the photon pulse shape as well as its bandwidth, reducing it to less than $\GammaoneD^{ef}$ (see ref. \cite{pichler2017universal} for more details).
We shape the pulse of the emitted photons by a tunable qubit-waveguide interaction strength for $Q_E$ via parametric flux modulation of the qubit frequency \cite{beaudoin2012first, strand2013first, silveri2017quantum}. Specifically, we apply an AC flux drive to the SQUID loop of $Q_E$ with frequency $\omegamod$, which generates a series of sidebands, spaced by $\omegamod$, for each transition of the transmon qubit (for more details, see Appendix \ref{App:fluxcon}). 

By judiciously choosing the qubit frequency and modulation frequency such that $\omegaef - \omegamod = \omega_p$, while $\omegage - \omegamod$ lies outside the passband due to the anharmoniciy of the qubit, we can ensure that only a first-order sideband of the $e-f$ transition overlaps with the passband. Meanwhile, all other relevant qubit transition frequencies and their sidebands do not fall into the passband. Thereby we achieve photon emission into the waveguide from the $\ketf$ state through the first order $e-f$ sideband, while retaining protection of the $\ketg$ and $\kete$ levels. This is shown schematically in Fig. \ref{fig:Emission}a,b; where in  Fig. \ref{fig:Emission}a we illustrate this emission process through a level diagram, whereas in Fig. \ref{fig:Emission}b we show a simplified frequency spectrum of the particular configuration of qubit frequencies and sideband frequencies used in our experiment.  By choosing $\omegage/2\pi = 5.55$ GHz and $\omegamod/2\pi = 450$ MHz, we situate the lower first sideband of the $\kete \rightarrow \ketf$ transition at $\omega_p/2\pi = 4.823$ GHz, while all other sidebands and bare qubit transition frequencies are sufficiently detuned from the passband as to negligibly contribute to qubit emission, as verified by separate measurements. 

We thereby achieve shaped emission by continuously varying the amplitude of the flux modulation AC drive during the emission time, which varies the strength of the aforementioned emission sideband and thus allows us to achieve arbitrary time-dependent modulation of $Q_E$'s emission rate
$\GammaoneD^{ef}(t)$ (see Appendix \ref{App:fluxcon} for further details on how we achieve pulse shaping of emitted photons in this manner). With this capability we achieve shaped emission of Gaussian shaped photons with excellent accuracy, as demonstrated in Fig. \ref{fig:Emission}c, which henceforth constitutes our photonic time-bin qubits. We plot the measured photon flux of three emitted Gaussian pulses with different bandwidths (along with their respective fits), demonstrating the flexibility in our shaped emission scheme (photon flux is plotted in normalized units, see Appendix \ref{App:tomo} for further details). This emission is also achieved with high-efficiency, and thus enables deterministic high-fidelity preparation of entangled photonic states (see Appenddix \ref{App:tomo} for more details). 

Further, in Fig. \ref{fig:Emission}d, we plot $Q_E$'s population dynamics during emission, as well as the integral of the photon fluxes plotted in Fig. \ref{fig:Emission}c which, in the absence of waveguide-induced distortion, would coincide with the population dynamics of $Q_E$. We find excellent agreement between the two, indicating that the effects of the slow-light waveguide dispersion are minimal for Gaussian pulses.
Finally, we stress that our large $\GammaoneD$ allows high-efficiency emission of pulses that are tightly confined to a time-bin window of length as small as 30 ns, which not only is an important attribute to achieve in order to increase the size of generated cluster states given a fixed $\taud$, but also demonstrates significant improvement in emission speed of shaped photons over previous shaped emission demonstrations in circuit QED systems\cite{pechal2014microwave, forn2017demand, ilves2020demand, reuer2021realization}. 

\begin{figure}[tbp]
\centering
\includegraphics[width = \columnwidth]{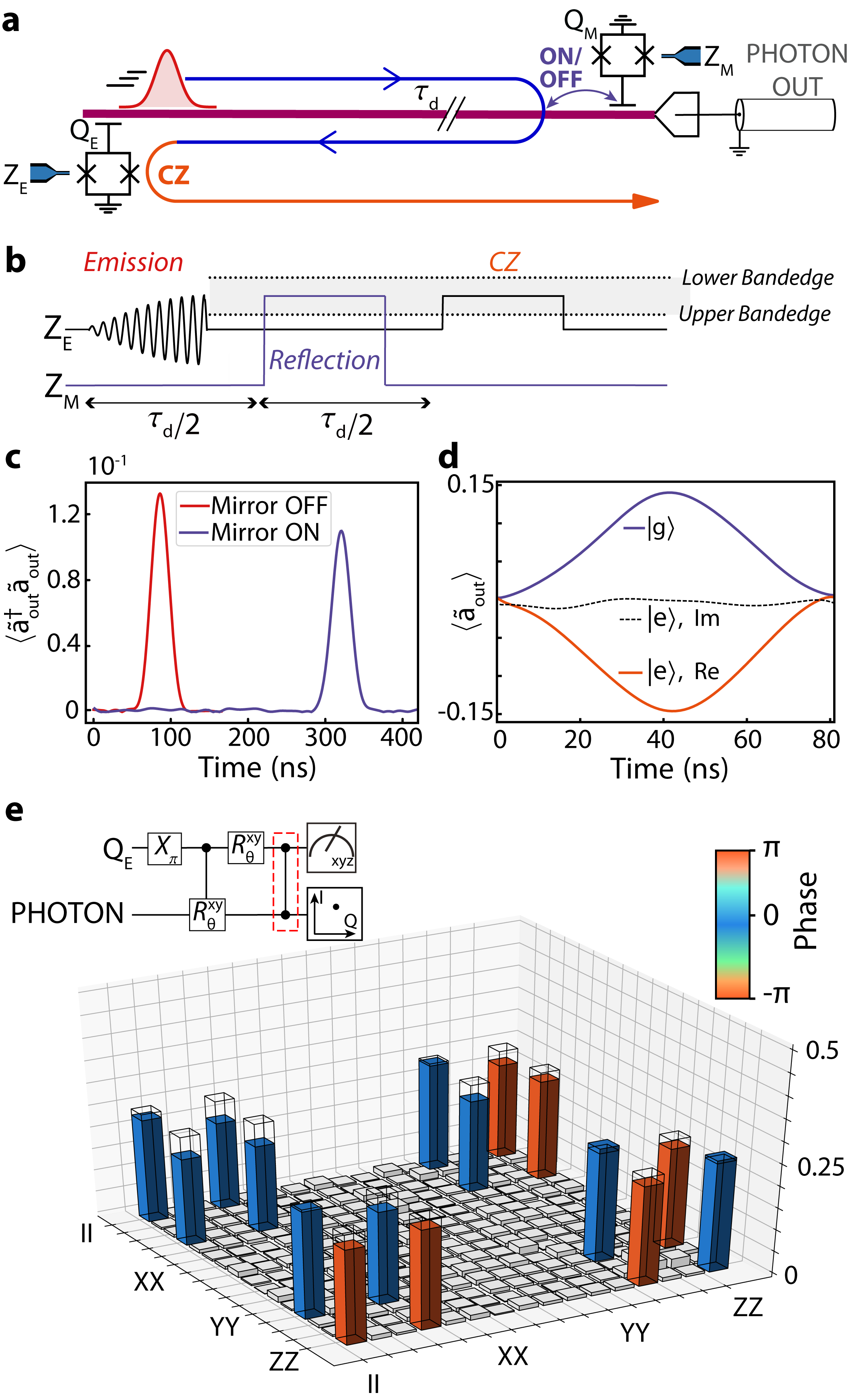}
\caption{\textbf{CZ Gate Between Emitter Qubit and Previously Emitted Photons via Time-Delayed Feedback.} \textbf{a,} Illustration of the time-delayed feedback process that realizes the $CZ$ gate between $Q_E$ and its emitted photon, where the photon undergoes a round-trip through the slow-light waveguide and re-scatters on $Q_E$. \textbf{b,}. Z-control of the qubits that implements the $CZ$ gate. Both square pulses on $Z_M$ and $Z_E$ tune their respective qubit frequencies to the middle of the passband. \textbf{c,} Measured photon flux of qubit emission with the $Z_M$ square pulse for mirror reflection ON vs OFF. \textbf{d,} Measured $\langle a_{\text{out}} \rangle$ of the reflected pulse after it re-interacts with $Q_E$ , where $Q_E$ is prepared in either the $\ketg$ or the $\kete$ state. The complex phase of $\langle a_{\text{out}} \rangle$ in both cases is normalized to the phase of the measurement where $Q_E$ is prepared in the $\ketg$ state. \textbf{e,} Quantum Process Tomography of the Pauli process matrix $\chi_{\text{CZ}}$ of the $CZ$ gate between $Q_E$ and its emitted photon, demonstrating a 90\% fidelity relative to the ideal gate; the procedure for performing the tomography is shown on the top left.} 
\label{fig:Mirror_Catch}
\end{figure}

\textit{Qubit-Photon CZ Gate Implementation -} In addition to high efficiency shaped photon preparation, we also demonstrate a high fidelity $CZ$ gate between $Q_E$ and previously emitted photonic qubits, which is effected by the time-delayed feedback. In Fig. \ref{fig:Mirror_Catch}a we show an schematic of the process, where an itinerant photon emitted by $Q_E$ propagates through the waveguide, is reflected by $Q_M$, and propagates back towards $Q_E$, whereupon photon scattering on $Q_E$ realizes the $CZ$ gate. Afterward, the photon propagates back toward the output, and is allowed to leave the slow-light waveguide by suitable $Q_M$ control. The fast flux control sequence necessary to implement this process is shown in Fig. \ref{fig:Mirror_Catch}b. An amplitude modulated AC pulse on the $Z_E$ line induces photon emission, while a square pulse is initiated in the $Z_M$ line at a time $t = \taud/2$, the single-trip time of the waveguide (see Appendix \ref{App:fluxcon} for details on flux-line distortion compensation). The square pulse amplitude is chosen such that $\omegage^M$ is tuned to the center of the passband, which reflects the emitted itinerant photon. 
At $t = \taud$, a square pulse is initiated on the $Z_E$ line, which tunes $\omegaef^E$ to the center of the passband in order to re-scatter the reflected photon and realize the $CZ$ gate. Note that while a flux-modulation sideband is used to emit the photon, decay rates induced by the sideband are at maximum less than 50\% the intrinsic $\GammaoneD^{ef}$ of the qubit.  Thus it is more suitable to rapidly tune the qubit frequency rather than modulate it to re-scatter the photon, given that larger $\GammaoneD$ increases the fidelity of the time-delayed feedback induced $CZ$ gate (see ref. \cite{pichler2017universal} for more details).

In Fig. \ref{fig:Mirror_Catch}c,d we illustrate the actions of reflection by $Q_M$ and photon re-scattering by $Q_E$ on an emitted photon. In Fig. \ref{fig:Mirror_Catch}c we show the emitted photon's measured photon flux at the digitizer with the square pulse on the $Z_M$ line turned on or turned off. With the $Z_M$ square pulse turned on, the photon's arrival at the digitizer is delayed by $\taud$, while negligible photon flux is measured at all prior times, demonstrating that the $Q_M$ reflects the itinerant photon with high efficiency. Note that the magnitude of the photon flux when the $Z_M$ square pulse is turned on is lower than when the pulse is turned off due to the additional 0.6 dB loss incurred by the itinerant photon during its round-trip. Furthermore, in Fig. \ref{fig:Mirror_Catch}d we show the emitted photon's average measured \textit{field} when $Q_E$ is prepared in either the $\ketg$ state or the $\kete$ state (where the phase of the field is referenced to the measurement in which $Q_E$ is in the $\ketg$ state). 
It is evident that the sign of the real part of the re-scattered photon's complex field changes when the state of $Q_E$ is changed from $\ketg$ to $kete$ (while the imaginary part of the re-scattered photon's complex field is negligible). This corresponds to a state-dependent $\pi$ difference in the phase of the photon, as desired for the $CZ$ gate implementation.

\begin{figure*}[tbp]
\centering
\includegraphics[width = \textwidth]{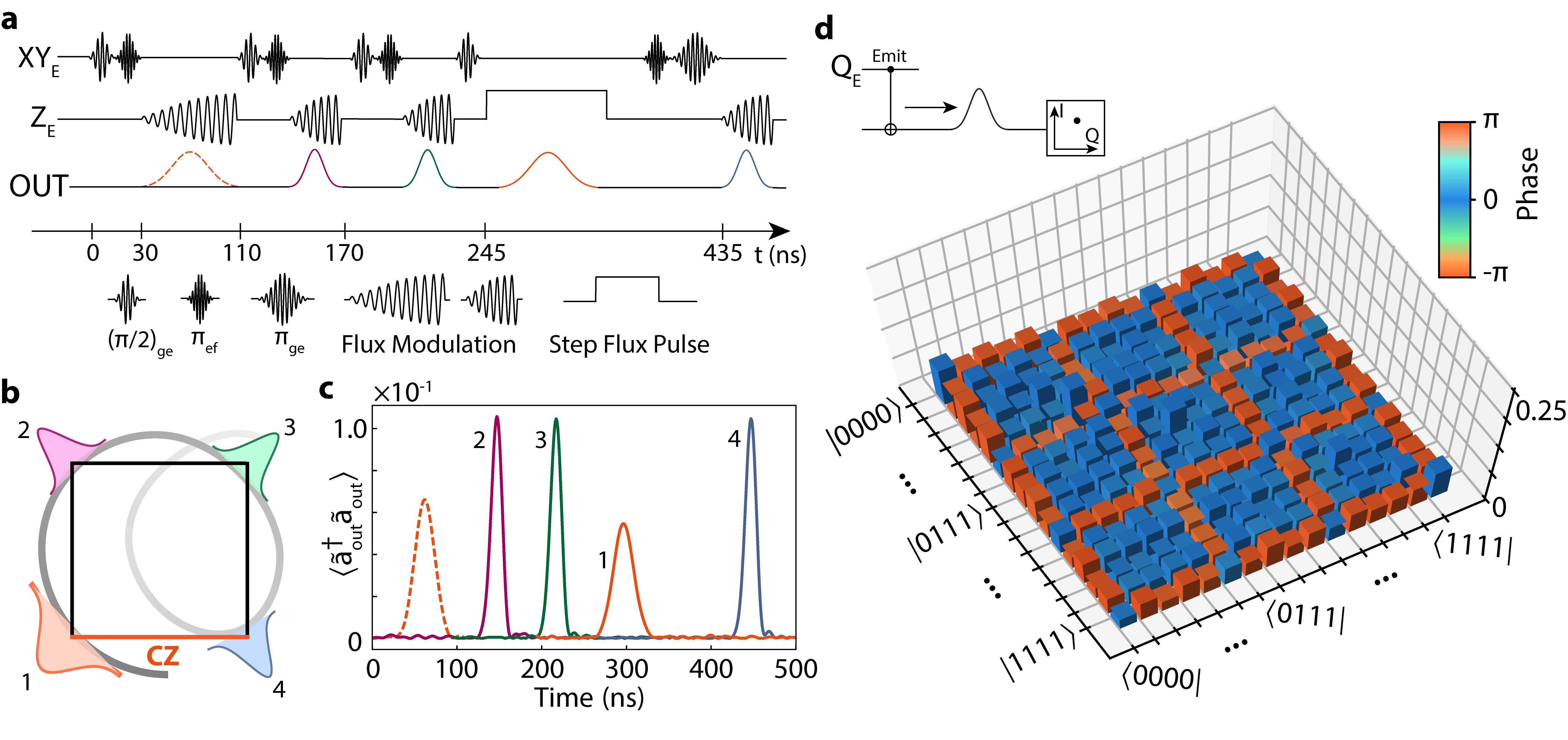}
\caption{\textbf{Deterministic Generation of a 4-photon 2D Cluster State} \textbf{a} Pulse sequence of $Q_E$ control lines, and illustration of outgoing photon flux from $Q_E$. Depiction of $Z_M$ line control can be found in Fig.~\ref{fig:Mirror_Catch}. Photons 1,2,3, and 4 are represented by the colors orange, purple, green, and blue, respectively. \textbf{b} Illustration of the generated entangled state. ``CZ" signifies the entanglement that arises due to the CZ gate between $Q_E$ and photon 1. \textbf{c} Photon flux of individual time-bin photonic qubits. The dotted orange line corresponds to the photon flux of the first emitted photon in the absence of reflection by the mirror qubit, and is only shown for illustration purposes. \textbf{d} Density matrix $\rho$ of the generated 2D cluster state obtained from photonic quantum state tomography. The height of the bars represent the magnitudes of the elements of $\rho$, while the color of the bars represent the phases of the elements of $\rho$. The fidelity of the generated state $F = \text{Tr}\left ( \sqrt{ \sqrt{\rho} \rho_{\text{ideal}} \sqrt{\rho} } \right )^2$ is 70\%. } 
\label{fig:Cluster_State}
\end{figure*}

In addition, we perform quantum process tomography in order to demonstrate the quantum character of the $CZ$ gate. The tomography procedure is shown in the top left of Fig. \ref{fig:Mirror_Catch}e: different input photonic states are first prepared via suitable $Q_E$ control, followed by qubit preparation into $Q_E$'s different cardinal states. The $CZ$ gate is then performed for each photon/qubit state combination, after which single shot measurements of both the qubit state and the photon field are carried out. The single-shot measurements of the time-dependent photon field, obtained via heterodyne detection of the field after suitable amplification, are post-processed to field quadratures $I$ and $Q$ of the photonic qubit, and are thereupon correlated with single-shot qubit readout measurements (for further details on the single-shot measurement of the field quadratures of the photonic qubits, see Appendix \ref{App:tomo}). Joint qubit-photon moments $\langle (a^\dagger )^n a^m \sigma_i \rangle$  are calculated from the processed single-shot data, which are finally used to reconstruct the process matrix $\chi_{\text{CZ}}$, show in Fig. \ref{fig:Mirror_Catch}e; we calculate a process fidelitiy of $\text{Tr}\left(\sqrt{ \sqrt{\chi_{\text{CZ}}} \chi_{\text{ideal}} \sqrt{\chi_{\text{CZ}}} } \right)^2$ of 90\%. We attribute most of the infidelity to dephasing and state preparation and measurement (SPAM) errors, given that a similar measurement of the $\mathbb{I} \bigotimes \mathbb{I}$ process matrix yields a process fidelity of 92.5\% (for further details of the process tomography, see Appendix \ref{App:processtomo}).

\textit{2D Cluster State Preparation -} Finally, with our high efficiency shaped photon preparation and high fidelity $CZ$ gate, we demonstrate generation of a 2D cluster state of four microwave photons. In Fig. \ref{fig:Cluster_State}a, we show the full $Q_E$ control we used to generate the cluster state, which results in the entangled state schematically shown in Fig. \ref{fig:Cluster_State}b. The control sequence essentially consists of four cycles of the aforementioned operation that generates a 1D cluster state: a $\pi_{ge}/2$ pulse followed by a $\pi_{ef}$ pulse and flux modulation induced photon emission. Additionally, the first emitted photon is reflected by $Q_M$ and re-scatters on $Q_E$ in between the fourth $\pi_{ge}/2$ and $\pi_{ef}$ pulses, thus entangling photon 1 to photon 4 once photon 4 is emitted. Notably, before the last $\pi_{ef}$ pulse we also apply a $\pi_{ge}$ to $Q_E$ in order to disentangle it from the photonic state upon its final emission. Also, we emit photon 1 with a lower bandwidth than other photons in order to maintain the high fidelity of the CZ gate between photon 1 and $Q_E$, while photons 2,3, and 4 are emitted more rapidly (within a 30ns time window) in order to more efficiently use the fixed $\taud$ delay available. The measured photon flux of the individual time-bin photonic qubits is shown in Fig. \ref{fig:Cluster_State}c, where their position in time corresponds to their arrival time at the digitizer.

In order to tomographically reconstruct the generated state, we once again obtain the single-shot field quadratures $I_i$ and $Q_i$ of each photonic time-bin qubit, and obtain their correlations through calculation of all joint moments of the photonic fields. With the moments, we obtain the density matrix $\rho$ of the generated state through a maximum likelihood (MLE) algorithm, shown in Fig. \ref{fig:Cluster_State}d (for a detailed description of the photonic state tomography process, refer to Appendix \ref{App:tomo}). When compared to the ideal expected density matrix $\rho_{\text{ideal}}$ shown in Fig. \ref{fig:clusterphase}b, we observe excellent agreement between the phase pattern of the density matrix elements, qualitatively indicating that we have achieved the desired entanglement structure of the state shown in Fig. \ref{fig:Cluster_State}b. Further, each photon has a finite weight-three stabilizer operator $\sigma_x^i \bigotimes_{j \in N(i)} \sigma_z^j $ expectation value, where $N(i)$ are the nearest neighbors of photon $i$ (assuming the connectivity of Figure \ref{fig:Cluster_State}b), with calculated values $0.73, 0.73, 0.8, 0.75$, consistent with a square entanglement connectivity rather than a linear entanglement connectivity.  

We calculate a fidelity $F = \text{Tr}\left ( \sqrt{ \sqrt{\rho} \rho_{\text{ideal}} \sqrt{\rho} } \right )^2$ of 70\% between the generated and ideal state, indicating achievement of genuine four-partite entanglement and successful implementation of the protocol of ref. \cite{pichler2017universal}. This is in good agreement with our estimated state fidelity limit of 76\%, which we calculate from contributions to preparation infidelity that include the dephasing of $Q_E$ (the primary source of infidelity), the round-trip loss suffered by photon 1, and measured qubit preparation and control errors (see Appendix \ref{App:improvements} for more details). 


\section{Conclusion}
\label{conclusion}

In conclusion, we successfully implemented a resource-efficient protocol for generation of multidimensional cluster states by utilizing a single superconducting qubit as a source of entangled photons, and a coupled resonator array as a slow-light waveguide for time-delayed feedback. We accomplished this by achieving rapid, shaped emission of single photons, as well as by implementing a high fidelity $CZ$ quantum gate between the quantum emitter and previously emitted photons through the controllable time-delayed feedback of our system. This allowed us to generate a 2D cluster state of four microwave photons, attaining a state fidelity of 70\% (95\% CI [69.1\%, 70.4\%]).

There are numerous avenues for straightforward improvements to our implementation of the cluster state generation protocol that would enable generation of significantly larger cluster states (which we discuss in detail in Appendix \ref{App:improvements}). For instance, by improving the dephasing times of the qubit ($T_2^* = 561$ ns in this work) and the quality factors of the unit cell resonators of the slow-light waveguide (approximately $\sim 90,000$) to state-of-the-art values \cite{megrant2012planar, calusine2018analysis, woods2019determining}, the major sources of infidelity we incurred could already be dispensed with. Furthermore, potentially increasing the anharmonicity $\eta$ of the qubit through different qubit design \cite{nguyen2019high, yurtalan2021characterization, yan2020engineering} would enable even larger $\GammaoneD$, allowing for a higher fidelity $CZ$ gate with high-bandwidth photons and even more rapid emission of shaped photon pulses. The round-trip delay, $\taud$, could also be increased by either further reducing the footprint of our unit cell resonators, for example by leveraging compact high kinetic inductance superconducting resonators \cite{shearrow2018atomic, grunhaupt2018loss},  or by incorporation of acoustic delay lines \cite{bienfait2019phonon, andersson2019non, dumur2021quantum}, increasing the photon(phonon)-pulse storage capacity of the delay line and the corresponding size of realizable cluster states.  

Not only would these discussed improvements substantially increase the realizable size of 2D cluster states, they would also allow for generation of more complex graph states such as 3D cluster states. Our time-delayed feedback based scheme for generating 2D cluster states can be easily extended to generate 3D cluster states by simply adding another time-delayed feedback event with a different delay for every photon \cite{wan2021fault, shi2021deterministic} (which is achievable simply by incorporation of another mirror qubit), where each photon would then be re-scattered by the emitter qubit twice at different times. Indeed, as a preliminary demonstration of this capability, in Appendix \ref{App:tomo} we demonstrate generation of a 5-photon tetrahedral-like cluster state where we implemented the time-delayed feedback process twice for one photon. 
3D cluster states, which have yet to be generated using even the probabilistic heralding techniques employed in optical systems, have been proposed as a resource for realizing fault-tolerance in measurement based quantum computation \cite{raussendorf2007topological}. We therefore expect the deterministic techniques presented here using the rich toolbox of circuit QED to not only improve upon the conventional optics-based approaches for realizing multidimensional cluster states, but to also broaden the scope and applicability of such states for quantum information processing.

\begin{acknowledgments}
We thank Eunjong Kim for helpful discussions regarding experimental setup, and we thank Mo Chen for his collaboration in fridge-related work. This work was supported by the AFOSR MURI Quantum Photonic Matter (grant 16RT0696), through a grant from the  Department of Energy (grant DE-SC0020152), and through a sponsored research agreement with Amazon Web Services. V.F gratefully acknowledges support from NSF GFRP Fellowship.
\end{acknowledgments}

\appendix

\section{Measurement Setup}
\label{App:Fab_Meas}

\subsection{Measurement Setup}

\begin{figure}[tbp]
\centering
\includegraphics[width = \columnwidth]{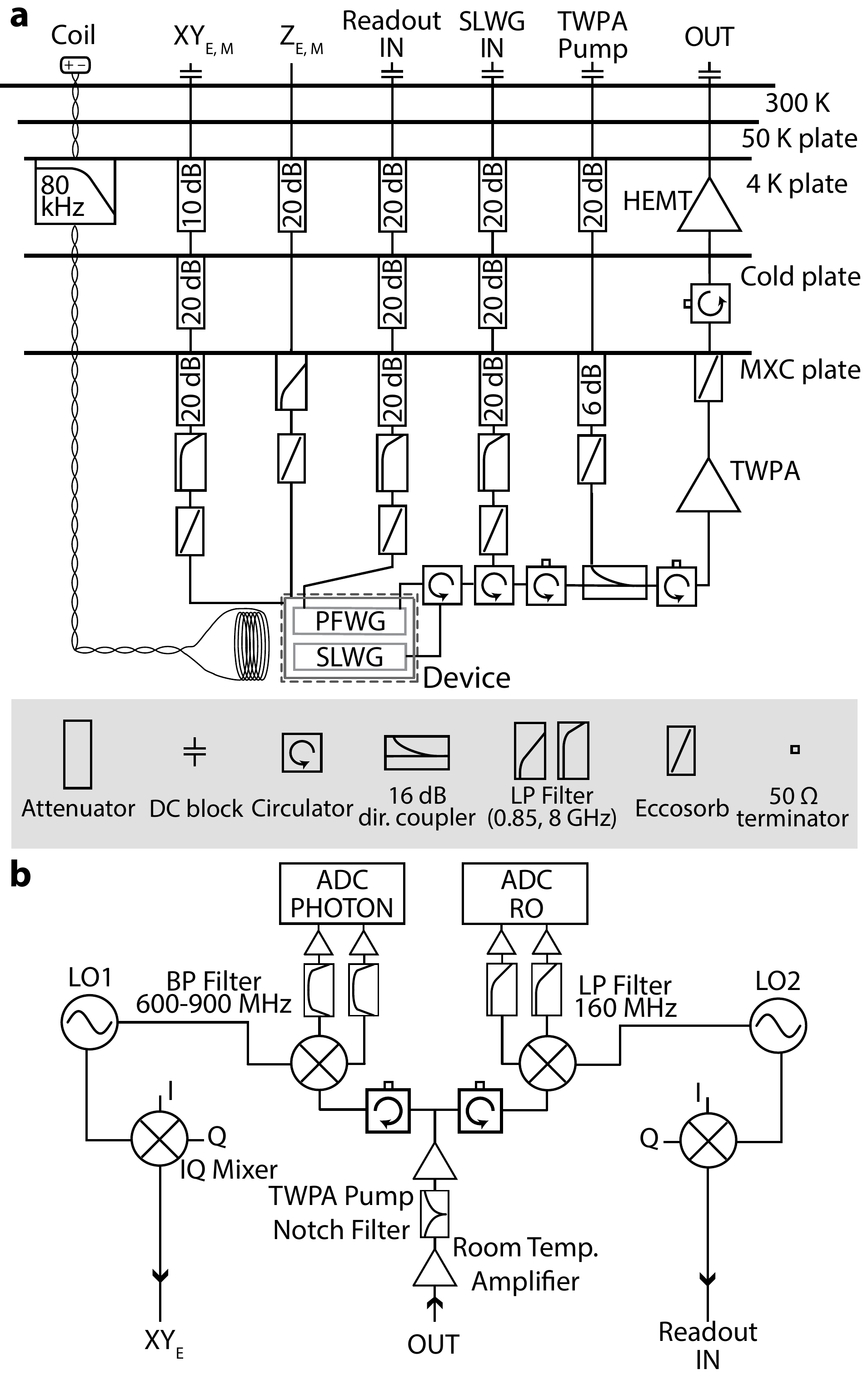}
\caption{\textbf{Measurement Setup} \textbf{a} Schematic of the measurement wiring inside the dilution refrigerator. See Appendix text for further details (``dir." is shorthand for ``directional", ``LP" is shorthand for ``Low Pass", and ``BP" is shorthand for ``Band Pass".). See Fig.~\ref{fig:Device} for electrical connections at the sample. \textbf{b} Simplified diagram of measurement wiring outside the dilution refrigerator.} 
\label{fig:meas_setup}
\end{figure}

A schematic of the fridge wiring and our room-temperature analog signal processing electronics is shown in Fig.~\ref{fig:meas_setup}. Measurements are performed in a 3He/4He dry dilution refrigerator. The sample is wirebonded to a CPW printed circuit board (PCB) with coaxial connectors,
and is housed inside a copper box that is mounted to the MXC plate of the fridge with $T_{MXC}$ = 7 mK. A coil is placed on top of the copper box for static flux tuning of the qubits, and the sample is enclosed in two layers of magnetic shielding to suppress effects of stray magnetic fields. See refs \cite{ferreira2021collapse, keller2017transmon, mirhosseini2019cavity} for more details on device fabrication.

Attenuators are placed at several temperature stages of the fridge to provide thermalization of the coaxial input lines and to reduce thermal microwave noise at the input to the sample. Our gigahertz microwave lines (XY\textsubscript{E}, XY\textsubscript{M}, Readout IN, SLWG IN, TWPA Pump) have significantly more attenuation than our fast flux lines (Z\textsubscript{E}, Z\textsubscript{M}) for reasons explained in ref. \cite{krinner2019engineering}. In addition, fast flux lines are filtered by an 850 MHz low-pass filter below the MXC plate, which suppresses thermal noise photons at higher frequencies while still maintaining short rise and fall times of square flux pulses, as well as allowing transmission of AC flux drives. The tuning coil is differentially biased by two DC input lines, with 80 kHz low-pass filters at the 4K stage to further suppress noise photons. Furthermore, Gigahertz microwave input lines are filtered by an 8GHz lowpass filter and all microwave lines have an Eccosorb filter, in order to ensure strong suppression of thermal noise photons at very high frequencies. Note also that all \fiftyOhm~terminations are thermalized to the MXC plate in order to suppress thermal noise from their resistive elements.

Output signals from the Purcell filter waveguide (PFWG) and slow-light waveguide (SLWG) device lines are merged to a single amplifier chain in the following manner. Their corresponding coaxial lines are connected to a circulator as shown in Fig.~\ref{fig:meas_setup}a, such that signals exiting the SLWG continue directly to the output chain, while signals exiting the Purcell filter are first routed to the SLWG device line and subsequently reflect off of the finite-bandwidth structure, thus finally routing them to the output chain. Note that input signals to the SLWG IN line undergo similar routing in order to arrive at the device. 

Our amplifier chain at the "OUT" line consists of a quantum-limited traveling-wave parametric amplifier (TWPA) \cite{Macklin2015} as the initial amplification stage, followed by a Low Noise Factory LNF-LNC4\_8C high mobility electron transistor (HEMT) amplifier mounted at the 4K plate. For operation of the TWPA, a microwave pump signal from Rohde \& Schwarz SMB100A is added to the amplifier via the coupled port of a 16 dB directional coupler, with its isolated port terminated in 50-$\mathrm{\Omega}$. We include two isolators between the directional coupler and the sample in order to shield the sample from the strong TWPA pump. 

Outside the fridge, we further amplify output signals with amplification that is suitable for the dynamic range of our ADC. We note that we use a Micro Lambda Wireless MLBFR-0212 tunable notch filter  to reject the TWPA pump signal in order to prevent saturation of the following room temperature electronics. Additionally, we use IF amplifiers (0-1GHz bandwidth) for downconverted signals due to IQ mixer saturation power limits. 

Due to their different frequencies, we route SLWG and PFWG signals to different downconversion stages via a 2-way power splitter, followed by a circulator at each branch to prevent crosstalk between the two branches. The ``PHOTON" branch is connected to a IQ mixer for downconversion of $\sim$ 4.8 GHz photonic signals, which are then measured by an Alazartech ATS9371 digitizer (ADC PHOTON); measurement of both photonic signal quadratures $I(t)$ and $Q(t)$ comprise the heterodyne measurement of time-dependent photon signals alluded to in Appendix \ref{App:tomo}. Meanwhile, the other branch of the power splitter is also connected to an IQ mixer for downconversion of $\sim$ 7.5GHz readout signals, which are then measured a Keysight M3102 digitizer (ADC RO). We note that downconversion mixers share LO signals (generated by Rohde \& Schwarz SMB100A microwave signal generators) with their upconversion counterparts (where a Zurich HDAWG is used) , in order to ensure phase drift/jitter of LO's during upconversion are cancelled out during downconversion. And crucially, we place additional filters before measurement at the ADC in order to suppress noise outside of the IF measurement band of interest. This not only allows for better utilization of the ADC dynamic range, but also rejects noise at irrelevant Nyquist bands that ``fold" over to the bandwidth of measured signals; we note that this effectively improved the $n_{\text{noise}}$ of our photon measurement chain by almost a factor of 2 (see Appendix \ref{App:tomo} for more details).

\section{Device Characterization}
\label{App:device}

\subsection{Slow-light Waveguide}
\begin{figure*}[tbp]
\centering
\includegraphics[width = \textwidth]{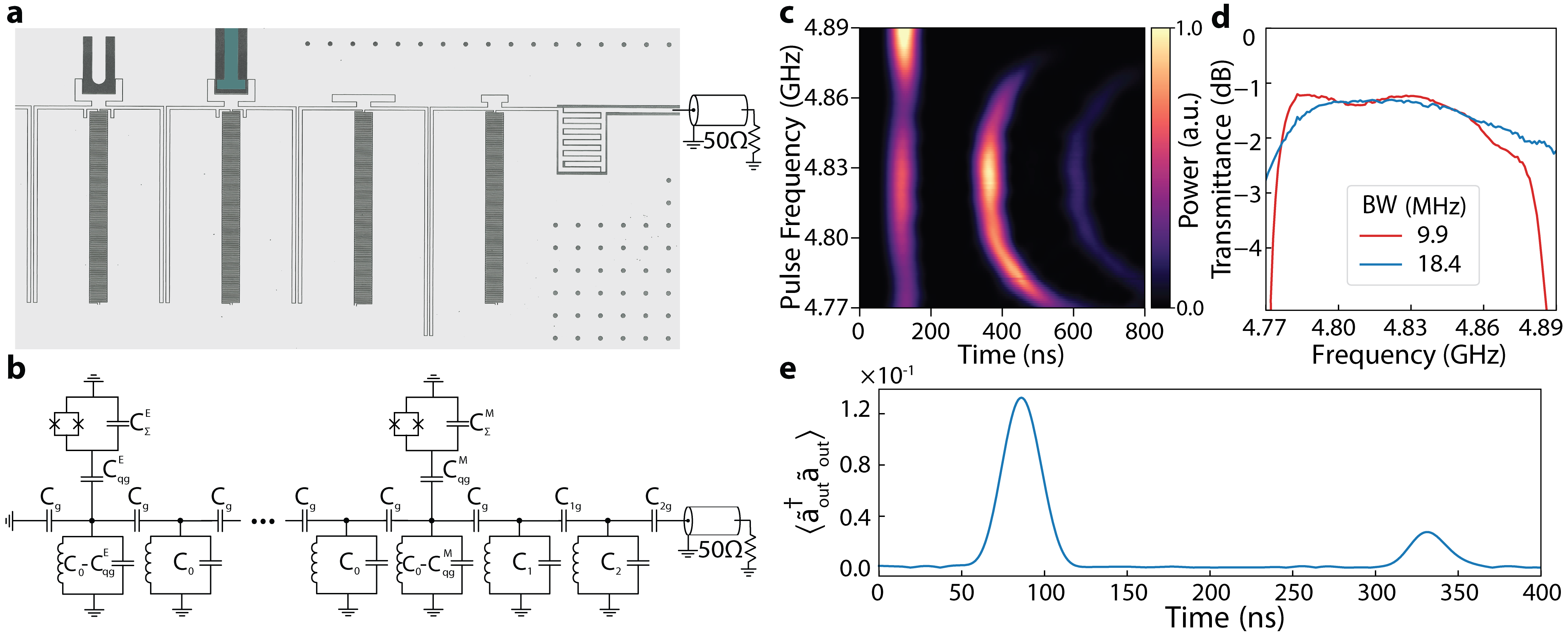}
\caption{\textbf{Metamaterial Slow-Light Waveguide Characterization} \textbf{a}, False-colored optical image of the end of the slow-light waveguide that is connected to the CPW output waveguide, including the ``tapered" boundary matching circuit consisting of the last two resonators. The mirror qubit shunt capacitance is false colored in green \textbf{b}, Full circuit model of the slow-light waveguide and coupled qubits. \textbf{c}, Transient response of slow-light waveguide with narrow-band input pulses of frequencies near around the passband. See Appendix text for further details. \textbf{d}, Transmittance of the tapered end of slow-light waveguide, calculated from the data in subfigure \textbf{c} and a separate measurement of round-trip loss. \textbf{e} Photon flux of emitter qubit emission, measured for 400 ns. After one round-trip delay of the slow-light waveguide, the initially non-transmitted portion of the qubit's emitted pulse can be observed.}
\label{fig:slowlight}
\end{figure*}

As discussed in the 2D cluster state generation protocol proposed in ref \cite{pichler2017universal}, one of the dimensions of the resultant cluster state is limited by the number of photons that can be held in the delay line simultaneously, necessitating a delay line with a sufficiently large round trip time $\taud$. In this work, we realize such a delay line via implementation of a slow-light waveguide (SLWG), which provides large group delay for time-delayed feedback. In addition, the SLWG also provides spectral constriction of propagating modes to a passband with a finite bandwidth, where the photonic density of states (DOS) sharply decreases at the bandedges and is negligible outside the passband, thus enabling selective emission of the $Q_E$'s $\ketf \xrightarrow{} \kete$ transition, as discussed in the main text. The SLWG is physically realized as a periodic array of capacitively coupled lumped-element superconducting microwave resonators, with low resonator loss and negligible resonator frequency disorder, as was demonstrated in our prior work \cite{ferreira2021collapse}. It can be shown that such a design allows for large group delay per resonator $\sim \frac{1}{2J}$, where $J$ is the photon hopping rate between adjacent resonators, as well as strong emission of transmon qubits only at qubit frequencies within the SLWG passband.

The SLWG is implemented by periodically placing $N = 50$ unit cells across the device as seen in Fig.\ref{fig:Device}b, where a unit cell consists of a lumped-element resonator realized with tightly meandered lines providing the majority of the inductance, wider rectangular features providing the majority of the capacitance, and with capacitive coupling between adjacent resonators achieved via their long capacitive wings, as shown in  Fig.\ref{fig:Device}c. At the output side of the SLWG, the Bloch impedance of the SLWG is matched to its output 50 $\mathrm{\Omega}$ CPW via a ``taper section" comprised of two lumped element resonators, where their coupling capacitances towards the output are gradually increased, and their capacitances to ground are correspondingly gradually decreased to compensate for resonance frequency changes. Crucially, in order to prevent distortion of $Q_E$ photon emission, at the terminated side of the single-ended SLWG a capacitance to ground via a long capacitive wing is placed at the left of the first unit cell resonator (Fig.\ref{fig:Device}c), thus maintaining the resonance frequency of the first resonator to be the same as the frequency of the other resonators, which ensures monotonic emission from $Q_E$ (as observed in separate modeling). 

The corresponding circuit model of the SLWG waveguide coupled to $Q_E$ and $Q_M$ is depicted in Fig.\ref{fig:slowlight}b. In the regime of $C_g \ll C_0$, the dispersion of the SLWG is approximately, 
\begin{equation}
    \omega_k = \omega_p + 2J\cos{(k)}
    \label{dispersion}
\end{equation}
where $\omega_0 = 1/\sqrt{L_0C_0}$ is the resonance frequency of unit cell resonators, $J = \omega_0 \frac{C_g}{2C_0}$, $\omega_p = \omega_0 - 2J$ is the center frequency of the passband, and the passband width is $4J$. To mitigate the deleterious effects in the time-domain shape of emitted photons emerging from the higher-order dispersion \cite{engelen2006theeffect}, a sufficiently large $J$ is required. On the other hand, our requirement for large group delay $\tau_d = \frac{N}{J}$ necessitates a sufficiently small $J$. In order to balance the conflicting requirements of large delay and manageable dispersion, we chose $J = 33.5$ MHz as a target parameter that corresponds to the round-trip delay of $\tau_d= 237$ ns. 


We thus aimed for the following target circuit parameters: $L_0$ = 3.1 nH, $C_0$ = 353 fF, $C_g$ = 5.05 fF, $C_{1}$ = 347 fF, $C_{1g}$ = 8.6 fF, $C_{2}$ = 267 fF, and $C_{2g}$ = 87 fF, yielding $J/{2\pi}$ = 33.5 MHz, $\omega_p/{2\pi}$ = 4.744 GHz, and the requisite impedance matching at the boundary. As seen in Fig.\ref{fig:slowlight}a, for the taper section the increasing coupling capacitances are implemented as longer capacitive wings or interdigitated capacitors, and adjustments to the resonance frequencies are achieved by both shortening the length of the meandered lines and modifying the head capacitances. In addition, the coupling capacitance of $Q_E$ and $Q_M$ to their respective unit cells, as depicted in Fig.\ref{fig:Device}c and Fig.\ref{fig:slowlight}a, were designed to be $2.41$ fF and $5.37$ fF, respectively. This yields the qubit-unit cell coupling $g_{uc} = 38.5$ MHz of $Q_E$ and $g^M_{uc} = 85.6$ MHz of $Q_M$ via the following relation:
\begin{equation}
    g_{uc} = \frac{C^E_{qg}}{2\sqrt{(C_0 + 2C_g)(C^E_{\Sigma} + C^E_{qg})}}\omega_p 
\end{equation}
where $g^M_{uc}$ is obtained by a similar calculation. As discussed in the next subsection of the appendix, these small coupling capacitances lead to large emission rates due to the slow-light nature of the SLWG, where a small group velocity $v_g = \frac{\partial{\omega}}{\partial{k}}$ is commensurate with a large density of states $\sim 1/ |v_g|$, which enhances emission rates. \cite{calajo2016atom, dirac1927quantum}

In order to characterize the SLWG, we investigated the transmittance of the SLWG boundary for an itinerant pulse by sending coherent gaussian pulses of variable carrier frequency through the SLWG IN line and measuring their outgoing intensity at ADC PHOTON after they pass through the device. The measurement result, comprising distinct features separated in time that correspond to different reflection events, is shown in \ref{fig:slowlight}c. First, when the pulses arrive at the SLWG boundary, due to the finite reflectance of the taper section, a fraction of the incident pulse is reflected (and thus does not enter the SLWG) and is measured as the first bright feature in Fig.\ref{fig:slowlight}c. Next, the transmitted fraction of the pulse propagates through the SLWG, completes a round-trip, and arrives at the SLWG boundary again. While a small fraction of the pulse is again reflected due to finite reflectance, most of the energy transmits through the boundary to constitute the second bright feature in Fig.\ref{fig:slowlight}c. Finally, this reflected fraction of the pulse completes a second round-trip, and is found as the last bright feature of in Fig.\ref{fig:slowlight}c. Note that this process continues with more round-trips, while the measured data up to the second round-trip is used for analysis. 

We estimated the transmittance $T$ of the SLWG boundary via comparing the energy contained in the second bright feature $E_2$ and the energy contained in the last bright feature $E_3$, where we define the energy of the ``feature" $E = \int |\langle V(t) \rangle |^2 dt$, where $V(t)$ is the measured voltage at the ADC for a particular ``feature". As discussed, the pulse corresponding to the last bright feature undergoes an additional incidence at the SLWG boundary and an additional round-trip in the SLWG relative to the pulse corresponding to the second bright feature. Thus, we can compare their energies via the following relation:
\begin{equation}
    E_3 = R(1-L)E_2
\end{equation}
where $R = 1-T$ is the reflectance of the boundary, and $L$ is the photon loss during a round-trip. By using $L \approx 0.13$, which is obtained from the measurement of Fig. \ref{fig:Mirror_Catch}, we estimate transmittance $T \approx -1.2$ dB at the center of the SLWG passband. The transmittance, shown in Fig.\ref{fig:slowlight}d, is measured for two different bandwidths of the incident Gaussian pulses, such that the slow pulses (red curve) have approximately the same bandwidth as photon 1 of the generated cluster state (see Fig. \ref{fig:Cluster_State}) and the fast pulses (blue curve) have approximately the same bandwidth as photon 2-4 of the generated cluster state. The difference of the transmittance between the two cases demonstrates the necessity of adjustment of the power calibration scaling factor $G$ of the output chain according to the bandwidth of the photons (see Appendix \ref{App:tomo} for further details). Note that we measured $T$ via the transient response of the SLWG because the transient response more directly captured the SLWG transmissivity for broadband itinerant signals, as well as because the transient response is less susceptible than the steady-state response to the compounding effects of multiple reflection events due to all impedance mismatches at the output of the SLWG and throughout the OUT line.

Additionally, we directly investigated the effect of reflection at the SLWG boundary on photon pulses emitted from the $Q_E$. For this measurement, shown in Fig. \ref{fig:slowlight}e, a photon pulse with bandwidth of 9.8 MHz is emitted from $Q_E$ prepared in the $\ketf$ state via shaped emission. This photon first propagates through the SLWG and is partially transmitted at the tapered boundary due to the finite transmissivity of the taper with transmittance $T$; this transmitted fraction then arrives at the ADC and the photon flux is measured. Meanwhile, the reflected fraction of the photon undergoes an additional round-trip in the SWLG, and thus arrives at the ADC time $t = \taud$ later, as seen in Fig. \ref{fig:slowlight}e. If this returning portion of the photon field interacts with the qubit during subsequent photon emissions, it can lead to qubit control errors as well as an overlap of our desired photon signal with this spurious reflected signal, which leads to measurement errors. Thus, when generating the four photon 2D cluster state presented in Fig. \ref{fig:Cluster_State}, and the 5 photon state presented in Fig. \ref{fig:tetra}, we had to ensure that photon emission did not overlap with the returning reflected portion of previously emitted pulses. Thus, for photons emitted after $t=\taud$ into the generation sequence, their emission time was judiciously chosen to avoid this overlap. This is why there is a gap in time between the measured photon flux of photon 1 and photon 4 in Fig. \ref{fig:Cluster_State}c.

\subsection{Qubits}

\begin{figure}[tbp]
\centering
\includegraphics[width =\columnwidth]{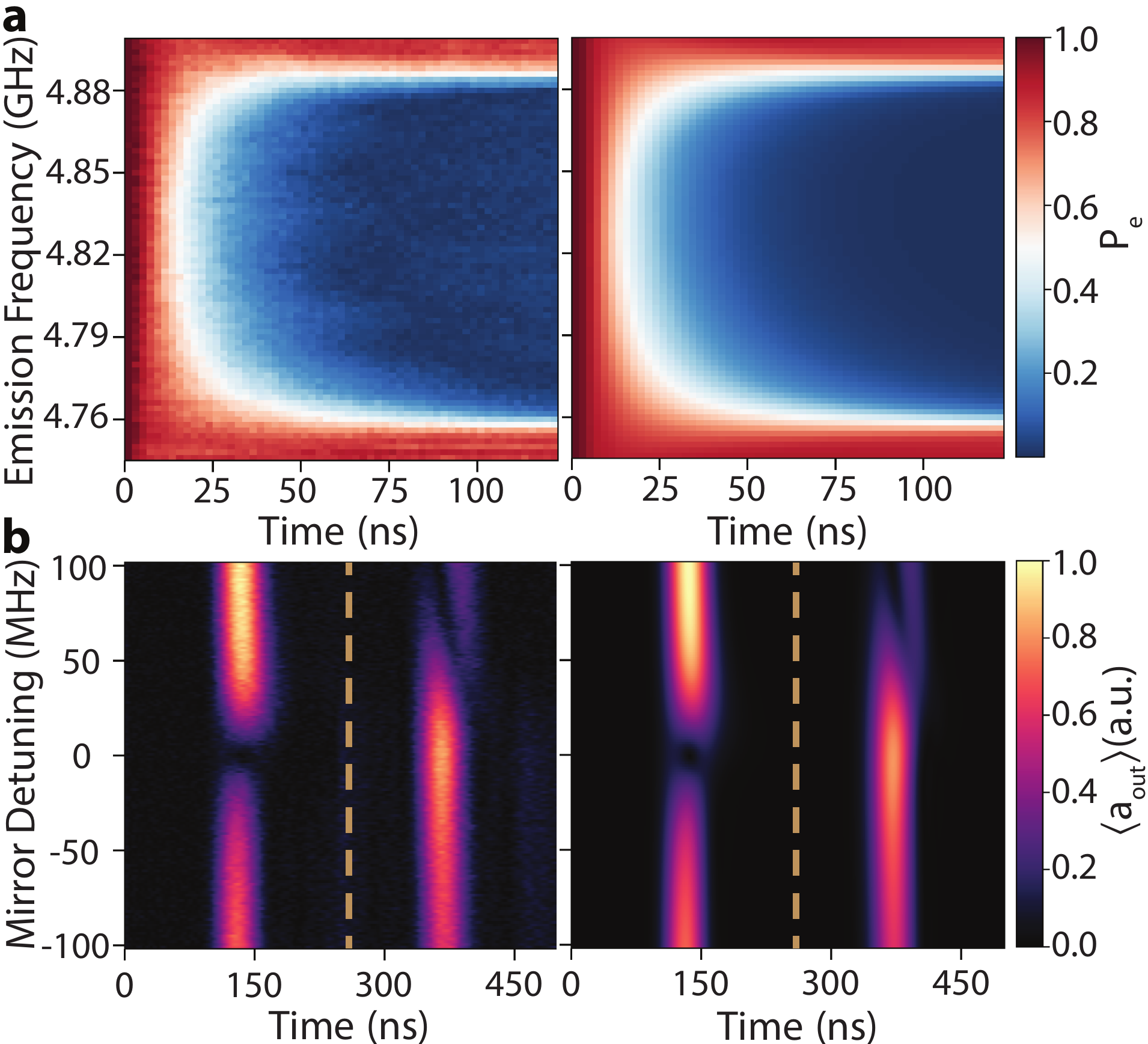}
\caption{\textbf{Emitter and Mirror Qubit Characterization.} \textbf{a}, Left: Measured emission dynamics of $Q_E$ prepared in $\kete$, where emission to the ground state is induced via flux modulation with $\omegamod = 450$ MHz. Right: fit to tight-binding model of equation \ref{tightH}. The fit yields SLWG center frequency of $\omega_p = 4.823$ GHz, $Q_E$ to first unit cell coupling of $g_{uc}/2\pi = 35.16$ MHz, $Q_E$ to second unit cell coupling of $g_{nuc}/2\pi = 2.27$ MHz, and unit cell to unit cell coupling $J/2\pi = 33.96$ MHz. \textbf{b}, Left: Measured averaged field of $Q_E$ emission with different mirror detuning from the center of the passband; right: fit to single-excitation Hamiltonian yielding an effective mirror qubit to unit cell coupling of $g_{uc}^M/2\pi = 57$ MHz. In both cases, the mirror is detuned away from the passband after the time indicated by the dashed yellow line.} 
\label{fig:emfits}
\end{figure}

To characterize the system consisting of $Q_E$ and $Q_M$ coupled to our SLWG, we performed multiple dynamical measurements. The central parameters of the system Hamiltonian, $\omega_p$, $J$, $g_{uc}$, and $g^M_{uc}$ were obtained via fitting the results from these measurements to the expected results from a time-domain simulation of a model Hamiltonian. In the following paragraphs, we discuss how we performed the measurements, and the simulation methods. 

In order to investigate the interaction between $Q_E$ and the SLWG, we measured the decay dynamics of $Q_E$ prepared in $\kete$ interacting with the SLWG, as found in Fig.\ref{fig:emfits}a (left). First, $Q_E$ is prepared in the first excited state $\ket{e}_E$, following which flux modulation of $Q_E$'s transition frequency induces an interaction between a sideband of $Q_E$ and the SLWG. This interaction time (during which the flux modulation is on) is varied, and the sideband frequency is swept across the passband, as indicated on the x- and y-axis of Fig.\ref{fig:emfits}a, respectively. Finally, the interaction is deactivated by turning off the flux modulation, followed by readout of $Q_E$ to measure the remaining population in $\ket{e}_E$. In this experiment, the flux modulation altered the effective qubit-unit cell coupling rate $g^{\text{eff}}_{uc} = \xi g_{uc}$, where $\xi$ is the sideband amplitude. We implemented $\xi = 0.22$ (see Appendix \ref{App:fluxcon} for details on flux modulation) in order to slow down $Q_E$'s intrinsic emission rate, such that we were able to perform time-resolved measurements of $Q_E$'s dynamics without being restricted by the limited sampling rate of our instruments. However, the resulting decay rate is sufficiently strong such that the population in $\ket{e_E}$ completely decays to ground state when the sideband is resonant with the passband of the SLWG, as seen in Fig.\ref{fig:emfits}a.

The measured decay dynamics are fit to the following tight-binding interaction picture Hamiltonian 

\begin{align}
\begin{split}
 \hat{H}^{E} &= (\omega^E_{1} - \omega_p)\ket{e}\bra{e}_E + g^{\text{eff}}_{uc}(\hat{\sigma}^{E}_+\hat{a}_{1} + \hat{\sigma}^{E}_-\hat{a}^\dagger_{1}) \\&+g^{\text{eff}}_{nuc}(\hat{\sigma}^{E}_+\hat{a}_{2} + \hat{\sigma}^{E}_-\hat{a}^\dagger_{2}) + J\sum_{x = 1}^{50} (\hat{a}^{\dagger}_{x}\hat{a}_{x+1} + \hat{a}_{x}\hat{a}^{\dagger}_{x+1})
\label{tightH}
\end{split}
\end{align}

\noindent where $\omega^E_{1}$ is the frequency of the sideband $Q_E$ that is resonant with the SLWG, $\hat{\sigma}^E_{+}$, $\hat{\sigma}^E_{-}$ are the raising and lowering operators of $Q_E$, $\hat{a}^{\dagger}_{x}$, $\hat{a}_{x}$ are the raising and lowering operators of the unit cell resonator at position $x$, $g_{nuc}$ is the parasitic coupling rate of $Q_E$ to the second unit cell resonator, and $\xi = 0.22$ is the sideband amplitude that renormalizes the following coupling rates to $g^{\text{eff}}_{uc} = \xi g_{uc}$ and $g^{\text{eff}}_{nuc} = \xi g_{nuc}$. Note that  $g^{\text{eff}}_{nuc}$ accounts for the asymmetry of the decay dynamics at frequencies near the upper bandedge and the lower bandedge of the SLWG that is observed in the data, as discussed in \cite{ferreira2021collapse}. Also note that the interaction time of $Q_E$ with the SLWG is shorter than $\taud$, and thus the Hamiltonian terms involving the boundary taper resonators of the SLWG and $Q_M$ can be neglected in this model. 

With this Hamiltonian, we simulated the decay dynamics of $Q_E$ initially prepared in $\ket{e}_E$ for various values of  $\omega^E_{1}$, as done in experiment. We restricted the simulation Hilbert space to the vacuum state and single-excitation manifold of the system, which is appropriate for simulation of the decay dynamics. The fit is performed with $\omega_p$, $J$, $g^{\text{eff}}_{uc}$, and $g^{\text{eff}}_{nuc}$ as fit parameters, yielding $\omega_p = 4.823$ GHz, $J = 33.96$ MHz, $g_{uc} = 35.16$ MHz, and $g_{nuc} = 2.27$ MHz, with the simulated dynamics shown in Fig. \ref{fig:emfits}a (right), demonstrating excellent agreement to the data. 

With these parameters, we calculate the intrinsic $\GammaoneD$ of $Q_E$ when it is tuned to the middle of the passband via the formula $2 \gvacMMuc^2 / J$ \cite{calajo2016atom, gonzalez2017markovian}, where $2J$ is the group velocity (per unit cell) in the middle of the passband, while $\gvacMMuc$ also corresponds to the coupling of the qubit to each propagating mode of the passband (note that this formula applies to a qubit end-coupled to a waveguide, while for a side-coupled qubit the effective $\GammaoneD$ is $\gvacMMuc^2 / J$). The dependence of $\GammaoneD$ on $J$ is reflective of the slow-light effect on the emission dynamics of the qubit, where a smaller $J$ leads to a smaller group velocity $v_g = \frac{\partial{\omega}}{\partial{k}}$, which in 1D systems corresponds to a large density of states $1/\pi |v_g|$. Per Fermi's Golden Rule, a large density of states boosts emission rates for a given coupling \cite{dirac1927quantum}. Thus, due to the slow group velocity of the SLWG, we are able to achieve strong emission rates without relying on bulky coupling capacitors of the qubit to the waveguide, and instead achieve sufficient coupling by simply bringing the qubit island within enough proximity to the unit cell of the SLWG. This allows us to hew to the qubit design principles outlined in ref. \cite{barends2013coherent} that ensure high qubit $T_1$. Note that we utilize this value of $\GammaoneD$ for absolute power calibration of measured field amplitudes (see App. \ref{App:gainscaling}).

In addition, the interaction of $Q_M$ with an incident photon pulse as a function of $Q_M$'s frequency was also investigated experimentally. The measurements consisted of emitting a Gaussian photon pulse from $Q_E$ with a bandwidth of 9.8 MHz and carrier frequency $\omega_p$ via shaped photon emission, followed by rapid tuning of $Q_M$'s frequency to the vicinity of the passband after the photon's one-way propagation time of $t=\taud/2$ through the waveguide. This tuning is maintained for the duration of the emitted pulse's interaction with the mirror and then is subsequently turned off. These measurements are performed for various $Q_M$ bias frequencies during the rapid tuning; the measured average SLWG output photon field $\langle a_\text{out} \rangle$ as a function of $Q_M$ frequency (see Appendix \ref{App:tomo} for details on measurement of $\langle a_\text{out} \rangle$) is plotted in Fig. \ref{fig:emfits}b (left).

The transmitted fraction of the photon pulse upon the first incidence at the SLWG boundary is measured as the first bright feature at time $140$ ns. When $Q_M$ is tuned close to the center of the passband (``Mirror ON"), $Q_M$ scatters the photon pulse with large $\GammaoneD$ and thus reflects most of the energy, which is observed as the disappearance of the first bright feature near zero detuning in Fig. \ref{fig:emfits}b. The second bright feature corresponds to the fraction of the photon pulse that was reflected at the SLWG boundary, traveled a round-trip through the waveguide, and subsequently exited the SLWG for measurement. Note that the yellow line in Fig. \ref{fig:emfits}b corresponds to the time when the $Q_M$ fast flux bias is turned off; thus turning off the interaction of $Q_M$ with subsequently incident photon fields. 

The measured data of Fig. \ref{fig:emfits}b are fit to the expected output photon field, which is simulated with the following model Hamiltonian

\begin{align}
\begin{split}
 \hat{H}^{EM} &= \hat{H}^{E}(t)\\&+ \Delta^M(t)\ket{e}\bra{e}_M + g^{M}_{uc}(\hat{\sigma}^{M}_+\hat{a}_{50}+ \hat{\sigma}^{M}_-\hat{a}^\dagger_{50}) \\ &+ \Delta_{1}\hat{a}^{\dagger}_{51}\hat{a}_{51} + \Delta_{2}\hat{a}^{\dagger}_{52}\hat{a}_{52}  + J_1(\hat{a}^{\dagger}_{51}\hat{a}_{52} + \hat{a}_{51}\hat{a}^{\dagger}_{52})
\end{split}
\label{completeH}
\end{align}

\noindent where $\Delta^M(t)$, $\Delta_1$, and $\Delta_2$ are the detunings of $Q_M$, the left taper cell resonator, and the right taper cell resonator from the center of the passband $\omega_p$ respectively, $\sigma^M_{+}$, $\sigma^M_{-}$ are the raising and lowering operators of $Q_M$, and $J_1$ is the photon hopping rate between the taper cell resonators. We replicate the described rapid tuning of $Q_M$ used in the experiment via the Hamiltonian time-dependent term $\Delta^M(t)$. In addition, the external loading of the system to the output $50$ $\Omega$ waveguide is implemented in the model via a dissipation collapse operator in the last taper resonator with rate $\kappa = 148$ MHz (calculated from circuit parameters of the system). $\hat{H}^{E}(t)$ corresponds to the Hamiltonian of equation \ref{tightH} where $\xi$ is time-dependent, which allows us to model shaped photon emission. The envelope of output field $|\langle a_{out}(t)\rangle|$ is obtained in the simulation via taking the time derivative of the accumulated population in the zero-excitation ground state. This output field obtained from the simulation is fit to the measured data by utilizing $\Delta_1$, $\Delta_2$, $g^M_{uc}$, and $J_1$ as fit parameters, yielding $\Delta_1 = -6$ MHz, $\Delta_2 = -70$ MHz, $g^M_{uc} = 57$ MHz, and $J_1 = 45.4$ MHz. The simulated dynamics, shown in Fig. \ref{fig:emfits}b (right), demonstrates excellent agreement to the data. 

In our modeling, the non-zero $\Delta_1$ and $\Delta_2$ fit values account for the asymmetry of the measured photon field at positive detunings of $Q_M$ and negative detunings of $Q_M$ that is observed in the data. Moreover, in our model we do not include parasitic couplings of $Q_M$ to neighboring resonators, and thus any effect of parasitic couplings on the overall $\GammaoneD$ and reflectance of $Q_M$ are incorporated into the one effective coupling rate $g^M_{uc}$. We note that the fitted value of $g^M_{uc}$ is consistent with the amount of transmitted energy from an incident photon that $Q_M$ does not reflect, calculated as 0.02 from the data in Fig. \ref{fig:Mirror_Catch}c; this corresponds to a ``mirror efficiency" of 0.98 as we have defined it. 

\subsection{Purcell Filter}

We perform conventional dispersive readout of the state of our qubits by probing $\lambda/4$ coplanar waveguide resonators that are capacitively coupled to the qubits in the dispersive regime. There is an implicit speed-fidelity tradeoff in such readout schemes due to the Purcell decay of the qubit into the readout lines mediated by the readout resonator to which it is coupled. Reducing the Purcell decay without adding auxiliary circuit components requires reducing the dispersive shift of the cavity, thus reducing readout SNR, or the readout resonator decay rate $\kappa$, thus reducing readout speed \cite{jeffrey2014fast}. The common method for bypassing the implicit speed-accuracy tradeoff of such a readout scheme is to add an extra layer of bath engineering via a Purcell filter that modifies the environmental impedance seen by the qubit-resonator system so as to maintain a desirably large $\kappa$ (for rapid information gain about the qubit state) while simultaneously suppressing decay at the qubit center frequency \cite{jeffrey2014fast,bronn2018high,sete2015quantum}.

A Purcell filter can be modeled by replacing the series impedance of the output CPW seen by the qubit-resonator system with a frequency-dependent environmental impedance $Z_{\text{ext}}(\omega)$. Within such a model the qubit Purcell decay is given by \cite{cleland2019mechanical}:
\begin{equation}
    \begin{gathered}
       \gamma_P^{\text{filt}} = \kappa \frac{g^2}{\Delta^2} \frac{\Re{Z_{\text{ext}}(\omega_q)}}{\Re{Z_{\text{ext}}(\omega_r)}}
    \end{gathered}
\end{equation}
which is just the bare Purcell decay weighted by the ratio of the real impedances of the external load at the qubit and readout resonator frequencies. Thus, by engineering the frequency-dependence of $Z_{\text{ext}}$ to be matched to the output CPW at $\omega_r$, while have negligible real part at $\omega_q$, we can surpress Purcell decay while efficiently probing the readout resonator. 

A coupled resonator array, such as the one we use to implement the SLWG used in our experiment, can be used as a Purcell filter due to its highly flexible impedance properties, allowing for a purely imaginary Bloch impedance at frequencies outside of its passband and a purely real Bloch impedance inside \cite{pozar2005microwave}. Ideally this allows for complete extinction of qubit Purcell decay by placing $\omega_Q$ outside the passband of the array, while maintaining high readout speed by placing $\omega_r$ within the passband. In essence, the coupled resonator array has a large nonzero density of states through which the readout resonator can decay if $\omega_r$ is within the passband, while having no density of states, and thus no available decay channels, at the qubit frequency \cite{ferreira2021collapse}.

\begin{figure}[tbp]
\centering
\includegraphics[width = \columnwidth]{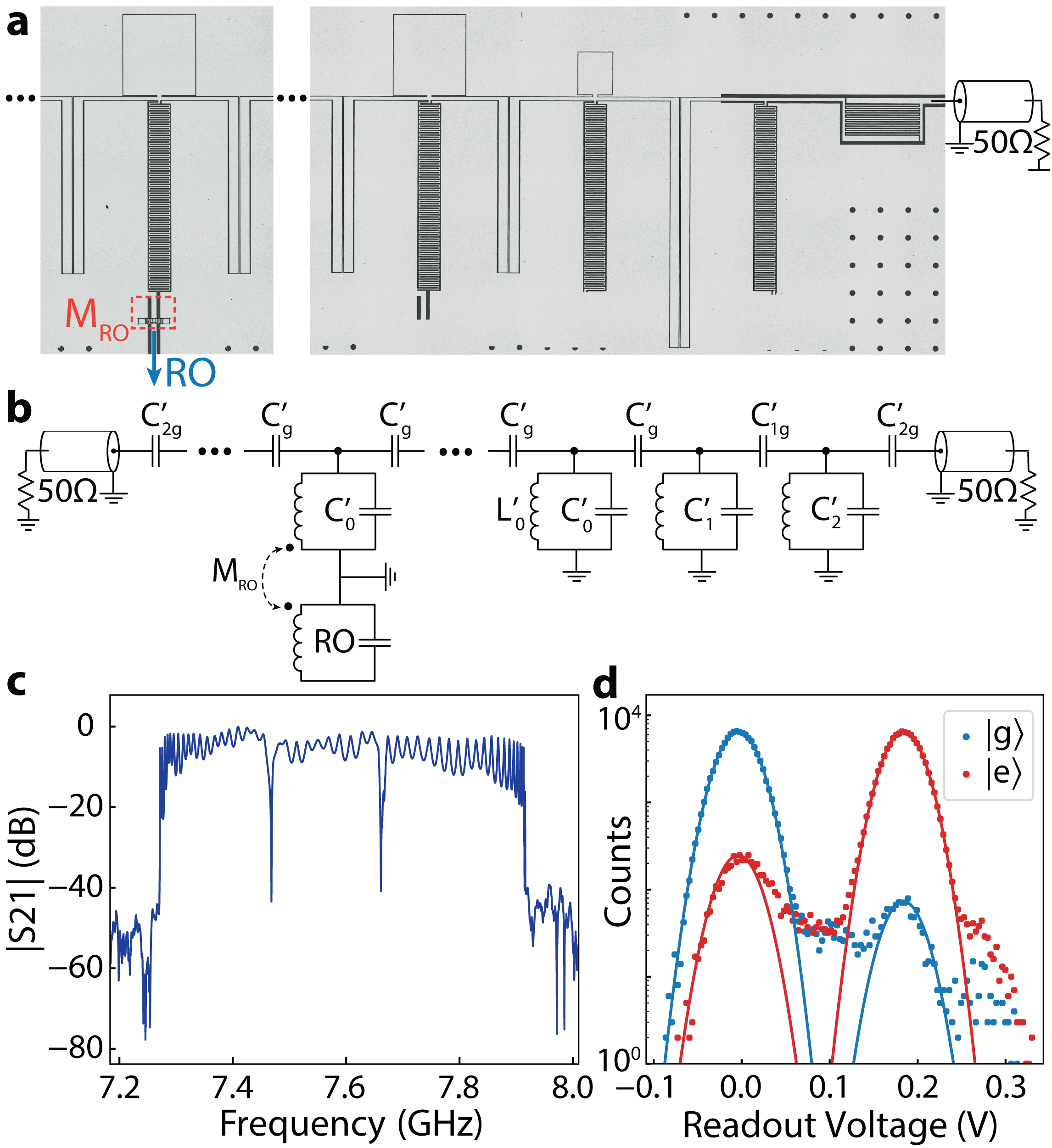}
\caption{\textbf{Purcell Filter Waveguide and Readout Characterization} \textbf{a}, False-colored optical image of the on-chip Purcell filter waveguide. The image on the right depicts one end of the Purcell filter waveguide, while the image on the left depicts the unit cell inductively coupled to one of the CPW readout resonators of the qubits \textbf{b}, Full circuit model of the purcell filter and an inductively coupled readout resonator. \textbf{c}, Transmission spectrum of the full purcell filter waveguide with two side-coupled readout resonators $R_E$ and $R_M$. \textbf{d}, Log-linear raw histogram of single-shot readout measurement results for 100,000 ground state preparations and 100,000 excited states preparations. Solid lines are fits to a bimodal normal distribution. Readout fidelity = 97.6\% was obtained from this histogram.} 
\label{fig:purcell}
\end{figure}

To that end, we engineered a Purcell Filter Waveguide (PFWG) serving as a single Purcell filter for the two readout resonators of both the emitter and mirror qubits. The Purcell filter, which can be seen in Fig. \ref{fig:purcell} a. and b., extends between two ports of our device and replaces the usual CPW readout lines to which readout resonators are coupled to on chip. It has the same circuit topology as the SLWG designed for the cluster state generation scheme, comprising 54 lumped element resonator unit cells coupled to their nearest neighbors capacitively. Referring to the model of Fig. \ref{fig:purcell}b., the following circuit parameters: $L'_0 = 1.2$ nH, $C'_0 = 323.5$ fF, $C'_g = 19.5$ fF, $C'_1 = 315.0$ fF, $C'_{1g} = 27.4$ fF, $C'_2 = 218.2$ fF,  $C'_{2g} = 126.4$ were targeted for the PFWG using the same principles employed in designing the SLWG.

The transmission spectrum of the PFWG, including the two resonances of the readout resonators, can be seen in Fig. \ref{fig:purcell} c. The passband of the PFWG is situated from $7.24$ GHz to $7.9$ GHz so as to safely  encompass the resonances of both $Q_E$ and $Q_M$ readout resonators centered at $\omega_{RM} \sim 7.4$ GHz and $\omega_{RE} \sim 7.7$ GHz respectively, while excluding the entire tuning ranges of the qubits and the frequency of a pump tone at $\sim 7.95$ GHz used for driving a Josephson Travelling Wave Parametric Amplifier (TWPA) for output signal amplification.

The readout resonators are inductively side-coupled to the PFWG by bringing the current antinode of the $\lambda / 4$ resonator into close proximity to the grounded end of a unit cell's meander trace, as can be seen in Fig \ref{fig:purcell}a, for a target resonator decay rate of $\kappa = 10$ MHz. Due to geometric constraints each resonator was coupled to one of the eleventh unit cells of the PFWG counted from its ends. Note that we chose inductive coupling to the PFWG via the current antinode of the resonator because that afforded strong coupling to the PFWG, while still allowing for capacitive coupling to the qubit at the resonator's charge antinode.

The readout-unit cell coupling strength was adjusted in design by changing the distance between the last airbridge of the readout resonator and the current antinode of the resonator near the meander trace of the PFWG unit cell. Moving the airbridge closer to the coupling point reduces the overall strength of the inductive coupling, while moving it away increases the strength. We believe the presence of the airbridge screens the extent of magnetic fields generated by the current near the coupling point and thus reduces the overall overlap volume of fields generated by the resonator and the PFWG unit cell. The fabricated resonator decay rate was found to be approximately $\kappa \sim 11$ MHz. Moreover, the dispersive shift of the readout resonator was measured to be $2 \chi = 4.2$ MHz for a qubit-readout resonator detuning of $\Delta = 1.45$ GHz, yielding a qubit-resonator coupling strength $g \sim 140$ MHz that agrees well with the design value. The measured Purcell-protected $T_1$ time of $Q_E$ at its upper sweet spot was measured to be $20 \mu s$, which is more than one order of magnitude larger than what would be expected in the absence of a Purcell filter; we believe this $T_1$ is ultimately limited by sample loss.

In order to optimize $Q_E$'s single shot readout, we first found the readout probe pulse carrier frequency and length that maximized the complex voltage contrast between the readout transmission when $Q_E$ was initialized to either $\ketg$ or $\kete$. Due to the distorting effects of the ripples in the PFWG transmission spectrum, the optimal frequency of the readout probe tone was found empirically. We also chose the optimal readout power by maximizing contrast while avoiding any powers that led to spurious features in the 2D single shot readout signal histograms in the IQ plane (which we attributed to readout-induced qubit transitions). To characterize the readout fidelity we prepared $Q_E$ in either the $\ketg$ or $\kete$ state, and measured histograms of demodulated single shot signals resulting from probing the readout resonator. These histograms were fit to a double-Gaussian model seen in Fig \ref{fig:purcell} d. from which a ground-excited discrimination boundary was determined. The readout fidelity with respect to this discrimination boundary was found to be 97.6\%; this high single shot readout fidelity was an important resource for the joint qubit-photon measurements required for the quantum process tomography of the $CZ$ gate used in the cluster state generation protocol.

\section{Flux Control for Shaped Photon Emission and Qubit-Photon CZ Gate}
\label{App:fluxcon}

As alluded to in the main text, sophisticated flux control techniques for dynamical control of the qubit frequency were critical in achieving both shaped photon emission as well as a high fidelity qubit-photon CZ gate. Below we present a summary of the techniques we employed in order to achieve distortion free square flux pulses at $Q_E$ and $Q_M$, and precise control of the time-dependent coupling between $Q_E$ and the SLWG via flux modulation.  

\subsection{Distortion  Pre-Compensation of Square Flux Pulses}

\begin{figure}[tbp]
\centering
\includegraphics[width = \columnwidth]{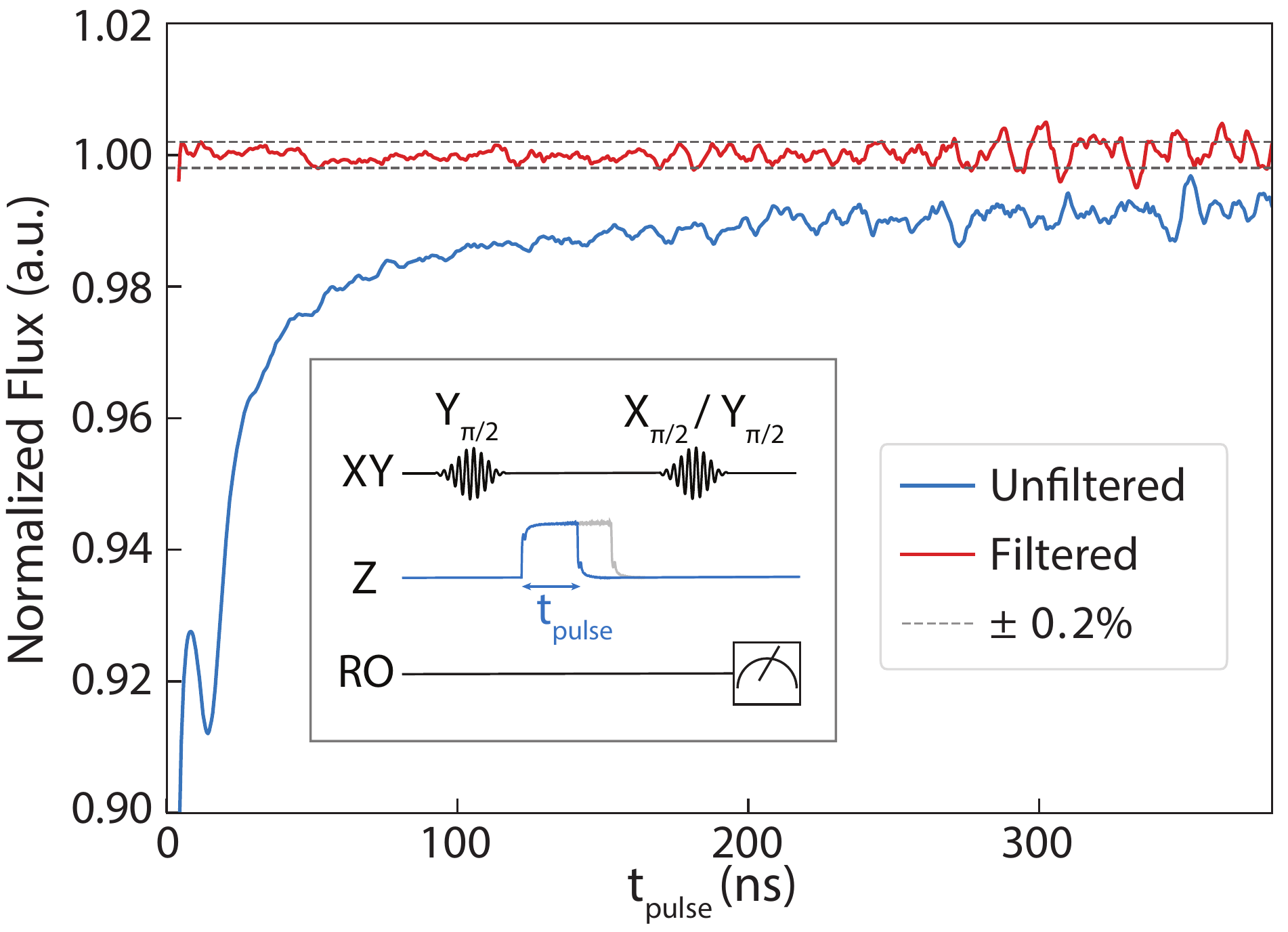}
\caption{Reconstructed step response of flux line with and without ``Cryoscope" distortion pre-compensation. The pulse sequence used for reconstructing the step response is illustrated as an inset; see Appendix text for further details.} 
\label{fig:cryoscope}
\end{figure}

Contributions from dilution refrigerator wiring to signal distortions from flux control lines are often temperature dependent, necessitating techniques for \textit{in situ} characterization of such distortions via the controlled qubit itself. We used the so-called ``Cryoscope" technique \cite{rol2020time}, consisting of Ramsey-type measurements to reconstruct the step-response of the flux line followed by iterative digital pre-compensation, to mitigate distortion in our $Z_E$ and $Z_M$ lines. With pre-compensation, we achieved a desired flat step response within $\pm 0.2\%$ of error, as depicted in Fig.\ref{fig:cryoscope}, for both qubits. The qubit measurements undertaken to reconstruct the step response of the flux line are shown in the inset of Fig.\ref{fig:cryoscope}. We refer the reader to ref. \cite{rol2020time} for a detailed description of the entire ``Cryoscope" process, and discuss small modifications to what is presented in ref. \cite{rol2020time} below.

Firstly, we observe that we did not require real-time digital filtering given that our pulse sequences were only $\sim$ 500 ns in length, and thus chose to use pre-compiled waveforms in order to have more computational flexibility for pre-distortion. Additionally, we note that we observed residual long-time transient responses when applying pre-compensated flux pulses, as discussed in ref. \cite{johnson2011controlling}. To address this problem, rather than waiting for decay of the transient response, a negative copy of the flux signal is appended at the end of every sequence. 

Moreover, when obtaining the reconstructed step response, we found it useful to digitally filter the $\langle X \rangle (t) + i\langle Y \rangle (t)$ data, in order to eliminate data contributions from phase errors in the gates or population offsets, which manifest themselves as spurious features in the spectrum of the data. Moreover, we apply oscillating decaying exponential IIR filters of the form $1 + Ae^{-t/\tau_{\text{IIR}}}\cos{(\omega_{\text{IIR}}t + \phi_{\text{IIR}})}$ in addition to solely decaying exponential IIR filters to achieve better pre-compensation. Finally, for the FIR short-scale precompensation, we mention that it is important to include the smoothing effects of the Savitzky-Golay filter in calculation of the predicted signal from the optimized FIR coefficients.

\subsection{Photonic Pulse Shaping}

As described in the main text, it is important to properly control the time-domain shape of emitted photon pulses in order to ameliorate the effects from the SLWG's non-linear dispersion and to improve the fidelity of the qubit-photon CZ gate. Arbitrary photon pulse shapes can be achieved by controlling the time-dependent decay rate of the $Q_E$, which necessitates a tunable interaction between $Q_E$ and the SLWG. 

\begin{figure}[tbp]
\centering
\includegraphics[width = \columnwidth]{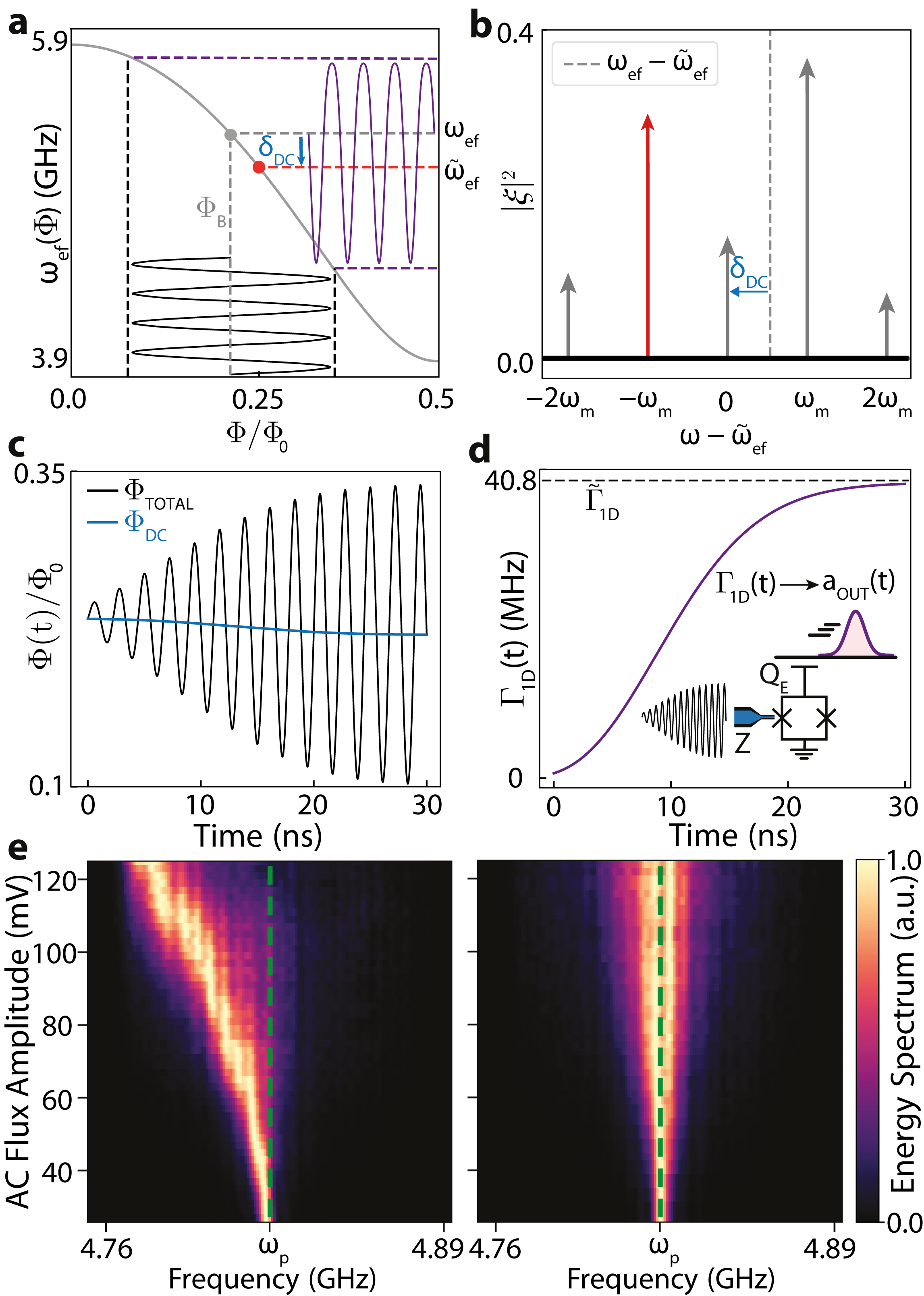}
\caption{ \textbf{Pulse Shaping via Flux Modulation of the Emitter Qubit} \textbf{a}, Illustration of flux modulation of the emitter qubit, when the qubit has a static flux bias $\Phi_B = 0.234\Phi_0$ and the AC flux amplitude $\Phi_{AC} = 0.152\Phi_0$, where $\Phi_0$ is the magnetic flux quantum. The gray curve is the $\omegaef$ emitter qubit tuning curve. The black line depicts typical flux modulation amplitudes in terms of flux quanta, while the purple curve depicts qubit frequency as a function of time under flux modulation. $\Tilde{\omega}_{ef}$ corresponds to the average qubit frequency under flux modulation. \textbf{b} Sideband spectrum of the emitter qubit under flux modulation. $|\xi|$ refers to sideband strength. The red colored arrow corresponds to the sideband used to effect emission into the SLWG in our experiment. \textbf{c} Flux-modulation waveform used in experiment to generate high-bandwidth photons 2, 3, and 4. The blue curve corresponds to a dynamic DC correction that is used to maintain the emission frequency constant; see text for details. \textbf{d}, Effective $\GammaoneD(t)$ obtained from the flux modulation waveform shown in \textbf{c}. \textbf{e}, Energy spectra of qubit emission for a constant flux modulation amplitude that is swept. Left: Energy spectra without DC shift compensation. Right: Energy spectra with DC shift compensation, where DC shifts are calculated theoretically assuming an specific insertion loss in the flux line for the used $\omegamod$.} 
\label{fig:sidebandcon}
\end{figure}

For flux-tunable transmon qubits, such tunable interaction can be attained via sinusoidal flux modulation of the qubit frequency (depicted in Fig.\ref{fig:sidebandcon}a) which induces a sideband-mediated interaction with the SLWG whose strength is controlled by the amplitude of the flux modulation AC flux drive \cite{beaudoin2012first, strand2013first, silveri2017quantum}. In this work, we utilize amplitude modulated AC flux pulses to dynamically control the sideband interaction strength between $Q_E$ and the SLWG, thereby achieving shaped photon pulses. In the following paragraphs, we discuss the theory and technical details of our flux modulation based pulse shaping technique. 

We first review the underlying physics of flux modulation by analyzing the following Hamiltonian of a qubit coupled to a waveguide 

\begin{equation}
\hat{H} = \frac{1}{2}\omega_{Q}(\Phi(t)) \hat{\sigma}_z + \int_k \omega_k \hat{a}^\dagger_k\hat{a}_k + g_{Q}\int_k (\hat{\sigma}_- \hat{a}^\dagger_k + \hat{\sigma}_+ \hat{a}_k) 
\label{emissionH}
\end{equation}

\noindent where $\omega_{Q}(\Phi(t))$ is the qubit frequency of $Q_E$ as a function of the time-dependent flux $\Phi(t)$, $\omega_k$ is the frequency of a propagating waveguide mode with wavevector $k$, $g_Q$ is the unit cell coupling of $Q_E$, $\hat{\sigma}_{+}$, $\hat{\sigma}_{-}$ are the raising and lowering operators of $Q_E$ (note that in this model, the two levels of the qubit correspond to the $\kete$ and $\ketf$ levels of $Q_E$ that participate in photon emission in our experiment) and $\hat{a}_{k}^\dagger$, $\hat{a}_{k}$ are the raising and lowering operators of mode $k$. By going into the interaction picture by the unitary transformation $U(t) = \mathrm{exp}[-i\int_0^t \frac{1}{2}\omega_{Q}(\Phi(t'))\hat{\sigma}_z dt' - it\int_k \omega_k \hat{a}^{\dagger}_{k}\hat{a}_{k}]$, we arrive at the following interaction Hamiltonian, 

\begin{equation}
\hat{H}_{int} = g_{Q} \int_k e^{-i(\phi(t) -\omega_k t)}\hat{\sigma}_- \hat{a}^\dagger_k  + \text{h.c.}
\label{beforeexpH}
\end{equation}

\noindent where $\phi(t) = \int_0^{t} \omega_{Q}(\Phi(t')) dt'$. Note that $g_Q$ here is independent of $k$, as is the case for a qubit coupled to a single unit cell of an infinite periodic array of coupled resonators \cite{calajo2016atom}. 

Under sinusoidal modulation of external flux $\Phi(t)$, we can write $\Phi(t) = \Phi_{B} + \Phi_{AC} \sin(\omegamod t)$, where $\Phi_{B}$ is the the static flux bias of $Q_E$, $\Phi_{AC}$ is the AC flux ampltiude, and $\omegamod$ is the modulation frequency. The periodicity of the flux signal allows for the $e^{-i\phi(t)}$ term to be expanded by the following Fourier series \cite{didier2018analytical},

\begin{equation}
\hat{H}_{int} = g_{Q} \int_k \sum_{s} \xi_s e^{-i(\tilde{\omega}_{Q} - s\omegamod - \omega_k)t}\hat{\sigma}_-\hat{a}^\dagger_k + h.c.
\label{Hfluxint}
\end{equation}

\noindent where $\tilde{\omega}_{Q}$ is the average of $\omega_Q$, and $\xi_s$ is the Fourier coefficient of the $s$-th term, which we refer to as the ``sideband amplitude". We note that because the tuning curve is non-linear (as depicted in Fig.\ref{fig:sidebandcon}a), sinusoidal flux modulation will result in an average DC shift to the static qubit frequency $\omega_{Q}(\Phi_B)$, which is captured by the term $\tilde{\omega}_{Q}$. Moreover, note that one can obtain the magnitudes of $\xi_s$ by simply taking the Fourier transform of $e^{-i\phi(t)}$, as shown in Fig. \ref{fig:sidebandcon}b for one set of flux modulation and qubit parameters. 

According to the RWA, we expect that only non-fast rotating terms of the Hamiltonian of equation \ref{Hfluxint} would appreciably contribute to the qubit dynamics; hence we seek terms where $\tilde{\omega}_{Q} - s\omegamod - \omega_k \approx 0$. Assuming that the waveguide has a finite bandwidth passband, and that only the first lower sideband ($s = 1$) is resonant with one of the passband modes, we can assume terms involving all other $s$ are fast-rotating terms and discard them. This results in the final Hamiltonian 

\begin{equation}
\hat{H}_{mod} = g_Q\xi\int_k e^{-i(\omega^1_{Q} - \omega_k)t}\hat{\sigma}_-\hat{a}^\dagger_k + \text{h.c.}
\label{fluxmodH}
\end{equation}

\noindent where $\omega^1_{Q} \equiv \tilde{\omega}_{Q} - \omega_m$ is the frequency of the first lower sideband, and $\xi \equiv \xi_1$. Thus, $\omega_k = \omega^1_{Q}$ will be the center frequency of emission, while photon emission will also occur at surrounding frequencies where $\omega^1_{Q} - \omega_k$ is small; thus imparting a finite bandwidth to any emitted photon. Note that the resultant Hamiltonian is in an equivalent form as Eq.~(\ref{beforeexpH}) up to a renormalization of the effective coupling rate. Thus, we can tune the strength of interaction between $Q_E$ and the SLWG by controlling the sideband amplitude $\xi$ and locating the first lower sideband inside the passband.

We are able to accurately predict the necessary AC flux drive amplitudes to achieve desired values of $\xi$ by numerical calculation of the ``sideband spectrum" of $Q_E$ under flux modulation. Remembering that we are concerned with emission from the $\ketf$ state, for this calculation, we require the functional form of the qubit tuning curve  $\omegaef (\Phi)$, the static flux bias $\Phi_B$, and the strength of the sinusoidal flux drive $\Phi_{AC}$.  We adopted the analytical form of the transmon tuning curve from ref \cite{didier2018analytical} for accurate calculation of $\omega_{ef}$ as a function of $\Phi$, as depicted in Fig.\ref{fig:sidebandcon}a (gray solid line), from measurement of the highest qubit frequency, the lowest qubit frequency, and the anharmonicity $\eta$ at the highest qubit frequency. 

The sideband spectrum is calculated via the  Fourier Transform of $e^{-i\phi(t)}$, with an example shown in Fig. \ref{fig:sidebandcon}b, where the $s=1$ sideband is highlighted in red. The spectrum yields the different $\xi_s$, as well as the average ``DC shift" of the qubit frequency $\delta_{DC} \equiv \tilde{\omega}_{Q}-\omega_{Q}(\Phi_B)$ which depends on both $\Phi_B$ as well as $\Phi_{AC}$. We can leverage this DC shift effect to obtain a mapping from AC flux amplitude at the qubit to input AC voltages to the fridge, as illustrated in Fig.\ref{fig:sidebandcon}e. By inducing $Q_E$ emission via flux modulation at various AC input voltages and measuring the carrier frequency of emitted photons, we observe the average DC shift of the qubit frequency via the changing carrier frequency of emitted photons. By comparing the change in photon carrier frequency to numerical predictions of $\delta_{DC}$, we can obtain the scaling factor for converting input AC voltages to $\Phi_{AC}$ at the qubit. Meanwhile, note that we obtain a similar scaling factor for converting static DC bias voltages to $\Phi_B$ at the qubit via measurements of the qubit tuning curve (note that the two scaling factors are different due to differing DC and AC losses of the flux line). 

Thus, we can achieve a desired time-dependent coupling between $Q_E$ and the SLWG via flux modulation, by effecting a time-dependent $\xi(t)$ via some specific $\Phi_{AC}(t)$. However, a time-dependent $\Phi_{AC}(t)$ will also lead to a time-dependent $\delta_{DC}(t)$, which necessitates a ``DC correction" signal to maintain the emission frequency constant. Therefore, we obtain the necessary flux drive $\Phi(t) = \Phi_{DC}(t) + \Phi_{AC}\sin(\omegamod t)$ that achieves a desired $\xi(t)$ while maintaining a constant emission frequency. This is achieved by considering a suitable range of AC flux amplitudes, and obtaining associated $\Phi_{DC}$ correction flux biases for each flux amplitude such that for a given $\Phi_{AC}$, overall static qubit bias $\Phi_B$, and the flux amplitude dependent correction bias $\Phi_{DC}$, the average qubit frequency $\tilde{\omega}_{ef}$ will be equal to $\omegaef (\Phi_B)$; see Fig. \ref{fig:sidebandcon}e (right) for demonstration of this DC correction procedure for various AC flux amplitudes. Then, the sideband amplitudes $\xi(\Phi_B, \Phi_{AC}, \Phi_{DC})$ are numerically calculated for each set of the aforementioned parameter values, with which a desired $\xi(t)$ can be mapped to the necessary $\Phi(t)$ signal; see Fig. \ref{fig:sidebandcon}c,d for an example. Finally, we note that under the flux drive $\Phi(t)$, the time-dependent decay rate $\Gamma_{1D}^{ef}(t)$ of $Q_E$ will be equal to $\Gamma_{1D}^{ef} \cdot |\xi(t)|^2$, where $\Gamma_{1D}^{ef}$ is the intrinsic decay rate of the $\ketf$ state given by $\sim 4g_{uc}^2/J$.

\begin{figure}[tbp]
\centering
\includegraphics[width = \columnwidth]{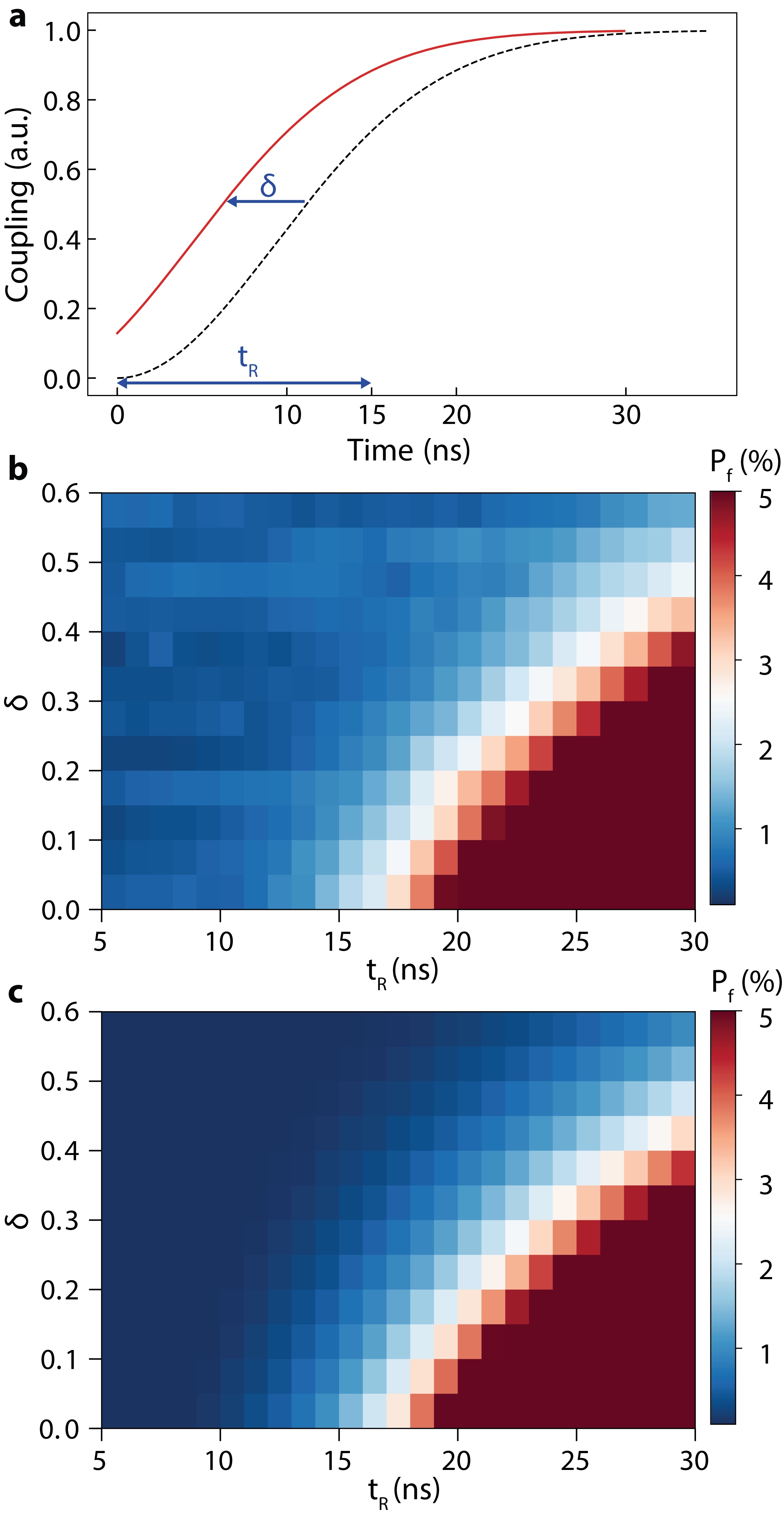}
\caption{\textbf{High Bandwidth Photon Emission} \textbf{a}, Shape of time-dependent coupling effected by flux modulation, given by the square of the erf function: $\left| \frac{2}{\sqrt{\pi}}\int^{t/t_R + \delta}e^{-t'^2} dt' \right|^2$, where $t_R$ and $\delta$ scale and shift the erf function with respect to time, respectively. \textbf{b} Residual $\ketf$ state population measurement after photon emission via flux modulation with various $t_R$ and $\delta$ parameters. \textbf{c} Simulation of the experiment done in part \textbf{b}, using the model of equation \ref{tightH}.}  
\label{fig:respop}
\end{figure}

As discussed in the main text, we sought to emit Gaussian shaped photons for our cluster state generation sequence, as illustrated in Fig. \ref{fig:Emission}. We observed, both numerically and experimentally, that shaped photons with Gaussian spectra could be emitted by realizing the following sideband amplitude time dependence $\xi(t)$:

\begin{equation}
    \begin{gathered}
    \xi(t) = \xi_M \text{erf}^2(\frac{t}{t_R} + \delta) \\
    \text{erf}(t) \equiv \frac{2}{\sqrt{\pi}}\int_0^t e^{-t'^2}dt'
    \end{gathered}
\label{xishape}
\end{equation}
\noindent where $t_R$ scales the erf function with respect to time, $\xi_M$ is the maximum attainable sideband amplitude at a given $\Phi_B$, and the second line defines the erf function whose square increases from $0$ and converges to $1$ smoothly. The spectral bandwidth of the resultant Gaussian pulse is controlled by $t_R$, where slow (fast) increase of $\GammaoneD(t)$ due to large (small) $t_R$ leads to small (large) bandwidth. Moreover, the $\delta$ parameter shifts the entire function with respect to time, such that it reduces the time needed to reach the maximum sideband amplitude for a given emission time and $t_R$; this is useful to further suppress residual $\ketf$ population after emission for short emission times. This parametrized time dependence is plotted in Fig.\ref{fig:respop}a. 

For the photonic pulses shown in Fig. \ref{fig:Cluster_State}4c, photon 1 was generated by realizing the time-dependent sideband amplitude $\xi(t)$ of equation ~\ref{xishape} with parameters $t_R = 50$ and $\delta = 0$, yielding a Gaussian pulse with 9.9 MHz bandwidth. However, for photons 2,3,4 we chose to utilize a finite $\delta$ in order to achieve a small $\ket{f}$ residual $\ketf$ population for the photons' short 30 ns emission time. In order to obtain the best $\delta, t_R$ combination, we modeled and measured experimentally this residual population after photon emission for a range of $\delta, t_R$ values, as depicted in Fig.\ref{fig:respop}b,c (see App. \ref{App:device} for modeling details). We found that the combination $t_R = 15$ ns, $\delta = 0.33$, suppresses residual $\ketf$ state population below $1\%$ and constricts emitted photon pulses to a short time-bin measurement window, and we chose this parameter combination for emission of photons 2,3,4 depicted in Fig. \ref{fig:Cluster_State}c. We note that while higher $\delta$ values in general result in less residual $\ketf$ population, large $\delta$ values also lead to distortions in the emitted Gaussian pulse; thus the parameter choice $t_R = 15$ ns, $\delta = 0.33$ strikes a balance between minimizing residual $\ketf$ state population and maintaining the approximately Gaussian shape of the emitted pulse with 17.9 MHz bandwidth.

We also note that we realize fast unconditional reset using flux modulation, where the $\kete$ and $\ketf$ state populations are emptied via induced photon emission. First, a constant flux modulation signal that induces emission of $\ketf$ to $\kete$ is applied to $Q_E$. Next, another constant flux modulation signal that induces emission from $\kete$ to $\ketg$ is applied to $Q_E$, bringing it to the ground state. Lastly, we wait approximately 3 $\mu$s after this reset before starting qubit control, in order to allow residual emitted fields trapped in the SLWG due to finite taper reflections to fully leave the waveguide. We note that this reset protocol effectively thermalizes $Q_E$ to the SLWG temperature; indeed we confirm via separate measurements that the resultant $\kete$ thermal population is $\sim 1\%$ (corresponding to an effective $\sim 50$ mK temperature). Using this unconditional reset protocol, we generate the 2D cluster state with a conservative repetition rate of 100 kHz.


\section{Radiation Field Quantum State Tomography}
\label{App:tomo}

In order to characterize generated multipartite entangled photonic states, we utilize tomography methods for itinerant microwave photons that were pioneered in Circuit QED systems by Eichler et. al. \cite{eichler2013thesis, eichler2011experimental} with suitable modifications when appropriate for us. Below we present a detailed summary of our entire data analysis and tomography procedure. We conclude by presenting additional photonic quantum state preparation and tomography results not presented in the main text.

\subsection{Measurement of $\aoutt$ and $\fluxoutt$ }

The average photon flux $\fluxoutt$ and average field $\aoutt$ of emitted photons are routinely measured in our experiment for the purposes of characterizing our shaped photon emission procedure, characterizing different aspects of our time-delayed feedback process, and obtaining mode matching functions $f(t)$ for the different photonic qubits in order to obtain time-independent statistics from their time-dependent fields (see rest of Appendix text for further details). Measurement of both of these quantities for an emitted photon starts with heterodyne measurement of both quadratures, $I(t)$ and $Q(t)$, of its time-dependent microwave field, via the output chain described in Appendix \ref{App:Fab_Meas}. Many measurements are performed and their results are averaged to compute the average field $\langle V(t) \rangle = \left \langle I(t) + iQ(t) \right \rangle$ and the the average photon flux $\langle V^2(t) \rangle = \left \langle |I(t)|^2 + |Q(t)|^2 \right \rangle$. Note that the calculation of $\langle V^2(t) \rangle$ results in a signal without any carrier frequency, but $\langle V(t) \rangle$ retains the carrier frequencies of $I(t)$ and $Q(t)$ which must be removed by digital demodulation.

Due to spurious DC shifts in the detection set-up, as well as imbalance and LO bleedthrough in the downconversion IQ mixer, the spectrum of $\langle V^2(t) \rangle$ and the demodulated $\langle V(t) \rangle$ will have spurious features outside of the baseband signal. In addition, these band-limited baseband signals will also have significant noise outside of their bandwidth. These undesirable features serve to obscure the time-dependent shape of the baseband signal that we wish to measure, and we remove them through digital low-pass filtering, with filter bandwidth set to be sufficiently high to capture all of the baseband signal. At this point, the resultant demodulated and filtered $\langle V(t) \rangle$ signal, followed by suitable normalization, can already be used as the mode-matching function $f(t)$, where the normalization is such that $\int |f(t)|^2 dt = 1$.

Further, with an absolute power calibration of our output chain that maps voltage measured at the ADC to field amplitude at the qubit  (see subsequent subsection for details on how to obtain this calibration), the digitally processed $\langle V^2(t) \rangle$ and $\langle V(t) \rangle$ signals can be suitably scaled to yield the true $\fluxoutt$ and $\aoutt$ in units of photon/s and $\sqrt{\text{photon/s}}$, respectively. However, for the purposes of plotting in this manuscript, we presented these quantities in terms of unitless, normalized values $\left \langle \tilde{a}_{\text{out}} \right \rangle = \aout/\tilde{\Gamma}_{1D}^{-1/2}$ and $\left \langle {\tilde{a}^\dagger}_{\text{out}} \tilde{a}_{\text{out}} \right \rangle = \fluxout/\tilde{\Gamma}_{1D}$, where $\tilde{\Gamma}_{1D}/2\pi = 40.8$ MHz is the maximum expected emission rate of the $\ket{f} \rightarrow \ket{e}$ transition under flux modulation induced emission (see Appendix \ref{App:fluxcon} for further details regarding $\tilde{\Gamma}_{1D})$. Thus, these normalized quantities express the time-dependent photon flux and field as a \textit{fraction} of the maximum expected photon flux and field, respectively, for an excited qubit with emission rate $\tilde{\Gamma}_{1D}$ (as an illustrative example, note that for constant flux modulation with flux amplitude that yields $\Gamma_{1D}^{ef}(t) = \tilde{\Gamma}_{1D}$, $\left \langle {\tilde{a}^\dagger}_{\text{out}} \tilde{a}_{\text{out}} \right \rangle$ at $t=0$ would be equal to 1).

\subsection{Absolute Power Calibration}

In order to perform quantum state tomography via heterodyne detection, we need an absolute power calibration that maps voltage measured at the ADC to field amplitude at the qubit's location on the device, given by some conversion factor $G$. This conversion factor $G$ includes the following contributions: the scaling from the quantum field $a$ to the physical voltage on the device, the gain of the output chain from the first amplifier forward, and the detection efficiency $\eta_{\text{det}}$. We define $\eta_{\text{det}}$ such that $(1-\eta_{\text{det}})$ corresponds to the fraction of the itinerant photon's energy that is lost before it reaches the first amplifier (which in our case is a quantum-limited TWPA), either due to the loss or spurious reflections that are suffered by the photon (see Fig. \ref{fig:slowlight}e for measurement of such reflections). Generically, $G$ is obtained by measuring a signal at the ADC whose power at the qubit can be independently verified. In our work, we rely on measurement of the AC Stark shift of the qubit frequency induced by an input pulse on the SLWG as our method for power calibration.

The procedure for obtaining $G$ via AC Stark shift measurements is the following. The qubit frequency $\omegage$ is detuned from $\omega_p$ by 740 MHz. A square pulse with carrier frequency $\omega_p$ is sent into the SLWG, for varying input powers. This square pulse induces an AC Stark shift $\Delta^{AC}$ on the qubit frequency whose magnitude is dependent on the SLWG input power; this $\Delta^{AC}$ is measured by determining the resonance excitation frequency of the qubit. The qubit's resonance frequency is measured by applying an excitation pulse to the qubit while the SLWG input pulse is off-resonantly driving the qubit; by sweeping the excitation pulse's frequency, measuring the qubit response, and fitting the resultant lineshape to a Gaussian, we obtain the resonance frequency via the Gaussian's mean. Repeating this procedure for all SLWG input powers, we experimentally obtain the dependence of $\Delta^{AC}$ on the power of the SLWG input pulse. Finally, the amplitude of the SLWG input pulse is measured at the ADC for all input powers used.  

The power dependence of $\Delta^{AC}$ is then fit to the following transmon AC Stark shift model involving five transmon levels \cite{Koch2007}:

\begin{align*}
\label{H_AC}
\hat{H} &= \sum_{j=0}^{N=4} \left (j\Delta + j(j-1)\eta \right ) \ket{j}\bra{j} \\
&+ \frac{\Omega}{2}\left (\sqrt{j+1}\ket{j + 1}\bra{j} + \text{h.c.} \right ) 
\end{align*}

\noindent where $\{j\}$ corresponds to the transmon levels, the SLWG drive frequency is $\omega_p$, $\Delta = \omegage - \omega_p$, the transmon anharmonicity is $\eta/2\pi = -277$ MHz, and $\Omega$ is the Rabi frequency of the SLWG drive. Note that this Hamiltonian is obtained from the full Hamiltonian of a transmon interacting with a classical drive by simply going into the rotating frame of the drive (via the unitary transformation $U = \text{exp} \left[ it\sum_j j\ket{j}\bra{j}\omega_p \right ]$) and discarding counter-rotating terms. Also note that our model includes multiple transmon levels because the presence of multiple transitions, along with their associated anharmonicities, quantitatively changes the theoretically predicted $\Delta^{AC}$. We found that we needed up to five transmon levels for the theoretically predicted $\Delta^{AC}$ to converge for our experimental parameters, whereas beyond five levels changes in the predicted $\Delta^{AC}$ were negligible.

For a qubit coupled to a single-ended waveguide, the Rabi Frequency is given by $\Omega = |\alpha|\sqrt{4\GammaoneD}$ \cite{hoi2013quantum}, where $|\alpha|$ is the field amplitude of the SLWG drive \textit{at the qubit}, and for our slow-light waveguide the qubit's emission rate into the waveguide (at the center of the passband) is given by $\GammaoneD = 2\gvacMMuc^2/J$ \cite{calajo2016atom}. The parameters $\gvacMMuc$ and $J$ of our device were obtained through device characterization experiments described in Appendix \ref{App:device}, and were directly used in this model. From this model Hamiltonian, $\Delta^{AC}$ is numerically calculated in the following manner: first, the Hamiltonian is diagonalized to obtain its eigenenergies. Then, the difference between the energies of the ``dressed" ground state and the ``dressed" excited state is obtained, and by subtracting $\Delta$ from this difference $\Delta^{AC}$ is finally obtained. 

The fit is performed simply by using $|\alpha| = V_{ADC} \cdot G$ in the model, where $V_{ADC}$ is the amplitude of the SLWG input pulse measured at the ADC. By obtaining $\Delta^{AC}$ with $G$ as a fit parameter, the fit of the model to the data is shown as the black curve in Fig. \ref{App:gainscaling}c, showing excellent agreement to the data. The obtained fit parameter $G$ was henceforth used to scale all radiation field voltages measured at the ADC. We note that by using a pulsed measurement, rather than a continuous wave (CW) SLWG input tone, the obtained $G$ more accurately captures the contribution of spurious reflections to the overall $\eta_{\text{det}}$ that is experienced by emitted pulses, and is significantly less sensitive to ripples in the output chain transfer function.

\label{App:gainscaling}
\begin{figure}[tbp]
\centering
\includegraphics[width = \columnwidth]{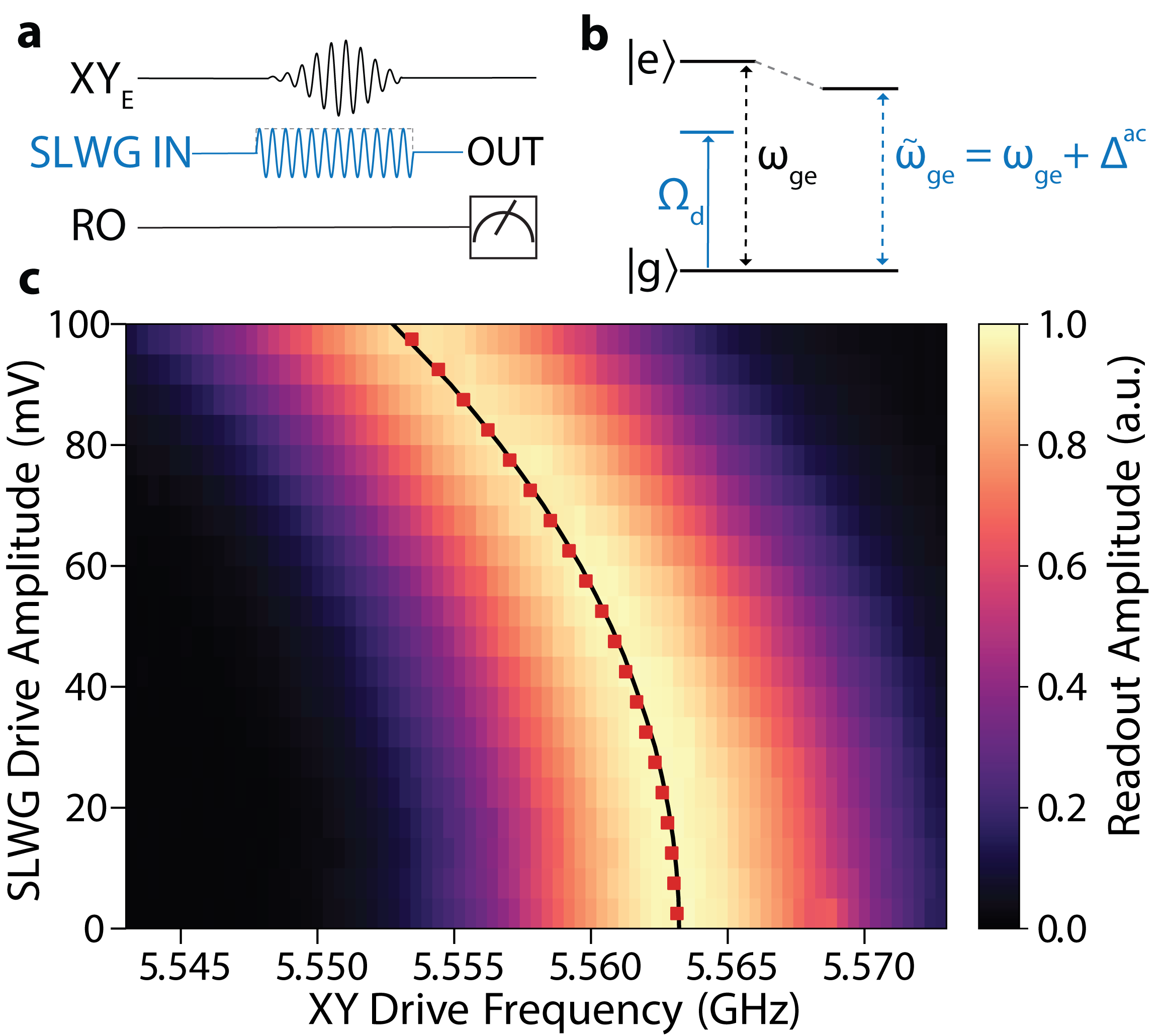}
\caption{\textbf{Absolute Power Calibration via AC Stark Shift of Emitter Qubit}. \textbf{a} Pulse sequence for the AC Stark Shift calibration experiment. An input square pulse with carrier frequency 740 MHz detuned from the qubit $\omegage$ is sent into slow-light waveguide (SLWG) and drives $Q_E$. Simultaneously, a qubit excitation pulse is sent into the $XY_E$ line and arrives at the qubit at the same time the input SLWG pulse is driving the qubit, for the purposes of determining the resonance frequency of the AC Stark Shifted qubit. This experiment is repeated for multiple SLWG input pulse powers. \textbf{b} Level diagram depicting the detuned drive on the qubit from the SLWG, and how that effects an AC Stark Shift of the qubit frequency. \textbf{c} AC Stark Shift measurement data. The square markers are the measured qubit frequencies at different SLWG drive amplitudes, while the black line is a fit to a model of the expected qubit frequency due to off-resonant driving. See the Appendix text for more details. } 
\label{fig:acstark}
\end{figure}

\subsection{Measuring Expectation Values of Radiation Field Moments}

The time-independent quantum statistics of emitted photons can be extracted from single shot measurements of their (properly scaled) time-dependent fields by integration with a suitable mode-matching function $f(t)$. This integration $\int f(t) a_{\text{out}}(t) dt = I+iQ = S$ can be shown to yield single-shot measurements of the complex quantity $S = a + h^\dagger$, where $a$ is the mode of interest, and $h$ is the noise mode of the detection chain. By taking many single-shot measurements, one gains access to the statistics of $a + h^\dagger$, and similarly, one can also perform many single-shot ``dark" measurements of the noise mode $h$ to obtain its statistics. By calculating the expectation values of moments of $a + h^\dagger$ and $h$ from their single shot measurements, the expectation values of moments of $a$ can thus be obtained, which is sufficient to reconstruct the density matrix of the mode of interest.

\begin{figure}[tbp]
\centering
\includegraphics[width = \columnwidth]{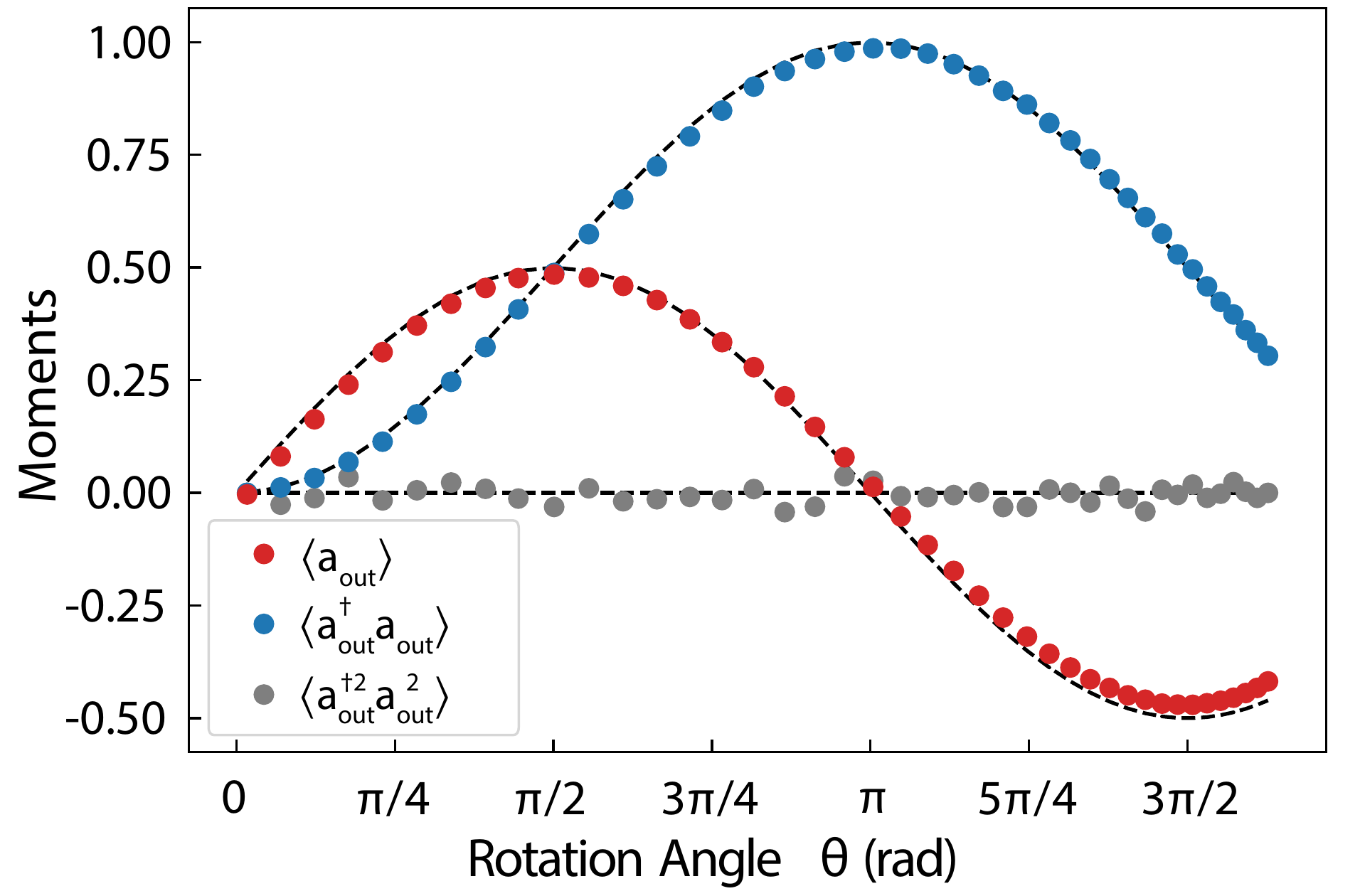}
\caption{Magnitude of measured moments $\langle a\rangle$ (red circles), $\langle a^\dagger a\rangle$ (blue circles), and $\langle(a^\dagger)^2a^2\rangle$ (gray circles), for emitted photon pulses from $Q_E$ when it is prepared in state $\ket{\psi}_{E} = \cos{(\theta/2)}\kete + \sin{(\theta/2)}\ketf$. Dashed black lines are ideal expected moments given a qubit rotation angle $\theta$.} 
\label{fig:ampmoments}
\end{figure}

The procedure described above can be straightforwardly extended to multiple modes. For our experiment, the mode-matching function $f_i(t)$ for each photonic time-bin qubit is obtained by direct measurement of the average pulse shape $\left \langle a_{i,\text{out}}(t) \right \rangle$. This allows for single shot measurements of $S_i = a_i + h_i^\dagger$ for every photon, which are then processed into joint moments $\mathcal{S}$ of the form $\left \langle (S_1^{\dagger})^{n_1} S_1^{m_1} (S_2^{\dagger})^{n_2} S_2^{m_2} ... (S_N^{\dagger})^{n_N} S_N^{m_N} \right \rangle$. Given that our emitter qubit is a single photon source, we take $n_i,m_i \in \{0,1\}$ by assuming that the Hilbert Space of the photonic modes can be restricted to the single-photon manifold subspace. Note that we have experimentally verified the single photon character of our emitted photons (for each time-bin photonic qubit) via measurements of $\left \langle (a^\dagger)^2 a^2 \right \rangle$ for various prepared photonic states, which are plotted in Fig. \ref{fig:ampmoments}. The measured $\left \langle (a^\dagger)^2 a^2 \right \rangle$ moments are close to 0 for all prepared photonic states, corresponding to a vanishing second-order correlation function $g^{(2)}(0)$ at zero time delay. 

In turn, the joint photon moments $\left \langle (a_1^{\dagger})^{n_1} a_1^{m_1} (a_2^{\dagger})^{n_2} a_2^{m_2} ... (a_N^{\dagger})^{n_N} a_N^{m_N} \right \rangle~ \forall~ n_i,m_i \in \{0,1\}$ can be calculated from algebraic formulas involving the measured joint moments $\mathcal{S}$ and the measured moments $\left \langle h_i^\dagger h_i \right \rangle $ under the following simplifying assumptions: the signal modes $a_i$ are uncorrelated from the noise modes $h_i$, the noise modes $h_i$ are not correlated to one another, and complex-valued moments of $h_i$ are taken to be zero. These assumptions are appropriate when the noise modes $h_i$ are in a thermal state, which is typically the case when the main added noise source of the output chain is amplifier noise (note that these assumptions were also verified experimentally). It can be shown that the expectation values of these joint photon moments is sufficient to uniquely reconstruct the density matrix of a multipartite state of $N$ photonic qubits. While algebraic formulas relating the density matrix elements to the joint photon moments can be derived, we instead reconstruct the density matrix of generated photonic states via a maximum-likelihood estimation (MLE) algorithm that uses the obtained joint photon moments as input (for more details, see the next subsection).

\subsection{MLE}
We reconstruct the quantum state of the entangled microwave photons using a maximum-likelihood estimation (MLE) state tomography technique under the following assumptions: (1) the Fock spaces of the bosonic modes representing individual time-bin photonic qubits can be restricted to the single excitation manifold (ie: the Hilbert space spanned by Fock states $\ket{0}$ and $\ket{1}$), and (2) the distributions of the sample means of the moments of the measured photonic fields are well approximated by normal distributions in the case of many samples (ie: the statistical central limit theorem holds for the distributions of the means of these moments).

In statistics a likelihood functional $\mathcal{L}(D|H)$ is a function on a set $D$ of observed statistical data sampled from a system which, assuming some underlying parameterization $H$ of the system, returns a value proportional to the probability that the assumed parameterization would result in the observed data. Thus $\mathcal{L}$ encapsulates how `likely' a set of observations is under certain assumptions on the system. Given a dataset of observations $D$, MLE techniques aim to explore the space of parameterizations of a system to find the one that maximizes a chosen likelihood functional. Our specified goal is to find the quantum state $\hat{\rho}$ of $N$ photons that best approximates the actual photonic state we have prepared with our protocol. Thus to proceed with an MLE approach to this state tomography problem we must identify a dataset $D$ we intend to collect and a likelihood functional $\mathcal{L}(D | \rho)$ over which we will optimize $\rho$.

To identify a sufficient dataset and associated likelihood functional for MLE state tomography of $N$ entangled photons, note that for a single bosonic mode $a$ constrained to the single excitation manifold, it suffices to know the expectation values $\langle a \rangle$, $\langle a^\dag \rangle$, and $\langle a^\dag a \rangle$ to uniquely reconstruct the quantum state of the mode. This is so because linear combinations of these operators along with the identity, when restricted to the Hilbert space of a two-level system, can reconstruct the single qubit operators $\sigma_x$, $\sigma_y$, and $\sigma_z$ whose expectation values uniquely determine an arbitrary single qubit state. In a similar way, unique reconstruction of the state of a joint system of $N$ bosonic modes each restricted to their single excitation manifold can be accomplished if all $2^{2N}$ expectations of the joint moments of the system of the form $\left \langle (a_1^{\dagger})^{n_1} a_1^{m_1} (a_2^{\dagger})^{n_2} a_2^{m_2} ... (a_N^{\dagger})^{n_N} a_N^{m_N} \right \rangle~ \forall~ n_i,m_i \in \{0,1\}$ are known.

Consider $A_j \in \{(a_1^{\dagger})^{n_1} a_1^{m_1} (a_2^{\dagger})^{n_2} a_2^{m_2} ... (a_N^{\dagger})^{n_N} a_N^{m_N}\}$ to be one of the $2^{2N}$ moments of interest for such an $N$ mode system. If we assume our system to be in the state $\rho$, then there will be an underlying distribution determined by $\rho$ governing the statistics of measured values of $A_j$ that will have some mean $\mu_j = \Tr(A_j \rho)$ and variance $v_j$. By the central limit theorem the sample mean of $N$ measurements of $A_j$ should, for large enough $N$, respect a normal distribution centered around $\Tr(A_j \rho)$ with variance $v_j / N$. This being the case, then the probability $p(\langle \bar{A}_j\rangle | \rho)$ of finding the sample mean of $N$ measurements of $A_j$ to be $\langle \bar{A}_j\rangle$ (we use the bar notation to emphasize that we are talking about a measured statistical value and not a calculated quantum mechanical expectation value), assuming a system state $\rho$, should obey \cite{chow2012universal}:
\begin{equation}
    \begin{split}
        p(\langle \bar{A}_j\rangle | \rho) \propto e^{-|\langle \bar{A}_j\rangle - \Tr(A_j \rho)|^2 / (v_j / N)}
    \end{split}
    \label{singlemomentlikelihood}
\end{equation}
Assuming the actual variance $v_j$ of the moment is very well approximated by the measured sample variance $\bar{v}_j$, which it should be for large $N$ by the law of large numbers, then $v_j$ can be safely replaced by the measured variance $\bar{v}_j$ in this expression.

Consequently we find that we can define a likelihood functional inspired by \eqref{singlemomentlikelihood} that takes the form \cite{eichler2012characterizing}:
\begin{equation}
\begin{gathered}
    \mathcal{L}(D | \rho) = \prod\limits_{j = 1}^{j = 2^{2N}} e^{-|\langle \bar{A}_j \rangle - \Tr(A_j \rho)|^2 / \bar{v}_j}
\end{gathered}
\label{likelihoodfunctional}
\end{equation}
This functional requires a dataset $D = \{(\langle \bar{A}_j\rangle, \, \bar{v}_j)\}_{j=1}^{j=2^{2N}}$ of measured sample means and variances of all the joint $N$ photon moments considered above

Because of the monotonically increasing nature of the logarithm, minimizing the negative log-likelihood is equivalent to maximizing the likelihood, and taking the negative of the logarithm of the likelihood yields:
\begin{equation}
\begin{gathered}
   -\log \mathcal{L}(D | \rho) = \sum\limits_{j = 1}^{j = 2^{2N}} |\langle \bar{A}_j \rangle - \Tr(A_j \rho)|^2 / \bar{v}_j
\end{gathered}
\label{loglikelihoodfunctional}
\end{equation}
Intuitively we see that minimizing this negative log-likelihood corresponds to finding the state $\rho$ whose moments minimize the mean-squared error of the measured moments, discounting the error associated with higher variance measured moments more than the error associated with low variance measured moments. This optimization problem has the form of a quadratic programming problem subject to physicality constraints on the quantum state $\rho$ (namely that $\rho$ be trace-one and positive semidefinite):

\begin{equation}
\begin{split}
\min_{\rho} \quad & -\log \mathcal{L}(D | \rho) \\
\textrm{s.t.} \quad & \Tr(\rho) = 1 \\
  & \rho \succ 0    \\
\end{split}
\end{equation}

To perform this optimization over valid states $\rho$ of an $N$ mode system we use the CVXPY python library \cite{diamond2016cvxpy}.

\subsection{Quantum State Tomography Results}

\begin{figure*}[tbp]
\centering
\includegraphics[width = \textwidth]{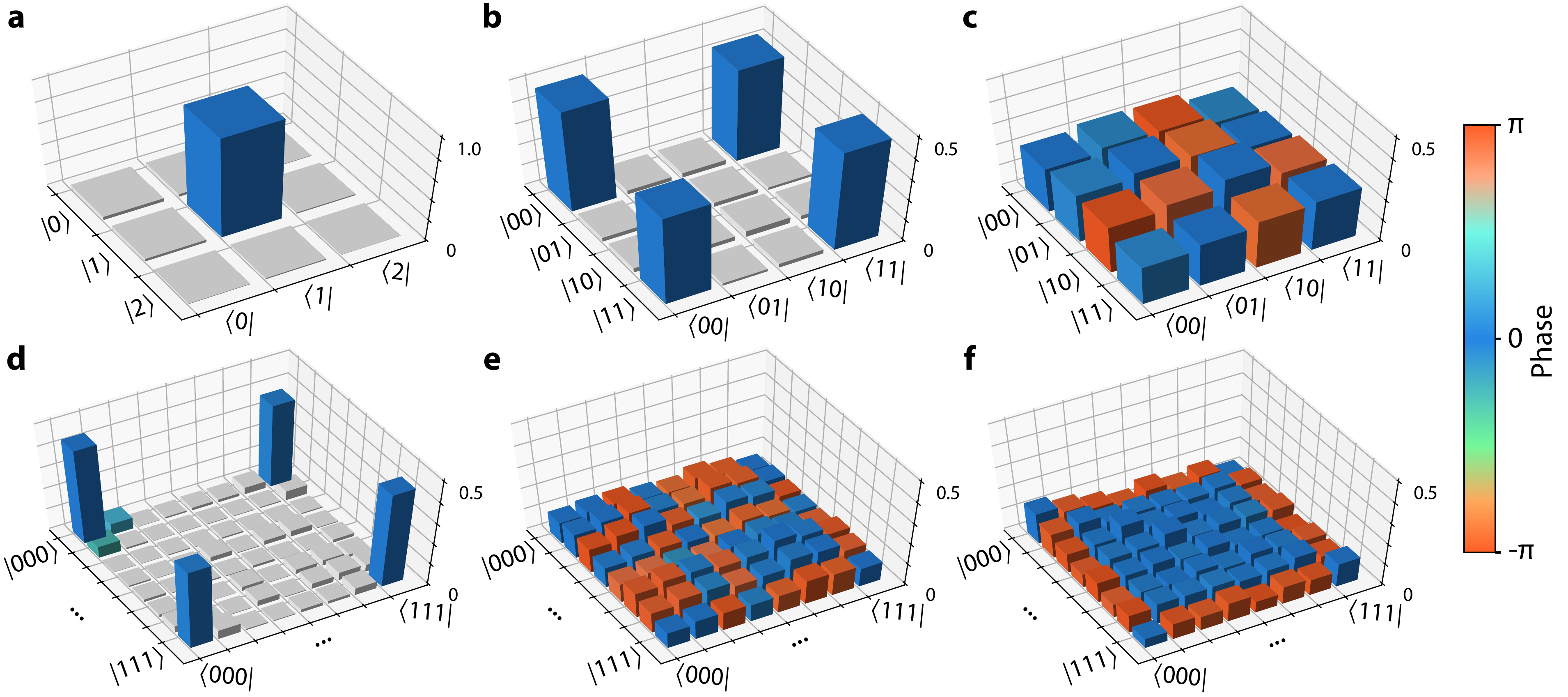}
\caption{\textbf{Reconstructed Density Matrices of Various Photonic States.} \textbf{a}, Reconstructed density matrix of a prepared single photon Fock state $\ket{1}$. ($F = 97\%$). Only here, the Hilbert space includes the two-photon manifold for a single mode in order to demonstrate the single photon character of photon emission. \textbf{b-f}, Reconstructed density matrices of a two-photon GHZ state ($F = 91\%$), a two-photon cluster state ($F = 91\%$), a three-photon GHZ state ($F = 83\%$), a three-photon 1D cluster state ($F = 90\%$), and a three-photon triangular cluster state (where there is all-to-all entanglement connectivity, $F = 73\%$). Note that time-delayed feedback was used to generate the three-photon triangular cluster state. The Hilbert space for each individual mode is truncated to the single-photon manifold. For each state, density matrix elements smaller than 10\% of the expected largest density matrix element are colored gray for ease of visualization. Note that global offset phases associated with each photon are adjusted via software in order to arrive at the density matrices plotted here; this amounts to local-Z corrections on the states.}
\label{fig:variousdm}
\end{figure*}

\begin{figure*}[tbp]
\centering
\includegraphics[width = \textwidth]{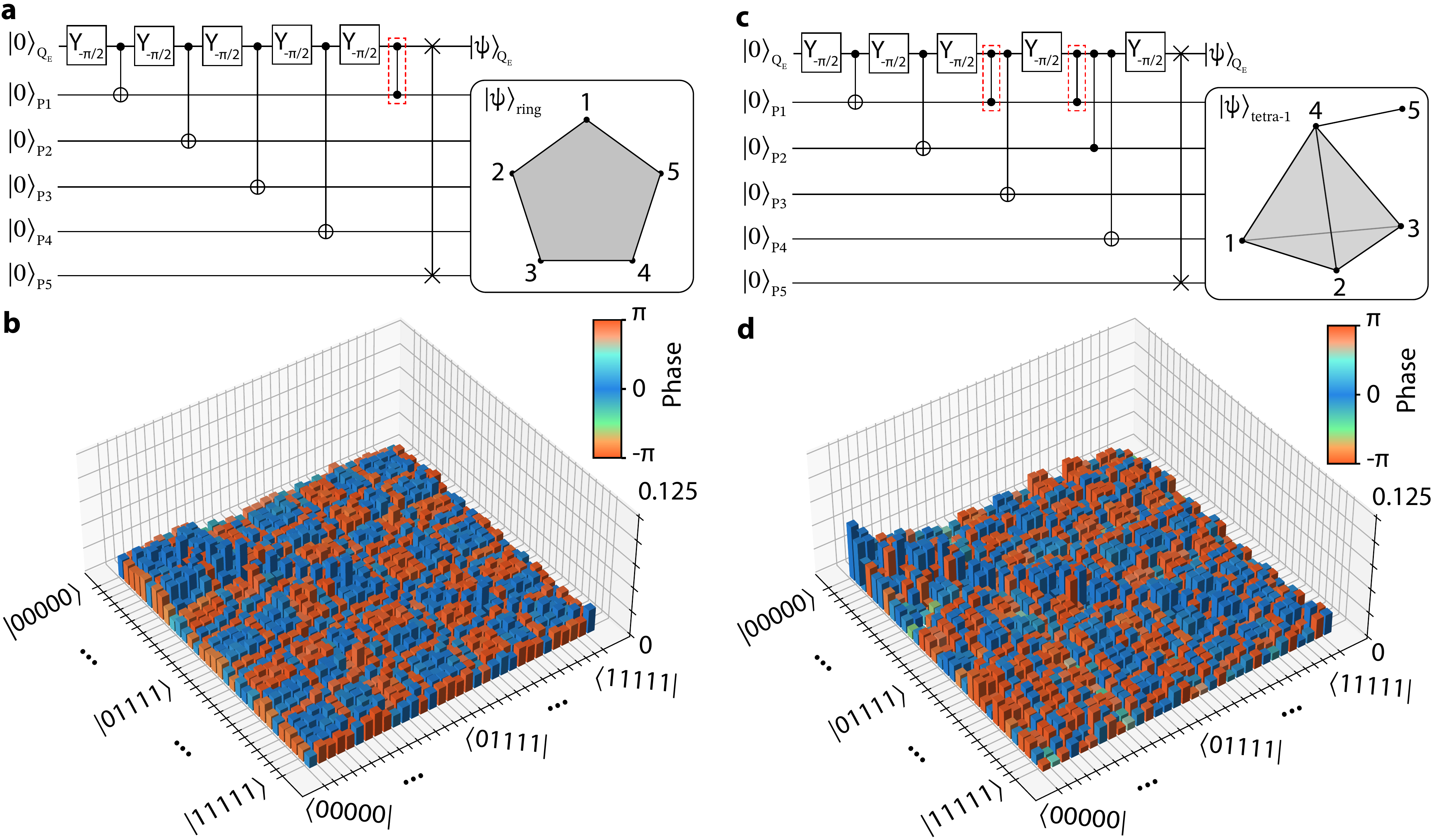}
\caption{\textbf{Five-Photon Cluster State with Multiple Time-Delayed Feedback Events.} \textbf{a}, Quantum circuit for the generation of the five-photon cluster state $\ket{\psi_{\text{ring}}}$. In this state, the entanglement structure forms a pentagon, as shown in the diagrams at the end of the quantum circuits. \textbf{b}, Reconstructed density matrix of the $\ket{\psi_{\text{ring}}}$ state. The fidelity of the generated state is $F = 61\%$. Note that the data for this state was taken in a separate cooldown, and due to techincal reasons, the absolute power calibration we used was acquired from the integrated flux of a prepared single photon state. \textbf{c}, Quantum circuit for the generation of the five-photon cluster state $\ket{\psi_{\text{tetra}-1}}$. In this state, the entanglement structure of the first four photons (photon 1-4) forms a tetrahedron, and the photon 5 is entangled to photon 4, as shown in the diagram on the right. Two time-delayed feedback events on photon 1 (corresponding the the highlighted CZ gates) entangle photon 1 with both photon 3 and photon 4. \textbf{d}, Reconstructed density matrix of the $\ket{\psi_{\text{tetra}-1}}$ state. The fidelity of the generated state is $F = 50\%$. Note that the global phases associated with each photon are adjusted via software in order to arrive at the plotted density matrices.} 
\label{fig:tetra}
\end{figure*}

In addition to the quantum state tomography results of Fig. \ref{fig:Cluster_State}, we present in Fig. \ref{fig:variousdm} and \ref{fig:tetra} reconstructed density matrices for other generated multipartite entangled photonic states (along with their associated fidelities), in order to illustrate the flexibility of our photonic state generation method. We bring particular attention to the 5 photon state illustrated in Fig. \ref{fig:tetra}c, with measured density matrix in Fig. \ref{fig:tetra}d. In order to generate this state, it was necessary to perform multiple CZ gate operations, including two CZ gate operations for photon 1. Such use of \textit{two} time-delayed feedback events for an emitted photon is the most fundamental prerequisite for extending our 2D cluster state generation scheme to generation of 3D cluster states \cite{raussendorf2007topological, wan2021fault, shi2021deterministic}. Thus, generation of the 5 photon state illustrated in Fig. \ref{fig:tetra}, via multiple time-delayed feedback events, constitutes a preliminary demonstration of the adaptability of our platform for future generation of 3D cluster states of microwave photonic qubits. We also note that the 97\% fidelity of the density matrix shown in Fig. \ref{fig:variousdm}a of a prepared single photon Fock state constitutes the quantum efficiency of our $Q_E$ single photon source (measurement errors notwithstanding), and that the reconstructed density matrix reveals our source's emission has negligible two-photon character. 

\begin{figure}[tbp]
\centering
\includegraphics[width = \columnwidth]{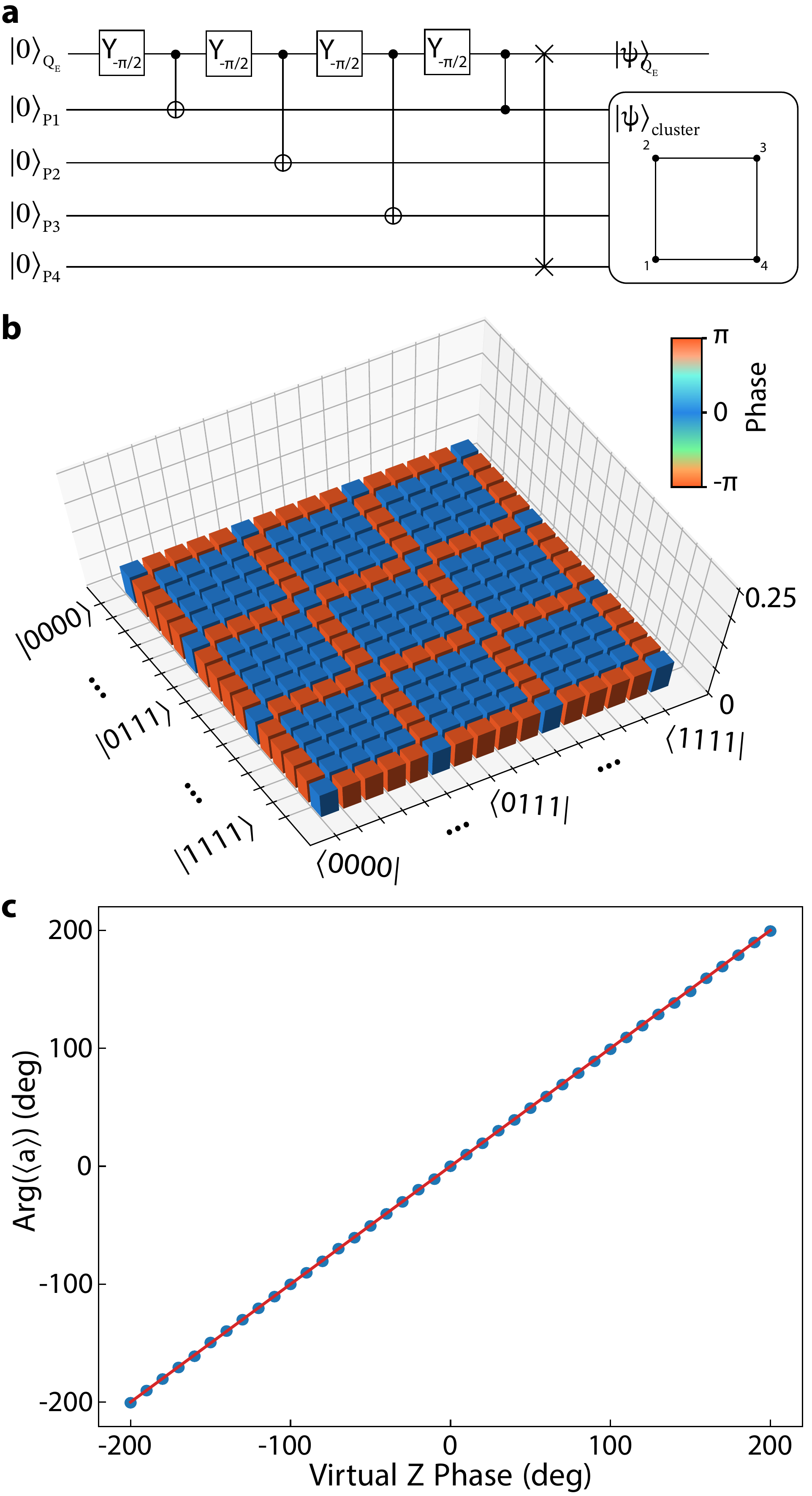}
\caption{\textbf{Phase Control of Cluster State} \textbf{a}, Quantum circuit of cluster state generation protocol, assuming $Q_E$ and photonic states prepared in ground state. \textbf{b}, Ideal cluster state obtained from the quantum circuit of \textbf{(a)}. \textbf{c}, Measured phase of single photon moment $\langle a \rangle$ as a function of the phase of virtual-Z gates applied to $Q_E$ during generation (blue circles). Red line shows ideal expected phase. Virtual-Z gates applied on $Q_E$ are realized by adding an offset phase to $\pi_{ef}$ pulses before photon emission.}  
\label{fig:clusterphase}
\end{figure}

We conclude this Appendix section by describing some technical details of our radiation field tomography measurements that may be of interest to the reader. Firstly, we note that when generating photonic states, lingering gate errors due to the AC Stark shifting of the $\kete \rightarrow \ketf$ transition, as well as the use of flux modulation, will result in spurious phases gained by the qubit, which will be imparted onto the phase of emitted photons. This phase, however, is deterministic, and thus can be compensated in hardware by suitable qubit Z-control. For the density matrix presented in Fig. \ref{fig:Cluster_State}d, these spurious phases were compensated for by the use of Virtual Z-gates \cite{mckay2017efficient} when performing $\pi_{ef}$ pulses before emission of every photon. Thus, we were able to generate the state whose ideal counterpart is shown in Fig. \ref{fig:clusterphase}b, and the 70\% state fidelity quoted in the main text is calculated with respect to this state. We also note that while these spurious phases correspond to local Z-gates for every photon, which can be removed from the processed tomography data, in practice they could hinder use of such cluster states in quantum information applications. Thus, we chose to demonstrate this additional photon phase control in our generation process, and we stress that the data presented in Fig. \ref{fig:Cluster_State}d did not have any post-processing phase modification.

\label{App:phasecorr}

In addition, the measured $\left \langle h_i^\dagger h_i \right \rangle$ moments reveal an effective added noise photon number of $n_{\text{noise}} \approx 3.5$, corresponding to a quantum measurement efficiency of the output chain of $\eta_{\text{meas}} = (1 + n_{\text{noise}})^{-1} \approx 0.22$. We note that for heterodyne detection, a finite detection efficiency $\eta_{\text{det}}$ can be shown to be equivalent to added noise in the output chain, and we believe that the majority of $n_{\text{noise}}$ can be attributed to losses and spurious reflections before the TWPA (where we estimate a total transmissivity for emitted photon pulses of -5.5dB). 

Furthermore, we note that the scaling factor $G$ will be slightly different for photons of different bandwidth. This is due in part to slight differences in the effective transmission coefficient of the tapered end of the SLWG for different bandwidth pulses (see Fig. \ref{fig:slowlight}d). Moreover, higher bandwidth pulses will have slightly higher mode-matching inefficiency due to dispersion-induced distortion, which results in a small fraction of the pulse being situated outside of its measurement time-bin window. When generating photonic states we use up to two different bandwidths, and we quantify the difference in their respective $G$ scaling factors by taking the ratio of their measured $\left \langle a^\dagger a\right \rangle$ moments when the qubit is fully excited to the $\ketf$ state before emission. We find a $\sim 5\%$ difference between the two $G$ scaling factors, which we take into account for calculation of joint photon moments (we use the $G$ obtained from Stark shift measurements for the lower bandwidth photon pulses) .

Finally, for generation of four photon and five photon states, we performed 500 million and 2 billion single shot measurements, respectively, in order to have sufficient averaging for higher order joint photon moments; these numbers are consistent with the predicted number of single shot measurements required from the statistical analysis presented in ref. \cite{da2010schemes}. This corresponded to measurement times of 6 hours and 24 hours, respectively; and due to the presence of slow qubit frequency drifts of $\sim 0.5$ MHz in our experimental setup, we recalibrated the qubit flux bias every hour during these long measurements. We expect that the use of GPU or FPGA based methods for data processing would significantly reduce these measurement times. Nevertheless, we note that full tomography of a photonic state of five itinerant microwave photons has hitherto never been demonstrated until now.

\section{Process Tomography of the Time-Delayed Feedback Operation}
\label{App:processtomo}

\subsection{QPT Experiment Design}

In order to characterize the qubit-photon CZ gate implemented with our time-delayed feedback protocol, we perform full quantum process tomography (QPT) of the qubit-photon interaction. We again limit the Hilbert space of the bosonic mode representing the itinerant photon to the single-excitation subspace, so the implemented CZ gate can be considered as a quantum process mapping the Hilbert space of an effective two qubit system to itself.

With this in mind, in characterizing our CZ implementation we are interested in the set of quantum processes that maps two-qubit states to two-qubit states. Such processes (outside of certain cases in which we are not concerned here, eg: projective measurements) are described by the set of completely-positive trace-preserving (CPTP) linear maps from two-qubit density matrices to two-qubit density matrices. \cite{nielsenchuang} 

Performing quantum process tomography requires identifying a complete set of `fiducial' input states of the system and an `informationally complete' set of measurement operators \cite{nielsenchuang, nielsen2021gate}. A complete fiducial set of input states on a $d$-dimensional Hilbert space $\mathcal{H}$ is a set of $d^2$ states whose density matrices span the space of density matrices on $\mathcal{H}$. An informationally complete set of measurement operators $\{M_j\}$ is a set of $d^2 - 1$ operators on $\mathcal{H}$ whose expectation values given a state $\rho$, $\{\Tr(M_j \rho)\}$, uniquely determine $\rho$.

As our set of $d^2$ fiducial input states we select all 16 possible unentangled states of the form $\ket{\psi_q}\otimes\ket{\phi_p}$ where $\ket{\psi_q} \in \{\ket{+z}_q, \ket{-z}_q, \ket{+x}_q, \ket{+y}_q\}$ and $\ket{\psi_p} \in \{\ket{+z}_p, \ket{-z}_p, \ket{+x}_p, \ket{-y}_p\}$ (where we are using the conventional names for eigenstates of the Pauli spin operators $\sigma_x$, $\sigma_y$, $\sigma_z$). For our informationally complete set of measurement operators we select the 15 non-identity joint qubit-photon correlators of the form $\sigma_i \otimes a^{\dag n} a^m$ (where $n, m \in \{0, 1\}$ and $\sigma_i \in \{\mathbb{I}, \sigma_x, \sigma_y, \sigma_z\}$). The expectations of these operators can be shown to uniquely specify any of the joint qubit-photon states we are considering in our effective two qubit Hilbert space \cite{eichler2012characterizing}. To measure the expectations of these operators for a given prepared state, we perform single-shot heterodyne measurements of the microwave field in conjunction with single shot measurements of the qubit polarization (after rotation to the appropriate basis). This allows us, on a shot-by-shot basis, to compute the correlations between the microwave field moments and the qubit polarization operators.

\begin{figure}[tbp]
\centering
\includegraphics[width = \columnwidth]{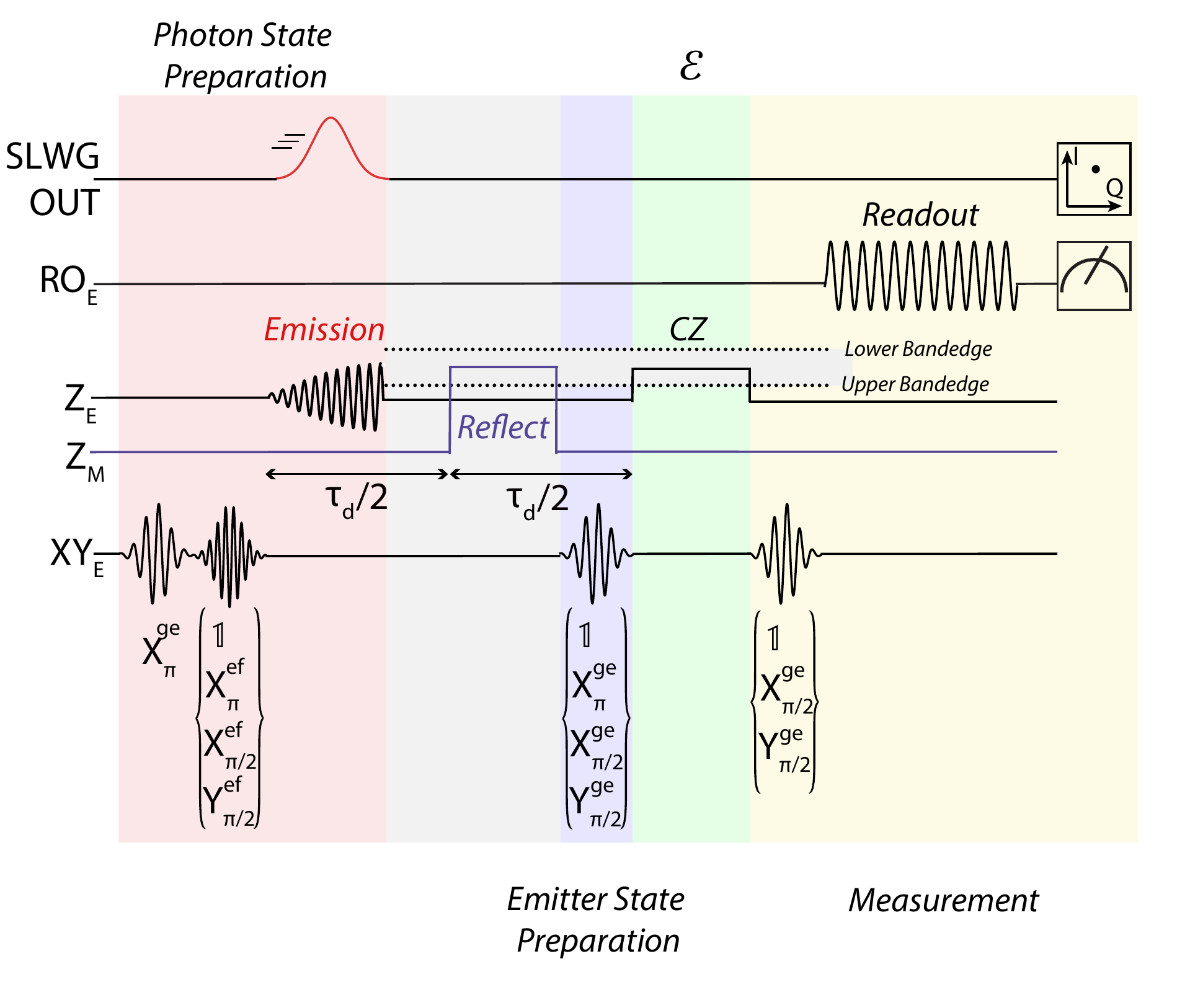}
\caption{\textbf{Experimental Pulse Sequence for QPT} A schematic for the pulse sequences that implement the QPT protocol for characterizing our $CZ$ implementation. The shaded regions of different coloration represent different subsequences of the QPT protocol indicated by the italic text above or below the figure (the grey region is time during which the photonic qubit propagates through one round-trip of the waveguide). The bracketed lists of gates below the pulses on the $\textrm{XY}_\textrm{E}$ line represent variable pulses corresponding to different photonic qubit state preparations (red region), $Q_E$ state preparations (blue region), and $Q_E$ measurement basis rotations (yellow region).}
\label{fig:QPT}
\end{figure}

The experimental sequence of our QPT implementation can be seen in Fig. \ref{fig:QPT}. We begin with $Q_E$ and the photonic qubit in their respective ground states, after which we prepare the state of the photonic qubit by performing an $X_\pi^{ge}$ pulse on $Q_E$, followed by one of an $X_\pi^{ef}$ pulse, an $X_{\pi/2}^{ef}$ pulse, an $Y_{\pi/2}^{ef}$ pulse, or no pulse ($\mathbb{I}^{ef}$), following which we use a flux modulation tone to induce shaped emission of the photonic time-bin qubit from $Q_E$. This results in the initialized photonic time-bin qubit state to be conditioned on the choice of $ef$ pulse, resulting in the states $\ket{1}$, $\ket{-y}$, $\ket{+x}$, or $\ket{0}$ respectively, while $Q_E$ ends in state $\kete$ after flux modulation. 

Before the photonic qubit finishes its propagation through one round trip of the waveguide, we prepare $Q_E$ in one of its four above specified cardinal states. When the photonic qubit finally returns to $Q_E$, both subsystems have been properly prepared and the $CZ$ gate between the two proceeds by way of our time-delayed feedback interaction. After the $CZ$ is completed the photonic time-bin qubit leaves the waveguide where it is amplified and its two independent $I$ and $Q$ quadratures are measured via heterodyne detection. The $Q_E$ state is also measured along one of the three chosen polarization axes defining which qubit polarization operator $\sigma_i$ we are measuring. Note that the emitter state preparation is deferred until immediately prior to the onset of time-delayed feedback, after the itinerant photon has travelled almost the entire round-trip length of the SLWG, in order to minimize the amount of dephasing suffered by $Q_E$ before the CZ gate. 

We perform the above control sequence for all $16 \times 3$ combinations of prepared states and possible values of $\sigma_i$. Note that, for example, while the experimental sequence corresponding to measuring $\sigma_x a$ and $\sigma_y a$ require different qubit basis rotation pulses, the nature of heterodyne measurement of the microwave field means a single experiment can be used to measure all four quantities $\{\sigma_x, \sigma_x a, \sigma_x a^\dag, \sigma_x a^\dag a\}$. We thus perform 48 different experiments (each repeated many times) to compute the $16 \times 15 = 240$ different expectations of the form $\langle\sigma_i \otimes a^{\dag n} a^m\rangle$ that uniquely specify each output state of the implemented CZ gate for each input state.

\subsection{MLE}
Once this data is collected, we perform a maximum-likelihood reconstruction of the time-delayed feedback operation to find the most likely quantum process approximating it. To do this we represent the quantum process $\mathcal{E}$ underlying the time-delayed feedback operation as its $\chi$ matrix in the Pauli Product basis of 2 qubits, whereby its action on a general input state is given by:

\begin{equation}
    \begin{split}
         \mathcal{E}(\rho) = \sum\limits_{n,m} \chi_{n,m} P_n \rho P_m^\dag
    \end{split}
\label{pauliproductrep}
\end{equation}

The quantity $\chi$ is a $16 \times 16$ Hermitian, positive-semidefinite matrix, and $\{P_j\}$ is some enumeration of the 2 qubit Pauli group. With this representation of the time-delayed feedback process, we use the python library CVXPY \cite{diamond2016cvxpy} to minimize the negative log-likelihood functional of the dataset of qubit-photon correlator expectations given the CZ process $\mathcal{E}$:
\begin{equation}
    \begin{split}
         -\log\mathcal{L}(D|\chi) = \sum\limits_{i=1, j=1}^{i=15, j=16} |\langle \bar{M}_{ij} \rangle - \Tr(M_j \; \mathcal{E} \circ \mathcal{B}_i(\rho_0))|^2 / v_{i,j}
    \end{split}
\label{qptloglikelihood}
\end{equation}
where  $\rho_0 = \ketbra{0}_q \otimes \ketbra{0}_p$ is the initial state of the joint qubit-photon system, assumed to be the ground state of both systems, $M_j$ is the $j^{th}$ qubit-photon correlator, $\mathcal{B}_i$ is the $i^{th}$ generalized state preparation superoperator (explained in more detail below), $\langle \bar{M}_{ij} \rangle$ is the measured sample mean of $M_j$ given state preparation $\mathcal{B}_i$, and $\bar{v}_{i, j}$ is the sample variance of $\langle \bar{M}_{i,j} \rangle$. The free parameters available to the optimization are the elements of $\chi$ implicitly contained in the computation of $\mathcal{E}\circ\mathcal{B}_i(\rho_0)$ above. 

The choice of this log-likelihood follows from exactly the same arguments as the state tomography log-likelihood functional equation \eqref{loglikelihoodfunctional} given in Appendix \ref{App:tomo}, extended to the case of $16$ simultaneous sets of state tomography data for the $16$ states $\mathcal{E}\circ\mathcal{B}_i(\rho_0)$. Enforcing the CPTP physicality constraints on $\mathcal{E}$ defines the quadratic programming optimization problem for finding the most-likely Pauli-Product representation of $\mathcal{E}$, $\hat{\chi}$:
\begin{equation}
\begin{split}
\min_{\chi} \quad & -\log \mathcal{L}(D | \chi) \\
\textrm{s.t.} \quad & \sum\limits_{n,m} \chi_{n,m} P_m^\dag P_n = \mathbb{I} \\
  & \chi \succ 0    \\
\end{split}
\end{equation}

After the $\chi$ matrix representing our $CZ$ implementation has been reconstructed we find the $CZ$ gate that it most closely approximates modulo any simultaneous local $Z$ operations on either $Q_E$ or the photonic time-bin qubit. These local $Z$ gates can be removed in software and do not quantitatively alter the entangling nature of the $CZ$ gate we are implementing, and thus have no impact on the $\text{Tr}\left(\sqrt{ \sqrt{\chi_{\text{CZ}}} \chi_{\text{ideal}} \sqrt{\chi_{\text{CZ}}} } \right)^2$ figure of merit we use to characterize the gate.

\subsection{Generalized state-preparation superoperator}
It is well-documented that QPT can suffer significantly from so-called state preparation and measurement (SPAM) errors, wherein errors during preparation of the QPT input states and errors during measurement are interpreted by the tomographic reconstruction method as errors on the process itself. We find two sources of state-preparation error that we are able to correct for systematically by substituting idealized state preparation unitaries for more general state preparation superoperators. These sources of error are the initial thermal population of $Q_E$ prior to the application of any state preparation control pulses, which by gate set tomography of $Q_E$ (we use the python library pyGSTi \cite{nielsen2020pyGSTi} for this) we estimate to be $\sim 1 \%$, and the round trip loss of $\sim 13 \%$ that the itinerant microwave photon suffers between emission and reinteraction with $Q_E$.

To account for these errors we define the generalized state-preparation super operators $\{\mathcal{B}_i\}$, each of which we factorize as the composition of three different processes:
\begin{equation}
\begin{split}
    \mathcal{B}_i = \mathcal{R} \circ \mathcal{U}_i \circ \mathcal{P}
\end{split}
\end{equation}
The process $\mathcal{U}_i$ corresponds to the $i^{th}$ \textbf{ideal} state preparation, the process $\mathcal{P}$ models the $1 \%$ initial thermal population of $Q_E$, and the process $\mathcal{R}$ models the $13 \%$ round trip loss of the itinerant microwave photon. $\mathcal{P}$ can be implemented by a pin map on the emitter qubit's Hilbert space: $\mathcal{P}(\rho_q) = 0.99\ketbra{0} + 0.01\ketbra{1}$ for all $\rho_q$. $\mathcal{R}$ can be modeled by a relaxation channel on the photonic qubit's Hilbert space with loss parameter $l = 0.13$. The preparation process $\mathcal{U}_i$ is given by the ideal processes implementing the pulse sequences in the red and blue shaded regions of Fig. \ref{fig:QPT} (i.e.: photonic qubit state preparation and $Q_E$ state preparation).

\subsection{Readout Error Correction}
In the same way that it is possible to correct for certain characterized state preparation errors, it is also possible to correct for qubit readout measurement errors that obey certain assumptions. By computing the confusion matrix of the single-shot $Q_E$ state measurement, we can correct for identifiable readout misclassification errors that give rise to erroneous values for qubit-photon correlator expectations.

Consider the quantity $\langle \sigma_z a \rangle$. This expectation can be computed from measured data in the following way:

\begin{equation}
    \begin{split}
        \langle \sigma_z a \rangle = \tilde{p}({+1})\langle \tilde{a} \rangle |_{\sigma_z = +1} - \tilde{p}({-1}) \langle \tilde{a} \rangle |_{\sigma_z = -1}
    \end{split}
\end{equation}

\noindent where $\tilde{p}({+1})$ denotes the proportion of single-shot measurements for which the qubit polarization along the $z$ quantization axis was found to be $+1$, and $\langle \tilde{a} \rangle |_{\sigma_z = +1}$ denotes the average value of the single shot field measurements in these same cases; with a similar computation for the $-1$ case.

Due to the fact that there are probabilities of mismeasurement of the qubit polarization, which we characterize in Appendix \ref{App:device}, the quantity $\tilde{p}({+1})\langle \tilde{a} \rangle |_{\sigma_z = +1}$ itself should be written as:
\begin{equation}
    \begin{split}
        \tilde{p}({+1})\langle \tilde{a} \rangle |_{\sigma_z = +1} = & \, p(+1|+1)p(+1)\langle a \rangle |_{\sigma_z = +1} \\ & + p(+1|-1)p(-1)\langle a \rangle |_{\sigma_z = -1}
    \end{split}
\end{equation}
where $p(+1|+1)$ corresponds to the probability of measuring a qubit polarization of $+1$ when the polarization was in fact $+1$, $p(+1)$ is the actual probability that an ideal measurement would have yielded a polarization of $+1$, and $\langle a \rangle|_{\sigma_z = +1}$ is the actual expected value of the field conditioned on the qubit polarization along $z$ being $+1$. There is a similar expression for $\tilde{p}({-1})\langle \tilde{a} \rangle |_{\sigma_z = -1}$:
\begin{equation}
    \begin{split}
        \tilde{p}({-1})\langle \tilde{a} \rangle |_{\sigma_z = -1} = & \, p(-1|+1)p(+1)\langle a \rangle |_{\sigma_z = +1} \\ & + p(-1|-1)p(-1)\langle a \rangle |_{\sigma_z = -1}
    \end{split}
\end{equation}
These two expressions can be combined into a simple linear relationship between the measured quantities $\tilde{p}(+1)\langle \tilde{a} \rangle |_{\sigma_z = +1}$ and $\tilde{p}(-1)\langle \tilde{a} \rangle |_{\sigma_z = -1}$ and the `premeasurement' undistorted quantities $p(+1)\langle a \rangle |_{\sigma_z = +1}$ and $p(-1)\langle a \rangle |_{\sigma_z = -1}$:
\begin{equation}
    \begin{split}
        \begin{pmatrix}
            \tilde{p}(+1)\langle \tilde{a} \rangle |_{\sigma_z = +1} \\
            \tilde{p}(-1)\langle \tilde{a} \rangle |_{\sigma_z = -1}
        \end{pmatrix} =
        C
        \begin{pmatrix}
            {p}(+1)\langle {a} \rangle |_{\sigma_z = +1} \\
            {p}(-1)\langle {a} \rangle |_{\sigma_z = -1}
        \end{pmatrix}
    \end{split}
\end{equation}
where $C$ is given by the confusion matrix:
\begin{equation}
    \begin{split}
        C = 
        \begin{pmatrix}
            p(+1|+1) & p(+1|-1) \\
            p(-1|+1) & p(-1|-1)
        \end{pmatrix}
    \end{split}
\end{equation}
This confusion matrix can be measured under the assumption of perfect state preparation by preparing the qubit many times in the ground or excited state, performing a single shot measurement of $Q_E$'s state, and counting the relative proportions of ground and excited measurements given a particular state preparation. This matrix $C$ can then be inverted and applied to the erroneous conditional photon moments to give the correct moments:
\begin{equation}
    \begin{split}
        \begin{pmatrix}
            {p}(+1)\langle {a} \rangle |_{\sigma_z = +1} \\
            {p}(-1)\langle {a} \rangle |_{\sigma_z = -1}
        \end{pmatrix}=
        C^{-1}
        \begin{pmatrix}
            \tilde{p}(+1)\langle \tilde{a} \rangle |_{\sigma_z = +1} \\
            \tilde{p}(-1)\langle \tilde{a} \rangle |_{\sigma_z = -1}
        \end{pmatrix} 
    \end{split}
\end{equation}
from which the correct qubit photon correlators can be computed.

\subsection{State Fidelity Confidence Intervals}
The 70\% state reconstruction fidelity relative to an ideal target cluster state is quoted with a 95\% confidence interval [69.1\%, 70.4\%] in the main text. We computed this confidence interval using a parametric bootstrapping protocol \cite{keith2018joint} that involved fitting the distributions of the measured photonic correlations and resampling the fit distributions to reconstruct 1000 bootstrap states. Each distribution of a measured photonic correlation contains 5000 points, where each point is computed from an average of 100,000 single shot field measurements. Thus corresponding to each moment was an approximately normal histogram of these 5000 values. These histograms could be fit and resampled to reconstruct bootstrapped versions of the generated cluster state. We reconstructed 1000 bootstrap copies this way, and for each copy we computed the fidelity relative to the target cluster state. Then we sorted these fidelities and approximated the 95\% confidence interval by taking the $25^{th}$ element of this sorted list fidelity as the lower bound of the 95\% confidence interval and the $975^{th}$ element as the upper bound.

\section{Sources of Infidelity and Future Improvements}
\label{App:improvements}

Here we examine the main sources of infidelity in our experiment that limited the fidelity of the generated 2D cluster state and other photonic states. We discuss how this infidelity could be ameliorated, and comment on other possible design and hardware improvements to considerably increase the size of the cluster state.  

\subsection{Infidelity Analysis}

The main source of infidelity in this work was the poor decoherence rate $T_2^* = 561$ ns of $Q_E$. We ascribe this low $T_2^*$ to excessive flux noise given our robust $T_1 = 34~\mu$s (and measured $T_2^*$ of over $15~\mu$s at its maximum frequency). The Ramsey decay time-dependence was strongly Gaussian, which suggests that our dephasing is limited by $1/f$ noise \cite{bylander2011noise, krantz2019quantum}. In order to properly model $1/f$ noise, typical Linbladian master-equation approaches, which assume a Markovian model of decoherence, do not suffice. We thus modeled the effect of $1/f$ noise in our cluster state generation sequence by simulating the state evolution of our joint system of $Q_E$ (here a 3-level system) and photonic qubits under the influence of a $\delta(t) \hat{c}^\dagger \hat{c}$ term in the Hamiltonian, where $\hat{c}$ here is the annihilation operator of $Q_E$'s anharmonic mode, and $\delta(t)$ is a random noise signal with noise power spectral density of $1/f$. The Hamiltonian for state evolution is thus comprised of this $\delta(t) \hat{c}^\dagger \hat{c}$ term, and the time-dependent Hamiltonian that realizes the pulse sequence depicted in Fig. \ref{fig:Cluster_State}b. 

Many different realizations of this noise are realized in the following manner: a random FFT spectrum is generated where each FFT bin is a normally distributed random complex value (and the spectrum is conjugate symmetric), the FFT bins are scaled according to a $1/f$ spectrum, and the inverse FFT is taken to arrive at a random noise signal. We ensure that the generated noise signals are long enough such that the center frequencies of FFT bins are as low as 50 Hz, in order for the resultant time signal to have significant power at very low frequencies. Consequently, we only use small portions of this long noisy time signal as the different realizations of $\delta(t)$ in our simulations (which are confirmed to have a $1/f$ power spectrum). We simulate state evolution of our system under different realizations of $\delta(t)$ and average the resultant states in order to obtain the ``average" effect of $1/f$ noise induced dephasing on the system. We confirm that this simulation approach reproduces Gaussian shaped Ramsey decay as well as the ``spin-echo" phenomenon; and with full simulation of our cluster state generation sequence of Fig. \ref{fig:Cluster_State}b under $1/f$ noise, we determine that our $Q_E$ dephasing results in an infidelity of $\sim$ 15\% for the final 2D cluster state.

Secondly, the second most significant source of infidelity in our generation scheme is the round-trip loss of the slow-light waveguide. As seen in Fig. \ref{fig:Mirror_Catch}c, the round-trip loss of the slow-light waveguide corresponds to 13\% energy loss for each photon undergoing a round-trip. Note that while a limited detection efficiency $\eta_{\text{det}}$ is compensated for in heterodyne based state tomography via the scaling factor $G$, loss that occurs before the $CZ$ gate, i.e. during state generation, is not considered part of $\eta_{\text{det}}$ and directly contributes to infidelity. We find that our photon loss contributes to $\sim$ 5\% infidelity. Further, we estimate that control and preparation errors, including qubit thermal population (measured to be 1\%), residual $\ketf$ state population after emission (measured to be 1\%), and $CZ$ gate infidelity contribute another total $\sim$ 4\% infidelity. In total, we estimate a 76\% fidelity limit for the generated state, which does not take into account measurement errors or other state preparation errors. This is in good agreement of our measured fidelity of 70\%. Finally, we note that while waveguide dispersion was not a serious impediment for the photonic state generation that we have presented, it would limit the use of higher bandwdith photons as compared to what we used.

\subsection{Scaling the Size of the Cluster State}

Although there were limiting technical issues in our experiment, we believe there is a straight forward path for mitigation of these issues, and realistic strategies for extending our generation scheme to synthesis of much larger cluster states. Firstly, we are confident that there is ample room to reduce the excessive flux noise in our setup to state-of-the-art values \cite{hutchings2017tunable}. Additionally, while our reliance on flux modulation for tunable coupling between $Q_E$ and the SLWG necessitates operation at ``flux sensitive" qubit frequencies, the use of tunable couplers \cite{chen2014qubit, yan2018tunable} between $Q_E$ and the SLWG would allow operation at $Q_E$'s ``sweet spot" frequencies, thereby increasing its resilience to flux noise. Given that $T_2^* \sim 100 \mu$s has been reported for transmon qubits in the past, we expect that the deleterious effects of dephasing could be nearly fully dispensed with. If even more emitter coherence were to be required, the use of an error-corrected ``logical qubit," with emission via an ancilla qubit, could be utilized to fully suppress emitter decoherence based errors.  

Furthermore, the 0.7 dB round-trip loss of our waveguide is another serious limiting factor to our fidelity, and arises due to the limited $Q$ of our unit cell resonators, which we estimate to be $90,000$. However, microwave superconducting resonators with $Q > 1,000,000$ have been fabricated in numerous previous works \cite{megrant2012planar, calusine2018analysis, woods2019determining}. Thus, with fabrication or materials improvement that have already been demonstrated, the waveguide loss could realistically be substantially reduced. We note that while our current compact unit cell design has sharp corners that likely induce strong electric fields that couple to TLS, this effect could be mitigated with different geometrical design or material improvement. 

Moreover, while the SLWG's dispersion hinders the use of higher bandwidth Gaussian photon pulses due to dispersion induced broadening, our tunable coupling capability or dispersion engineering allows for use of well-established \cite{killey2006electronic, ramachandran2007fiber} or novel \cite{huang2021nondispersing} dispersion mitigation techniques. By compensating for the dispersion through signal pre-distortion, or via dispersion cancelling elements post SLWG propagation, one could ensure that photon pulses arriving for re-scattering or for measurement are well-confined in time, thus alleviating the problem of overlapping broadened pulses. Such techniques would in principle enable even faster emission of photon pulses, and thus would allow for better utilization of limited round-trip delays.

Furthermore, achieving higher anharmonicity of $Q_E$ would allow for realization of even larger $\GammaoneD$, without compromising qubit coherence, than what was already achieved in this work. Increasing $\GammaoneD$ would allow for even more rapid emission of shaped photon pulses, as well as less residual $\ketf$ population after emission, and a high fidelity $CZ$ gate with high bandwidth photons. Analysis in ref. \cite{pichler2017universal} shows that higher $\GammaoneD$ improves the fidelity of CZ gates, due to decreased ``dispersion" of the $Q_E$ induced reflection phase near $Q_E$'s resonance frequency that is commensurate with the broadening of $Q_E$'s lineshape. This reduced ``dispersion" of the reflection phase results in the overall phase gained by the photon pulse during re-scattering to be closer to $\pi$, and reduces the distortion of the photon pulse imparted by the re-scattering process (which improves mode-matching efficiency). Limited anharmonicity $\eta$ of $Q_E$ is the main limiting factor to the magnitude of $\GammaoneD$, as a $\GammaoneD$ significantly larger than $\eta$ would lead to more substantial leakage of the $\kete$ population into the SLWG. However, by using a different superconducting qubit design that has higher  anharmonicity than the transmon \cite{nguyen2019high, yurtalan2021characterization, yan2020engineering}, the magnitude of $\GammaoneD$ could be substantially increased without compromising other aspects of cluster state generation. Also, achieving higher anhamonicity would allow for a larger waveguide passband, which would lower higher-order dispersion (albeit at the cost of less delay per resonator).

Additionally, the round-trip delay could be substantially increased in several ways. One straightforward way is simply by increasing the number of unit cells of the SLWG. Although that would increase the size of the device, which could introduce spurious box modes to the sample, recent advances in microwave packaging techniques could ameliorate the impact of larger device size \cite{huang2021microwave, bronn2018high}. Furthermore, the unit cell size could be reduced by leveraging compact high kinetic inductance superconducting resonators \cite{shearrow2018atomic, grunhaupt2018loss}, allowing for more delay per area. And looking forward even further, incorporation of an acoustic delay line into our system could allow for longer round-trip delays without additional dispersion or susceptibility to microwave packaging box modes \cite{bienfait2019phonon, andersson2019non, dumur2021quantum}, increasing the possible size of generated cluster states even further. 

We stress that in addition to increasing cluster state size, cluster state dimensionality could be increased to 3D by coupling of another mirror qubit somewhere along the delay line rather than at the end, which would impart the ability to perform time-delayed feedback with two different delays. The ability to perform two time-delayed feedback events with two different delays is the pre-requisite to generating 3D cluster states via sequential photon emission and time-delayed feedback \cite{, wan2021fault, shi2021deterministic}, because it allows for sufficient non-nearest neighbor entanglement between photons of the emitted pulse train such that the entanglement topology is that of a 3D cluster state. While this extension of our scheme would necessitate significantly larger delays, we believe achieving such delays is possible. Thus, we believe there is a viable path to measurement-based quantum computation with microwave photons via photonic resource state generation as we have described. We conclude by observing that, even if the size of generated photonic states were to hit some practical limitation, there are other measurement-based quantum computation schemes, such as fusion based quantum computation \cite{bartolucci2021fusion},  that only require the repeated synthesis of smaller photonic resource states which are later ``fused" into larger photonic states via linear optical elements. Our cluster state generation scheme would be well suited to be incorporated into such approaches, and could provide a ``bridge" towards the goal of photonic resource state generation via linear optical elements. 

Finally, we note that previous works \cite{wan2021fault, shi2021deterministic} have performed analysis of errors in cluster state generation in the context of fault-tolerant quantum computing. They find that for a gate error rate of $\sim 10^{-3}$ (where ``gates" in this context includes $Q_E$ single qubit gates, photon emission, and the qubit-photon CZ gate) and delay line losses of $\sim 3 \cdot 10^{-5}$ dB/ns, one can achieve a fault-tolerance ``break-even" point where the logical error rate is lower than the gate error rate, and beyond which logical errors are exponentially suppressed as delay line loss is decreased. State-of-the-art single qubit gates can routinely achieve such gate error rates of $10^{-3}$, while photon emission should in principle also achieve such error rates if there is enough time for full emission from the $\ketf$ state, and sufficient protection of the $\kete$ state. Furthermore as mentioned in the main text, although our reported CZ gate fidelity was 90 \%, we are able to ascribe most of that infidelity to SPAM errors via separate measurements. From separate simulations, we expected a ~97 \% fidelity for the CZ gate, and this fidelity could be increased further simply by increasing the $\GammaoneD$ of the emitter qubit. In addition, while necessary delay line losses are around $\sim 100$ times smaller than our current losses in our experiment, recent and future advances in superconducting circuit fabrication are expected to allow for 100 times (or more) lower losses in superconducting resonators \cite{place2021new}. Thus, we foresee that with realistic device and fabrication improvements, generating 3D cluster states for fault-tolerant measurement based quantum computation with negligible logical error rates should be feasible. Further, while single shot photon measurements along arbitrary basis would also be necessary for quantum computation with itinerant microwave photons, such single photon detection could be achieved with a ``detector" qubit. With such a detector, an itinerant photon's state would be mapped to the detector qubit state's via suitable time-dependent control of the detector qubit's coupling to the waveguide, as demonstrated in previous works \cite{wenner2014catching, kurpiers2018deterministic}; the ``detector" qubit could then be measured in an arbitrary basis. Lastly, we also note that with different photon qubit encodings, different generation and measurement protocols are also possible.

\vfill


\begin{thebibliography}{98}%
\makeatletter
\providecommand \@ifxundefined [1]{%
 \@ifx{#1\undefined}
}%
\providecommand \@ifnum [1]{%
 \ifnum #1\expandafter \@firstoftwo
 \else \expandafter \@secondoftwo
 \fi
}%
\providecommand \@ifx [1]{%
 \ifx #1\expandafter \@firstoftwo
 \else \expandafter \@secondoftwo
 \fi
}%
\providecommand \natexlab [1]{#1}%
\providecommand \enquote  [1]{``#1''}%
\providecommand \bibnamefont  [1]{#1}%
\providecommand \bibfnamefont [1]{#1}%
\providecommand \citenamefont [1]{#1}%
\providecommand \href@noop [0]{\@secondoftwo}%
\providecommand \href [0]{\begingroup \@sanitize@url \@href}%
\providecommand \@href[1]{\@@startlink{#1}\@@href}%
\providecommand \@@href[1]{\endgroup#1\@@endlink}%
\providecommand \@sanitize@url [0]{\catcode `\\12\catcode `\$12\catcode
  `\&12\catcode `\#12\catcode `\^12\catcode `\_12\catcode `\%12\relax}%
\providecommand \@@startlink[1]{}%
\providecommand \@@endlink[0]{}%
\providecommand \url  [0]{\begingroup\@sanitize@url \@url }%
\providecommand \@url [1]{\endgroup\@href {#1}{\urlprefix }}%
\providecommand \urlprefix  [0]{URL }%
\providecommand \Eprint [0]{\href }%
\providecommand \doibase [0]{https://doi.org/}%
\providecommand \selectlanguage [0]{\@gobble}%
\providecommand \bibinfo  [0]{\@secondoftwo}%
\providecommand \bibfield  [0]{\@secondoftwo}%
\providecommand \translation [1]{[#1]}%
\providecommand \BibitemOpen [0]{}%
\providecommand \bibitemStop [0]{}%
\providecommand \bibitemNoStop [0]{.\EOS\space}%
\providecommand \EOS [0]{\spacefactor3000\relax}%
\providecommand \BibitemShut  [1]{\csname bibitem#1\endcsname}%
\let\auto@bib@innerbib\@empty
\bibitem [{\citenamefont {Wootters}(1998)}]{wootters1998quantum}%
  \BibitemOpen
  \bibfield  {author} {\bibinfo {author} {\bibfnamefont {W.~K.}\ \bibnamefont
  {Wootters}},\ }\bibfield  {title} {\bibinfo {title} {Quantum entanglement as
  a quantifiable resource},\ }\href@noop {} {\bibfield  {journal} {\bibinfo
  {journal} {Phil. Trans. R. Soc. A.}\ }\textbf {\bibinfo {volume} {356}},\
  \bibinfo {pages} {1717} (\bibinfo {year} {1998})}\BibitemShut {NoStop}%
\bibitem [{\citenamefont {Horodecki}\ \emph {et~al.}(2009)\citenamefont
  {Horodecki}, \citenamefont {Horodecki}, \citenamefont {Horodecki},\ and\
  \citenamefont {Horodecki}}]{horodecki2009quantum}%
  \BibitemOpen
  \bibfield  {author} {\bibinfo {author} {\bibfnamefont {R.}~\bibnamefont
  {Horodecki}}, \bibinfo {author} {\bibfnamefont {P.}~\bibnamefont
  {Horodecki}}, \bibinfo {author} {\bibfnamefont {M.}~\bibnamefont
  {Horodecki}},\ and\ \bibinfo {author} {\bibfnamefont {K.}~\bibnamefont
  {Horodecki}},\ }\bibfield  {title} {\bibinfo {title} {Quantum entanglement},\
  }\href@noop {} {\bibfield  {journal} {\bibinfo  {journal} {Rev. Mod. Phys.}\
  }\textbf {\bibinfo {volume} {81}},\ \bibinfo {pages} {865} (\bibinfo {year}
  {2009})}\BibitemShut {NoStop}%
\bibitem [{\citenamefont {Bennett}(1998)}]{bennett1998quantum}%
  \BibitemOpen
  \bibfield  {author} {\bibinfo {author} {\bibfnamefont {C.~H.}\ \bibnamefont
  {Bennett}},\ }\bibfield  {title} {\bibinfo {title} {Quantum information},\
  }\href@noop {} {\bibfield  {journal} {\bibinfo  {journal} {Phys. Scr.}\
  }\textbf {\bibinfo {volume} {1998}},\ \bibinfo {pages} {210} (\bibinfo {year}
  {1998})}\BibitemShut {NoStop}%
\bibitem [{\citenamefont {Kimble}(2008)}]{kimble2008quantum}%
  \BibitemOpen
  \bibfield  {author} {\bibinfo {author} {\bibfnamefont {H.~J.}\ \bibnamefont
  {Kimble}},\ }\bibfield  {title} {\bibinfo {title} {The quantum internet},\
  }\href@noop {} {\bibfield  {journal} {\bibinfo  {journal} {Nature}\ }\textbf
  {\bibinfo {volume} {453}},\ \bibinfo {pages} {1023} (\bibinfo {year}
  {2008})}\BibitemShut {NoStop}%
\bibitem [{\citenamefont {Jozsa}(1997)}]{jozsa1997entanglement}%
  \BibitemOpen
  \bibfield  {author} {\bibinfo {author} {\bibfnamefont {R.}~\bibnamefont
  {Jozsa}},\ }\href@noop {} {\bibinfo {title} {Entanglement and quantum
  computation, appearing in geometric issues in the foundations of science,
  huggett s et. al., eds}} (\bibinfo {year} {1997})\BibitemShut {NoStop}%
\bibitem [{\citenamefont {Gisin}\ and\ \citenamefont
  {Thew}(2007)}]{gisin2007quantum}%
  \BibitemOpen
  \bibfield  {author} {\bibinfo {author} {\bibfnamefont {N.}~\bibnamefont
  {Gisin}}\ and\ \bibinfo {author} {\bibfnamefont {R.}~\bibnamefont {Thew}},\
  }\bibfield  {title} {\bibinfo {title} {Quantum communication},\ }\href@noop
  {} {\bibfield  {journal} {\bibinfo  {journal} {Nat. Phot.}\ }\textbf
  {\bibinfo {volume} {1}},\ \bibinfo {pages} {165} (\bibinfo {year}
  {2007})}\BibitemShut {NoStop}%
\bibitem [{\citenamefont {Kempe}(1999)}]{kempe1999multiparticle}%
  \BibitemOpen
  \bibfield  {author} {\bibinfo {author} {\bibfnamefont {J.}~\bibnamefont
  {Kempe}},\ }\bibfield  {title} {\bibinfo {title} {Multiparticle entanglement
  and its applications to cryptography},\ }\href@noop {} {\bibfield  {journal}
  {\bibinfo  {journal} {Phys. Rev. A}\ }\textbf {\bibinfo {volume} {60}},\
  \bibinfo {pages} {910} (\bibinfo {year} {1999})}\BibitemShut {NoStop}%
\bibitem [{\citenamefont {Raussendorf}\ and\ \citenamefont
  {Briegel}(2001)}]{raussendorf2001one}%
  \BibitemOpen
  \bibfield  {author} {\bibinfo {author} {\bibfnamefont {R.}~\bibnamefont
  {Raussendorf}}\ and\ \bibinfo {author} {\bibfnamefont {H.~J.}\ \bibnamefont
  {Briegel}},\ }\bibfield  {title} {\bibinfo {title} {A one-way quantum
  computer},\ }\href@noop {} {\bibfield  {journal} {\bibinfo  {journal} {Phys.
  Rev. Lett.}\ }\textbf {\bibinfo {volume} {86}},\ \bibinfo {pages} {5188}
  (\bibinfo {year} {2001})}\BibitemShut {NoStop}%
\bibitem [{\citenamefont {Raussendorf}\ \emph {et~al.}(2003)\citenamefont
  {Raussendorf}, \citenamefont {Browne},\ and\ \citenamefont
  {Briegel}}]{raussendorf2003measurement}%
  \BibitemOpen
  \bibfield  {author} {\bibinfo {author} {\bibfnamefont {R.}~\bibnamefont
  {Raussendorf}}, \bibinfo {author} {\bibfnamefont {D.~E.}\ \bibnamefont
  {Browne}},\ and\ \bibinfo {author} {\bibfnamefont {H.~J.}\ \bibnamefont
  {Briegel}},\ }\bibfield  {title} {\bibinfo {title} {Measurement-based quantum
  computation on cluster states},\ }\href@noop {} {\bibfield  {journal}
  {\bibinfo  {journal} {Phys. Rev. A}\ }\textbf {\bibinfo {volume} {68}},\
  \bibinfo {pages} {022312} (\bibinfo {year} {2003})}\BibitemShut {NoStop}%
\bibitem [{\citenamefont {Raussendorf}\ \emph {et~al.}(2007)\citenamefont
  {Raussendorf}, \citenamefont {Harrington},\ and\ \citenamefont
  {Goyal}}]{raussendorf2007topological}%
  \BibitemOpen
  \bibfield  {author} {\bibinfo {author} {\bibfnamefont {R.}~\bibnamefont
  {Raussendorf}}, \bibinfo {author} {\bibfnamefont {J.}~\bibnamefont
  {Harrington}},\ and\ \bibinfo {author} {\bibfnamefont {K.}~\bibnamefont
  {Goyal}},\ }\bibfield  {title} {\bibinfo {title} {Topological fault-tolerance
  in cluster state quantum computation},\ }\href@noop {} {\bibfield  {journal}
  {\bibinfo  {journal} {New J. Phys.}\ }\textbf {\bibinfo {volume} {9}},\
  \bibinfo {pages} {199} (\bibinfo {year} {2007})}\BibitemShut {NoStop}%
\bibitem [{\citenamefont {Briegel}\ \emph {et~al.}(2009)\citenamefont
  {Briegel}, \citenamefont {Browne}, \citenamefont {D{\"u}r}, \citenamefont
  {Raussendorf},\ and\ \citenamefont {Van~den Nest}}]{briegel2009measurement}%
  \BibitemOpen
  \bibfield  {author} {\bibinfo {author} {\bibfnamefont {H.~J.}\ \bibnamefont
  {Briegel}}, \bibinfo {author} {\bibfnamefont {D.~E.}\ \bibnamefont {Browne}},
  \bibinfo {author} {\bibfnamefont {W.}~\bibnamefont {D{\"u}r}}, \bibinfo
  {author} {\bibfnamefont {R.}~\bibnamefont {Raussendorf}},\ and\ \bibinfo
  {author} {\bibfnamefont {M.}~\bibnamefont {Van~den Nest}},\ }\bibfield
  {title} {\bibinfo {title} {Measurement-based quantum computation},\
  }\href@noop {} {\bibfield  {journal} {\bibinfo  {journal} {Nat. Phys.}\
  }\textbf {\bibinfo {volume} {5}},\ \bibinfo {pages} {19} (\bibinfo {year}
  {2009})}\BibitemShut {NoStop}%
\bibitem [{\citenamefont {Friis}\ \emph {et~al.}(2017)\citenamefont {Friis},
  \citenamefont {Orsucci}, \citenamefont {Skotiniotis}, \citenamefont
  {Sekatski}, \citenamefont {Dunjko}, \citenamefont {Briegel},\ and\
  \citenamefont {D{\"u}r}}]{friis2017flexible}%
  \BibitemOpen
  \bibfield  {author} {\bibinfo {author} {\bibfnamefont {N.}~\bibnamefont
  {Friis}}, \bibinfo {author} {\bibfnamefont {D.}~\bibnamefont {Orsucci}},
  \bibinfo {author} {\bibfnamefont {M.}~\bibnamefont {Skotiniotis}}, \bibinfo
  {author} {\bibfnamefont {P.}~\bibnamefont {Sekatski}}, \bibinfo {author}
  {\bibfnamefont {V.}~\bibnamefont {Dunjko}}, \bibinfo {author} {\bibfnamefont
  {H.~J.}\ \bibnamefont {Briegel}},\ and\ \bibinfo {author} {\bibfnamefont
  {W.}~\bibnamefont {D{\"u}r}},\ }\bibfield  {title} {\bibinfo {title}
  {Flexible resources for quantum metrology},\ }\href@noop {} {\bibfield
  {journal} {\bibinfo  {journal} {New J. Phys.}\ }\textbf {\bibinfo {volume}
  {19}},\ \bibinfo {pages} {063044} (\bibinfo {year} {2017})}\BibitemShut
  {NoStop}%
\bibitem [{\citenamefont {Shettell}\ and\ \citenamefont
  {Markham}(2020)}]{shettell2020graph}%
  \BibitemOpen
  \bibfield  {author} {\bibinfo {author} {\bibfnamefont {N.}~\bibnamefont
  {Shettell}}\ and\ \bibinfo {author} {\bibfnamefont {D.}~\bibnamefont
  {Markham}},\ }\bibfield  {title} {\bibinfo {title} {Graph states as a
  resource for quantum metrology},\ }\href@noop {} {\bibfield  {journal}
  {\bibinfo  {journal} {Phys. Revi. Lett.}\ }\textbf {\bibinfo {volume}
  {124}},\ \bibinfo {pages} {110502} (\bibinfo {year} {2020})}\BibitemShut
  {NoStop}%
\bibitem [{\citenamefont {Briegel}\ and\ \citenamefont
  {Raussendorf}(2001)}]{briegel2001persistent}%
  \BibitemOpen
  \bibfield  {author} {\bibinfo {author} {\bibfnamefont {H.~J.}\ \bibnamefont
  {Briegel}}\ and\ \bibinfo {author} {\bibfnamefont {R.}~\bibnamefont
  {Raussendorf}},\ }\bibfield  {title} {\bibinfo {title} {Persistent
  entanglement in arrays of interacting particles},\ }\href@noop {} {\bibfield
  {journal} {\bibinfo  {journal} {Phys. Rev. Lett.}\ }\textbf {\bibinfo
  {volume} {86}},\ \bibinfo {pages} {910} (\bibinfo {year} {2001})}\BibitemShut
  {NoStop}%
\bibitem [{\citenamefont {Muralidharan}\ and\ \citenamefont
  {Panigrahi}(2008)}]{muralidharan2008quantum}%
  \BibitemOpen
  \bibfield  {author} {\bibinfo {author} {\bibfnamefont {S.}~\bibnamefont
  {Muralidharan}}\ and\ \bibinfo {author} {\bibfnamefont {P.~K.}\ \bibnamefont
  {Panigrahi}},\ }\bibfield  {title} {\bibinfo {title} {Quantum-information
  splitting using multipartite cluster states},\ }\href@noop {} {\bibfield
  {journal} {\bibinfo  {journal} {Physical Review A}\ }\textbf {\bibinfo
  {volume} {78}},\ \bibinfo {pages} {062333} (\bibinfo {year}
  {2008})}\BibitemShut {NoStop}%
\bibitem [{\citenamefont {Schlingemann}\ and\ \citenamefont
  {Werner}(2001)}]{schlingemann2001quantum}%
  \BibitemOpen
  \bibfield  {author} {\bibinfo {author} {\bibfnamefont {D.}~\bibnamefont
  {Schlingemann}}\ and\ \bibinfo {author} {\bibfnamefont {R.~F.}\ \bibnamefont
  {Werner}},\ }\bibfield  {title} {\bibinfo {title} {Quantum error-correcting
  codes associated with graphs},\ }\href@noop {} {\bibfield  {journal}
  {\bibinfo  {journal} {Phys. Rev. A}\ }\textbf {\bibinfo {volume} {65}},\
  \bibinfo {pages} {012308} (\bibinfo {year} {2001})}\BibitemShut {NoStop}%
\bibitem [{\citenamefont {Bell}\ \emph {et~al.}(2014)\citenamefont {Bell},
  \citenamefont {Herrera-Mart{\'\i}}, \citenamefont {Tame}, \citenamefont
  {Markham}, \citenamefont {Wadsworth},\ and\ \citenamefont
  {Rarity}}]{bell2014experimental}%
  \BibitemOpen
  \bibfield  {author} {\bibinfo {author} {\bibfnamefont {B.}~\bibnamefont
  {Bell}}, \bibinfo {author} {\bibfnamefont {D.}~\bibnamefont
  {Herrera-Mart{\'\i}}}, \bibinfo {author} {\bibfnamefont {M.}~\bibnamefont
  {Tame}}, \bibinfo {author} {\bibfnamefont {D.}~\bibnamefont {Markham}},
  \bibinfo {author} {\bibfnamefont {W.}~\bibnamefont {Wadsworth}},\ and\
  \bibinfo {author} {\bibfnamefont {J.}~\bibnamefont {Rarity}},\ }\bibfield
  {title} {\bibinfo {title} {Experimental demonstration of a graph state
  quantum error-correction code},\ }\href@noop {} {\bibfield  {journal}
  {\bibinfo  {journal} {Nat. Comm.}\ }\textbf {\bibinfo {volume} {5}},\
  \bibinfo {pages} {1} (\bibinfo {year} {2014})}\BibitemShut {NoStop}%
\bibitem [{\citenamefont {Nielsen}(2004)}]{nielsen2004optical}%
  \BibitemOpen
  \bibfield  {author} {\bibinfo {author} {\bibfnamefont {M.~A.}\ \bibnamefont
  {Nielsen}},\ }\bibfield  {title} {\bibinfo {title} {Optical quantum
  computation using cluster states},\ }\href@noop {} {\bibfield  {journal}
  {\bibinfo  {journal} {Phys. Rev. Lett.}\ }\textbf {\bibinfo {volume} {93}},\
  \bibinfo {pages} {040503} (\bibinfo {year} {2004})}\BibitemShut {NoStop}%
\bibitem [{\citenamefont {Browne}\ and\ \citenamefont
  {Rudolph}(2005)}]{browne2005resource}%
  \BibitemOpen
  \bibfield  {author} {\bibinfo {author} {\bibfnamefont {D.~E.}\ \bibnamefont
  {Browne}}\ and\ \bibinfo {author} {\bibfnamefont {T.}~\bibnamefont
  {Rudolph}},\ }\bibfield  {title} {\bibinfo {title} {Resource-efficient linear
  optical quantum computation},\ }\href@noop {} {\bibfield  {journal} {\bibinfo
   {journal} {Phys. Rev. Lett.}\ }\textbf {\bibinfo {volume} {95}},\ \bibinfo
  {pages} {010501} (\bibinfo {year} {2005})}\BibitemShut {NoStop}%
\bibitem [{\citenamefont {Kok}\ \emph {et~al.}(2007)\citenamefont {Kok},
  \citenamefont {Munro}, \citenamefont {Nemoto}, \citenamefont {Ralph},
  \citenamefont {Dowling},\ and\ \citenamefont {Milburn}}]{kok2007linear}%
  \BibitemOpen
  \bibfield  {author} {\bibinfo {author} {\bibfnamefont {P.}~\bibnamefont
  {Kok}}, \bibinfo {author} {\bibfnamefont {W.~J.}\ \bibnamefont {Munro}},
  \bibinfo {author} {\bibfnamefont {K.}~\bibnamefont {Nemoto}}, \bibinfo
  {author} {\bibfnamefont {T.~C.}\ \bibnamefont {Ralph}}, \bibinfo {author}
  {\bibfnamefont {J.~P.}\ \bibnamefont {Dowling}},\ and\ \bibinfo {author}
  {\bibfnamefont {G.~J.}\ \bibnamefont {Milburn}},\ }\bibfield  {title}
  {\bibinfo {title} {Linear optical quantum computing with photonic qubits},\
  }\href@noop {} {\bibfield  {journal} {\bibinfo  {journal} {Rev. Mod. Phys.}\
  }\textbf {\bibinfo {volume} {79}},\ \bibinfo {pages} {135} (\bibinfo {year}
  {2007})}\BibitemShut {NoStop}%
\bibitem [{\citenamefont {Lindner}\ and\ \citenamefont
  {Rudolph}(2009)}]{lindner2009proposal}%
  \BibitemOpen
  \bibfield  {author} {\bibinfo {author} {\bibfnamefont {N.~H.}\ \bibnamefont
  {Lindner}}\ and\ \bibinfo {author} {\bibfnamefont {T.}~\bibnamefont
  {Rudolph}},\ }\bibfield  {title} {\bibinfo {title} {Proposal for pulsed
  on-demand sources of photonic cluster state strings},\ }\href@noop {}
  {\bibfield  {journal} {\bibinfo  {journal} {Phys. Rev. Lett.}\ }\textbf
  {\bibinfo {volume} {103}},\ \bibinfo {pages} {113602} (\bibinfo {year}
  {2009})}\BibitemShut {NoStop}%
\bibitem [{\citenamefont {Schwartz}\ \emph {et~al.}(2016)\citenamefont
  {Schwartz}, \citenamefont {Cogan}, \citenamefont {Schmidgall}, \citenamefont
  {Don}, \citenamefont {Gantz}, \citenamefont {Kenneth}, \citenamefont
  {Lindner},\ and\ \citenamefont {Gershoni}}]{schwartz2016deterministic}%
  \BibitemOpen
  \bibfield  {author} {\bibinfo {author} {\bibfnamefont {I.}~\bibnamefont
  {Schwartz}}, \bibinfo {author} {\bibfnamefont {D.}~\bibnamefont {Cogan}},
  \bibinfo {author} {\bibfnamefont {E.~R.}\ \bibnamefont {Schmidgall}},
  \bibinfo {author} {\bibfnamefont {Y.}~\bibnamefont {Don}}, \bibinfo {author}
  {\bibfnamefont {L.}~\bibnamefont {Gantz}}, \bibinfo {author} {\bibfnamefont
  {O.}~\bibnamefont {Kenneth}}, \bibinfo {author} {\bibfnamefont {N.~H.}\
  \bibnamefont {Lindner}},\ and\ \bibinfo {author} {\bibfnamefont
  {D.}~\bibnamefont {Gershoni}},\ }\bibfield  {title} {\bibinfo {title}
  {Deterministic generation of a cluster state of entangled photons},\
  }\href@noop {} {\bibfield  {journal} {\bibinfo  {journal} {Science}\ }\textbf
  {\bibinfo {volume} {354}},\ \bibinfo {pages} {434} (\bibinfo {year}
  {2016})}\BibitemShut {NoStop}%
\bibitem [{\citenamefont {Pichler}\ \emph {et~al.}(2017)\citenamefont
  {Pichler}, \citenamefont {Choi}, \citenamefont {Zoller},\ and\ \citenamefont
  {Lukin}}]{pichler2017universal}%
  \BibitemOpen
  \bibfield  {author} {\bibinfo {author} {\bibfnamefont {H.}~\bibnamefont
  {Pichler}}, \bibinfo {author} {\bibfnamefont {S.}~\bibnamefont {Choi}},
  \bibinfo {author} {\bibfnamefont {P.}~\bibnamefont {Zoller}},\ and\ \bibinfo
  {author} {\bibfnamefont {M.~D.}\ \bibnamefont {Lukin}},\ }\bibfield  {title}
  {\bibinfo {title} {Universal photonic quantum computation via time-delayed
  feedback},\ }\href@noop {} {\bibfield  {journal} {\bibinfo  {journal} {PNAS}\
  }\textbf {\bibinfo {volume} {114}},\ \bibinfo {pages} {11362} (\bibinfo
  {year} {2017})}\BibitemShut {NoStop}%
\bibitem [{\citenamefont {Wan}\ \emph {et~al.}(2021)\citenamefont {Wan},
  \citenamefont {Choi}, \citenamefont {Kim}, \citenamefont {Shutty},\ and\
  \citenamefont {Hayden}}]{wan2021fault}%
  \BibitemOpen
  \bibfield  {author} {\bibinfo {author} {\bibfnamefont {K.}~\bibnamefont
  {Wan}}, \bibinfo {author} {\bibfnamefont {S.}~\bibnamefont {Choi}}, \bibinfo
  {author} {\bibfnamefont {I.~H.}\ \bibnamefont {Kim}}, \bibinfo {author}
  {\bibfnamefont {N.}~\bibnamefont {Shutty}},\ and\ \bibinfo {author}
  {\bibfnamefont {P.}~\bibnamefont {Hayden}},\ }\bibfield  {title} {\bibinfo
  {title} {Fault-tolerant qubit from a constant number of components},\
  }\href@noop {} {\bibfield  {journal} {\bibinfo  {journal} {PRX Quant.}\
  }\textbf {\bibinfo {volume} {2}},\ \bibinfo {pages} {040345} (\bibinfo {year}
  {2021})}\BibitemShut {NoStop}%
\bibitem [{\citenamefont {Shi}\ and\ \citenamefont
  {Waks}(2021)}]{shi2021deterministic}%
  \BibitemOpen
  \bibfield  {author} {\bibinfo {author} {\bibfnamefont {Y.}~\bibnamefont
  {Shi}}\ and\ \bibinfo {author} {\bibfnamefont {E.}~\bibnamefont {Waks}},\
  }\bibfield  {title} {\bibinfo {title} {Deterministic generation of
  multidimensional photonic cluster states using time-delay feedback},\
  }\href@noop {} {\bibfield  {journal} {\bibinfo  {journal} {Phys. Rev. A}\
  }\textbf {\bibinfo {volume} {104}},\ \bibinfo {pages} {013703} (\bibinfo
  {year} {2021})}\BibitemShut {NoStop}%
\bibitem [{\citenamefont {Zhan}\ and\ \citenamefont
  {Sun}(2020)}]{zhan2020deterministic}%
  \BibitemOpen
  \bibfield  {author} {\bibinfo {author} {\bibfnamefont {Y.}~\bibnamefont
  {Zhan}}\ and\ \bibinfo {author} {\bibfnamefont {S.}~\bibnamefont {Sun}},\
  }\bibfield  {title} {\bibinfo {title} {Deterministic generation of
  loss-tolerant photonic cluster states with a single quantum emitter},\
  }\href@noop {} {\bibfield  {journal} {\bibinfo  {journal} {Phys. Rev. Lett.}\
  }\textbf {\bibinfo {volume} {125}},\ \bibinfo {pages} {223601} (\bibinfo
  {year} {2020})}\BibitemShut {NoStop}%
\bibitem [{\citenamefont {Xu}\ and\ \citenamefont
  {Fan}(2018)}]{xu2018generate}%
  \BibitemOpen
  \bibfield  {author} {\bibinfo {author} {\bibfnamefont {S.}~\bibnamefont
  {Xu}}\ and\ \bibinfo {author} {\bibfnamefont {S.}~\bibnamefont {Fan}},\
  }\bibfield  {title} {\bibinfo {title} {Generate tensor network state by
  sequential single-photon scattering in waveguide qed systems},\ }\href@noop
  {} {\bibfield  {journal} {\bibinfo  {journal} {APL Phot.}\ }\textbf {\bibinfo
  {volume} {3}},\ \bibinfo {pages} {116102} (\bibinfo {year}
  {2018})}\BibitemShut {NoStop}%
\bibitem [{\citenamefont {Goban}\ \emph {et~al.}(2014)\citenamefont {Goban},
  \citenamefont {Hung}, \citenamefont {Yu}, \citenamefont {Hood}, \citenamefont
  {Muniz}, \citenamefont {Lee}, \citenamefont {Martin}, \citenamefont
  {McClung}, \citenamefont {Choi}, \citenamefont {Chang} \emph
  {et~al.}}]{goban2014atom}%
  \BibitemOpen
  \bibfield  {author} {\bibinfo {author} {\bibfnamefont {A.}~\bibnamefont
  {Goban}}, \bibinfo {author} {\bibfnamefont {C.-L.}\ \bibnamefont {Hung}},
  \bibinfo {author} {\bibfnamefont {S.-P.}\ \bibnamefont {Yu}}, \bibinfo
  {author} {\bibfnamefont {J.}~\bibnamefont {Hood}}, \bibinfo {author}
  {\bibfnamefont {J.}~\bibnamefont {Muniz}}, \bibinfo {author} {\bibfnamefont
  {J.}~\bibnamefont {Lee}}, \bibinfo {author} {\bibfnamefont {M.}~\bibnamefont
  {Martin}}, \bibinfo {author} {\bibfnamefont {A.}~\bibnamefont {McClung}},
  \bibinfo {author} {\bibfnamefont {K.}~\bibnamefont {Choi}}, \bibinfo {author}
  {\bibfnamefont {D.~E.}\ \bibnamefont {Chang}}, \emph {et~al.},\ }\bibfield
  {title} {\bibinfo {title} {Atom--light interactions in photonic crystals},\
  }\href@noop {} {\bibfield  {journal} {\bibinfo  {journal} {Nat. Comm.}\
  }\textbf {\bibinfo {volume} {5}},\ \bibinfo {pages} {1} (\bibinfo {year}
  {2014})}\BibitemShut {NoStop}%
\bibitem [{\citenamefont {Corzo}\ \emph {et~al.}(2019)\citenamefont {Corzo},
  \citenamefont {Raskop}, \citenamefont {Chandra}, \citenamefont {Sheremet},
  \citenamefont {Gouraud},\ and\ \citenamefont {Laurat}}]{corzo2019waveguide}%
  \BibitemOpen
  \bibfield  {author} {\bibinfo {author} {\bibfnamefont {N.~V.}\ \bibnamefont
  {Corzo}}, \bibinfo {author} {\bibfnamefont {J.}~\bibnamefont {Raskop}},
  \bibinfo {author} {\bibfnamefont {A.}~\bibnamefont {Chandra}}, \bibinfo
  {author} {\bibfnamefont {A.~S.}\ \bibnamefont {Sheremet}}, \bibinfo {author}
  {\bibfnamefont {B.}~\bibnamefont {Gouraud}},\ and\ \bibinfo {author}
  {\bibfnamefont {J.}~\bibnamefont {Laurat}},\ }\bibfield  {title} {\bibinfo
  {title} {Waveguide-coupled single collective excitation of atomic arrays},\
  }\href@noop {} {\bibfield  {journal} {\bibinfo  {journal} {Nature}\ }\textbf
  {\bibinfo {volume} {566}},\ \bibinfo {pages} {359} (\bibinfo {year}
  {2019})}\BibitemShut {NoStop}%
\bibitem [{\citenamefont {Blais}\ \emph {et~al.}(2004)\citenamefont {Blais},
  \citenamefont {Huang}, \citenamefont {Wallraff}, \citenamefont {Girvin},\
  and\ \citenamefont {Schoelkopf}}]{blais2004cavity}%
  \BibitemOpen
  \bibfield  {author} {\bibinfo {author} {\bibfnamefont {A.}~\bibnamefont
  {Blais}}, \bibinfo {author} {\bibfnamefont {R.-S.}\ \bibnamefont {Huang}},
  \bibinfo {author} {\bibfnamefont {A.}~\bibnamefont {Wallraff}}, \bibinfo
  {author} {\bibfnamefont {S.~M.}\ \bibnamefont {Girvin}},\ and\ \bibinfo
  {author} {\bibfnamefont {R.~J.}\ \bibnamefont {Schoelkopf}},\ }\bibfield
  {title} {\bibinfo {title} {Cavity quantum electrodynamics for superconducting
  electrical circuits: An architecture for quantum computation},\ }\href@noop
  {} {\bibfield  {journal} {\bibinfo  {journal} {Phys. Rev. A}\ }\textbf
  {\bibinfo {volume} {69}},\ \bibinfo {pages} {062320} (\bibinfo {year}
  {2004})}\BibitemShut {NoStop}%
\bibitem [{\citenamefont {Eichler}\ \emph {et~al.}(2011)\citenamefont
  {Eichler}, \citenamefont {Bozyigit}, \citenamefont {Lang}, \citenamefont
  {Steffen}, \citenamefont {Fink},\ and\ \citenamefont
  {Wallraff}}]{eichler2011experimental}%
  \BibitemOpen
  \bibfield  {author} {\bibinfo {author} {\bibfnamefont {C.}~\bibnamefont
  {Eichler}}, \bibinfo {author} {\bibfnamefont {D.}~\bibnamefont {Bozyigit}},
  \bibinfo {author} {\bibfnamefont {C.}~\bibnamefont {Lang}}, \bibinfo {author}
  {\bibfnamefont {L.}~\bibnamefont {Steffen}}, \bibinfo {author} {\bibfnamefont
  {J.}~\bibnamefont {Fink}},\ and\ \bibinfo {author} {\bibfnamefont
  {A.}~\bibnamefont {Wallraff}},\ }\bibfield  {title} {\bibinfo {title}
  {Experimental state tomography of itinerant single microwave photons},\
  }\href@noop {} {\bibfield  {journal} {\bibinfo  {journal} {Phys. Rev. Lett.}\
  }\textbf {\bibinfo {volume} {106}},\ \bibinfo {pages} {220503} (\bibinfo
  {year} {2011})}\BibitemShut {NoStop}%
\bibitem [{\citenamefont {Hoi}\ \emph {et~al.}(2012)\citenamefont {Hoi},
  \citenamefont {Palomaki}, \citenamefont {Lindkvist}, \citenamefont
  {Johansson}, \citenamefont {Delsing},\ and\ \citenamefont
  {Wilson}}]{hoi2012generation}%
  \BibitemOpen
  \bibfield  {author} {\bibinfo {author} {\bibfnamefont {I.-C.}\ \bibnamefont
  {Hoi}}, \bibinfo {author} {\bibfnamefont {T.}~\bibnamefont {Palomaki}},
  \bibinfo {author} {\bibfnamefont {J.}~\bibnamefont {Lindkvist}}, \bibinfo
  {author} {\bibfnamefont {G.}~\bibnamefont {Johansson}}, \bibinfo {author}
  {\bibfnamefont {P.}~\bibnamefont {Delsing}},\ and\ \bibinfo {author}
  {\bibfnamefont {C.}~\bibnamefont {Wilson}},\ }\bibfield  {title} {\bibinfo
  {title} {Generation of nonclassical microwave states using an artificial atom
  in 1d open space},\ }\href@noop {} {\bibfield  {journal} {\bibinfo  {journal}
  {Phys. Rev. Lett.}\ }\textbf {\bibinfo {volume} {108}},\ \bibinfo {pages}
  {263601} (\bibinfo {year} {2012})}\BibitemShut {NoStop}%
\bibitem [{\citenamefont {Lang}\ \emph {et~al.}(2013)\citenamefont {Lang},
  \citenamefont {Eichler}, \citenamefont {Steffen}, \citenamefont {Fink},
  \citenamefont {Woolley}, \citenamefont {Blais},\ and\ \citenamefont
  {Wallraff}}]{lang2013correlations}%
  \BibitemOpen
  \bibfield  {author} {\bibinfo {author} {\bibfnamefont {C.}~\bibnamefont
  {Lang}}, \bibinfo {author} {\bibfnamefont {C.}~\bibnamefont {Eichler}},
  \bibinfo {author} {\bibfnamefont {L.}~\bibnamefont {Steffen}}, \bibinfo
  {author} {\bibfnamefont {J.}~\bibnamefont {Fink}}, \bibinfo {author}
  {\bibfnamefont {M.~J.}\ \bibnamefont {Woolley}}, \bibinfo {author}
  {\bibfnamefont {A.}~\bibnamefont {Blais}},\ and\ \bibinfo {author}
  {\bibfnamefont {A.}~\bibnamefont {Wallraff}},\ }\bibfield  {title} {\bibinfo
  {title} {Correlations, indistinguishability and entanglement in
  hong--ou--mandel experiments at microwave frequencies},\ }\href@noop {}
  {\bibfield  {journal} {\bibinfo  {journal} {Nat. Phys.}\ }\textbf {\bibinfo
  {volume} {9}},\ \bibinfo {pages} {345} (\bibinfo {year} {2013})}\BibitemShut
  {NoStop}%
\bibitem [{\citenamefont {Eichler}\ \emph {et~al.}(2015)\citenamefont
  {Eichler}, \citenamefont {Mlynek}, \citenamefont {Butscher}, \citenamefont
  {Kurpiers}, \citenamefont {Hammerer}, \citenamefont {Osborne},\ and\
  \citenamefont {Wallraff}}]{eichler2015exploring}%
  \BibitemOpen
  \bibfield  {author} {\bibinfo {author} {\bibfnamefont {C.}~\bibnamefont
  {Eichler}}, \bibinfo {author} {\bibfnamefont {J.}~\bibnamefont {Mlynek}},
  \bibinfo {author} {\bibfnamefont {J.}~\bibnamefont {Butscher}}, \bibinfo
  {author} {\bibfnamefont {P.}~\bibnamefont {Kurpiers}}, \bibinfo {author}
  {\bibfnamefont {K.}~\bibnamefont {Hammerer}}, \bibinfo {author}
  {\bibfnamefont {T.~J.}\ \bibnamefont {Osborne}},\ and\ \bibinfo {author}
  {\bibfnamefont {A.}~\bibnamefont {Wallraff}},\ }\bibfield  {title} {\bibinfo
  {title} {Exploring interacting quantum many-body systems by experimentally
  creating continuous matrix product states in superconducting circuits},\
  }\href@noop {} {\bibfield  {journal} {\bibinfo  {journal} {Phys. Rev. X}\
  }\textbf {\bibinfo {volume} {5}},\ \bibinfo {pages} {041044} (\bibinfo {year}
  {2015})}\BibitemShut {NoStop}%
\bibitem [{\citenamefont {Kannan}\ \emph {et~al.}(2020)\citenamefont {Kannan},
  \citenamefont {Campbell}, \citenamefont {Vasconcelos}, \citenamefont {Winik},
  \citenamefont {Kim}, \citenamefont {Kjaergaard}, \citenamefont {Krantz},
  \citenamefont {Melville}, \citenamefont {Niedzielski}, \citenamefont {Yoder}
  \emph {et~al.}}]{kannan2020generating}%
  \BibitemOpen
  \bibfield  {author} {\bibinfo {author} {\bibfnamefont {B.}~\bibnamefont
  {Kannan}}, \bibinfo {author} {\bibfnamefont {D.~L.}\ \bibnamefont
  {Campbell}}, \bibinfo {author} {\bibfnamefont {F.}~\bibnamefont
  {Vasconcelos}}, \bibinfo {author} {\bibfnamefont {R.}~\bibnamefont {Winik}},
  \bibinfo {author} {\bibfnamefont {D.}~\bibnamefont {Kim}}, \bibinfo {author}
  {\bibfnamefont {M.}~\bibnamefont {Kjaergaard}}, \bibinfo {author}
  {\bibfnamefont {P.}~\bibnamefont {Krantz}}, \bibinfo {author} {\bibfnamefont
  {A.}~\bibnamefont {Melville}}, \bibinfo {author} {\bibfnamefont {B.~M.}\
  \bibnamefont {Niedzielski}}, \bibinfo {author} {\bibfnamefont
  {J.}~\bibnamefont {Yoder}}, \emph {et~al.},\ }\bibfield  {title} {\bibinfo
  {title} {Generating spatially entangled itinerant photons with waveguide
  quantum electrodynamics},\ }\href@noop {} {\bibfield  {journal} {\bibinfo
  {journal} {Science Advances}\ }\textbf {\bibinfo {volume} {6}},\ \bibinfo
  {pages} {eabb8780} (\bibinfo {year} {2020})}\BibitemShut {NoStop}%
\bibitem [{\citenamefont {Besse}\ \emph {et~al.}(2020)\citenamefont {Besse},
  \citenamefont {Reuer}, \citenamefont {C.~Collodo}, \citenamefont {Wulff},
  \citenamefont {Wernli}, \citenamefont {Copetudo}, \citenamefont {Malz},
  \citenamefont {Magnard}, \citenamefont {Akin}, \citenamefont {Gabureac},
  \citenamefont {Norris}, \citenamefont {Cirac}, \citenamefont {Wallraff},\
  and\ \citenamefont {Eichler}}]{besse2020realizing}%
  \BibitemOpen
  \bibfield  {author} {\bibinfo {author} {\bibfnamefont {J.-C.}\ \bibnamefont
  {Besse}}, \bibinfo {author} {\bibfnamefont {K.}~\bibnamefont {Reuer}},
  \bibinfo {author} {\bibfnamefont {M.}~\bibnamefont {C.~Collodo}}, \bibinfo
  {author} {\bibfnamefont {A.}~\bibnamefont {Wulff}}, \bibinfo {author}
  {\bibfnamefont {L.}~\bibnamefont {Wernli}}, \bibinfo {author} {\bibfnamefont
  {A.}~\bibnamefont {Copetudo}}, \bibinfo {author} {\bibfnamefont
  {D.}~\bibnamefont {Malz}}, \bibinfo {author} {\bibfnamefont {P.}~\bibnamefont
  {Magnard}}, \bibinfo {author} {\bibfnamefont {A.}~\bibnamefont {Akin}},
  \bibinfo {author} {\bibfnamefont {M.}~\bibnamefont {Gabureac}}, \bibinfo
  {author} {\bibfnamefont {G.~J.}\ \bibnamefont {Norris}}, \bibinfo {author}
  {\bibfnamefont {J.~I.}\ \bibnamefont {Cirac}}, \bibinfo {author}
  {\bibfnamefont {A.}~\bibnamefont {Wallraff}},\ and\ \bibinfo {author}
  {\bibfnamefont {C.}~\bibnamefont {Eichler}},\ }\bibfield  {title} {\bibinfo
  {title} {Realizing a deterministic source of multipartite-entangled photonic
  qubits},\ }\href@noop {} {\bibfield  {journal} {\bibinfo  {journal} {Nat.
  Commun.}\ }\textbf {\bibinfo {volume} {11}} (\bibinfo {year}
  {2020})}\BibitemShut {NoStop}%
\bibitem [{\citenamefont {Ferreira}\ \emph {et~al.}(2021)\citenamefont
  {Ferreira}, \citenamefont {Banker}, \citenamefont {Sipahigil}, \citenamefont
  {Matheny}, \citenamefont {Keller}, \citenamefont {Kim}, \citenamefont
  {Mirhosseini},\ and\ \citenamefont {Painter}}]{ferreira2021collapse}%
  \BibitemOpen
  \bibfield  {author} {\bibinfo {author} {\bibfnamefont {V.~S.}\ \bibnamefont
  {Ferreira}}, \bibinfo {author} {\bibfnamefont {J.}~\bibnamefont {Banker}},
  \bibinfo {author} {\bibfnamefont {A.}~\bibnamefont {Sipahigil}}, \bibinfo
  {author} {\bibfnamefont {M.~H.}\ \bibnamefont {Matheny}}, \bibinfo {author}
  {\bibfnamefont {A.~J.}\ \bibnamefont {Keller}}, \bibinfo {author}
  {\bibfnamefont {E.}~\bibnamefont {Kim}}, \bibinfo {author} {\bibfnamefont
  {M.}~\bibnamefont {Mirhosseini}},\ and\ \bibinfo {author} {\bibfnamefont
  {O.}~\bibnamefont {Painter}},\ }\bibfield  {title} {\bibinfo {title}
  {Collapse and revival of an artificial atom coupled to a structured photonic
  reservoir},\ }\href@noop {} {\bibfield  {journal} {\bibinfo  {journal} {Phys.
  Rev. X}\ }\textbf {\bibinfo {volume} {11}},\ \bibinfo {pages} {041043}
  (\bibinfo {year} {2021})}\BibitemShut {NoStop}%
\bibitem [{\citenamefont {Shen}\ and\ \citenamefont
  {Fan}(2005)}]{shen2005coherent}%
  \BibitemOpen
  \bibfield  {author} {\bibinfo {author} {\bibfnamefont {J.-T.}\ \bibnamefont
  {Shen}}\ and\ \bibinfo {author} {\bibfnamefont {S.}~\bibnamefont {Fan}},\
  }\bibfield  {title} {\bibinfo {title} {Coherent single photon transport in a
  one-dimensional waveguide coupled with superconducting quantum bits},\
  }\href@noop {} {\bibfield  {journal} {\bibinfo  {journal} {Phys. Rev. Lett.}\
  }\textbf {\bibinfo {volume} {95}},\ \bibinfo {pages} {213001} (\bibinfo
  {year} {2005})}\BibitemShut {NoStop}%
\bibitem [{\citenamefont {Beaudoin}\ \emph {et~al.}(2012)\citenamefont
  {Beaudoin}, \citenamefont {da~Silva}, \citenamefont {Dutton},\ and\
  \citenamefont {Blais}}]{beaudoin2012first}%
  \BibitemOpen
  \bibfield  {author} {\bibinfo {author} {\bibfnamefont {F.}~\bibnamefont
  {Beaudoin}}, \bibinfo {author} {\bibfnamefont {M.~P.}\ \bibnamefont
  {da~Silva}}, \bibinfo {author} {\bibfnamefont {Z.}~\bibnamefont {Dutton}},\
  and\ \bibinfo {author} {\bibfnamefont {A.}~\bibnamefont {Blais}},\ }\bibfield
   {title} {\bibinfo {title} {First-order sidebands in circuit qed using qubit
  frequency modulation},\ }\href@noop {} {\bibfield  {journal} {\bibinfo
  {journal} {Phys. Rev. A}\ }\textbf {\bibinfo {volume} {86}},\ \bibinfo
  {pages} {022305} (\bibinfo {year} {2012})}\BibitemShut {NoStop}%
\bibitem [{\citenamefont {Strand}\ \emph {et~al.}(2013)\citenamefont {Strand},
  \citenamefont {Ware}, \citenamefont {Beaudoin}, \citenamefont {Ohki},
  \citenamefont {Johnson}, \citenamefont {Blais},\ and\ \citenamefont
  {Plourde}}]{strand2013first}%
  \BibitemOpen
  \bibfield  {author} {\bibinfo {author} {\bibfnamefont {J.}~\bibnamefont
  {Strand}}, \bibinfo {author} {\bibfnamefont {M.}~\bibnamefont {Ware}},
  \bibinfo {author} {\bibfnamefont {F.}~\bibnamefont {Beaudoin}}, \bibinfo
  {author} {\bibfnamefont {T.}~\bibnamefont {Ohki}}, \bibinfo {author}
  {\bibfnamefont {B.}~\bibnamefont {Johnson}}, \bibinfo {author} {\bibfnamefont
  {A.}~\bibnamefont {Blais}},\ and\ \bibinfo {author} {\bibfnamefont
  {B.}~\bibnamefont {Plourde}},\ }\bibfield  {title} {\bibinfo {title}
  {First-order sideband transitions with flux-driven asymmetric transmon
  qubits},\ }\href@noop {} {\bibfield  {journal} {\bibinfo  {journal} {Physical
  Review B}\ }\textbf {\bibinfo {volume} {87}},\ \bibinfo {pages} {220505}
  (\bibinfo {year} {2013})}\BibitemShut {NoStop}%
\bibitem [{\citenamefont {Silveri}\ \emph {et~al.}(2017)\citenamefont
  {Silveri}, \citenamefont {Tuorila}, \citenamefont {Thuneberg},\ and\
  \citenamefont {Paraoanu}}]{silveri2017quantum}%
  \BibitemOpen
  \bibfield  {author} {\bibinfo {author} {\bibfnamefont {M.}~\bibnamefont
  {Silveri}}, \bibinfo {author} {\bibfnamefont {J.}~\bibnamefont {Tuorila}},
  \bibinfo {author} {\bibfnamefont {E.}~\bibnamefont {Thuneberg}},\ and\
  \bibinfo {author} {\bibfnamefont {G.}~\bibnamefont {Paraoanu}},\ }\bibfield
  {title} {\bibinfo {title} {Quantum systems under frequency modulation},\
  }\href@noop {} {\bibfield  {journal} {\bibinfo  {journal} {Reports on
  Progress in Physics}\ }\textbf {\bibinfo {volume} {80}},\ \bibinfo {pages}
  {056002} (\bibinfo {year} {2017})}\BibitemShut {NoStop}%
\bibitem [{\citenamefont {Pechal}\ \emph {et~al.}(2014)\citenamefont {Pechal},
  \citenamefont {Huthmacher}, \citenamefont {Eichler}, \citenamefont
  {Zeytino{\u{g}}lu}, \citenamefont {Abdumalikov~Jr}, \citenamefont {Berger},
  \citenamefont {Wallraff},\ and\ \citenamefont
  {Filipp}}]{pechal2014microwave}%
  \BibitemOpen
  \bibfield  {author} {\bibinfo {author} {\bibfnamefont {M.}~\bibnamefont
  {Pechal}}, \bibinfo {author} {\bibfnamefont {L.}~\bibnamefont {Huthmacher}},
  \bibinfo {author} {\bibfnamefont {C.}~\bibnamefont {Eichler}}, \bibinfo
  {author} {\bibfnamefont {S.}~\bibnamefont {Zeytino{\u{g}}lu}}, \bibinfo
  {author} {\bibfnamefont {A.}~\bibnamefont {Abdumalikov~Jr}}, \bibinfo
  {author} {\bibfnamefont {S.}~\bibnamefont {Berger}}, \bibinfo {author}
  {\bibfnamefont {A.}~\bibnamefont {Wallraff}},\ and\ \bibinfo {author}
  {\bibfnamefont {S.}~\bibnamefont {Filipp}},\ }\bibfield  {title} {\bibinfo
  {title} {Microwave-controlled generation of shaped single photons in circuit
  quantum electrodynamics},\ }\href@noop {} {\bibfield  {journal} {\bibinfo
  {journal} {Phys. Rev. X}\ }\textbf {\bibinfo {volume} {4}},\ \bibinfo {pages}
  {041010} (\bibinfo {year} {2014})}\BibitemShut {NoStop}%
\bibitem [{\citenamefont {Forn-Diaz}\ \emph {et~al.}(2017)\citenamefont
  {Forn-Diaz}, \citenamefont {Warren}, \citenamefont {Chang}, \citenamefont
  {Vadiraj},\ and\ \citenamefont {Wilson}}]{forn2017demand}%
  \BibitemOpen
  \bibfield  {author} {\bibinfo {author} {\bibfnamefont {P.}~\bibnamefont
  {Forn-Diaz}}, \bibinfo {author} {\bibfnamefont {C.}~\bibnamefont {Warren}},
  \bibinfo {author} {\bibfnamefont {C.}~\bibnamefont {Chang}}, \bibinfo
  {author} {\bibfnamefont {A.}~\bibnamefont {Vadiraj}},\ and\ \bibinfo {author}
  {\bibfnamefont {C.}~\bibnamefont {Wilson}},\ }\bibfield  {title} {\bibinfo
  {title} {On-demand microwave generator of shaped single photons},\
  }\href@noop {} {\bibfield  {journal} {\bibinfo  {journal} {Phys. Rev. Appl.}\
  }\textbf {\bibinfo {volume} {8}},\ \bibinfo {pages} {054015} (\bibinfo {year}
  {2017})}\BibitemShut {NoStop}%
\bibitem [{\citenamefont {Ilves}\ \emph {et~al.}(2020)\citenamefont {Ilves},
  \citenamefont {Kono}, \citenamefont {Sunada}, \citenamefont {Yamazaki},
  \citenamefont {Kim}, \citenamefont {Koshino},\ and\ \citenamefont
  {Nakamura}}]{ilves2020demand}%
  \BibitemOpen
  \bibfield  {author} {\bibinfo {author} {\bibfnamefont {J.}~\bibnamefont
  {Ilves}}, \bibinfo {author} {\bibfnamefont {S.}~\bibnamefont {Kono}},
  \bibinfo {author} {\bibfnamefont {Y.}~\bibnamefont {Sunada}}, \bibinfo
  {author} {\bibfnamefont {S.}~\bibnamefont {Yamazaki}}, \bibinfo {author}
  {\bibfnamefont {M.}~\bibnamefont {Kim}}, \bibinfo {author} {\bibfnamefont
  {K.}~\bibnamefont {Koshino}},\ and\ \bibinfo {author} {\bibfnamefont
  {Y.}~\bibnamefont {Nakamura}},\ }\bibfield  {title} {\bibinfo {title}
  {On-demand generation and characterization of a microwave time-bin qubit},\
  }\href@noop {} {\bibfield  {journal} {\bibinfo  {journal} {npj Quantum Inf.}\
  }\textbf {\bibinfo {volume} {6}},\ \bibinfo {pages} {1} (\bibinfo {year}
  {2020})}\BibitemShut {NoStop}%
\bibitem [{\citenamefont {Reuer}\ \emph {et~al.}(2022)\citenamefont {Reuer},
  \citenamefont {Besse}, \citenamefont {Wernli}, \citenamefont {Magnard},
  \citenamefont {Kurpiers}, \citenamefont {Norris}, \citenamefont {Wallraff},\
  and\ \citenamefont {Eichler}}]{reuer2021realization}%
  \BibitemOpen
  \bibfield  {author} {\bibinfo {author} {\bibfnamefont {K.}~\bibnamefont
  {Reuer}}, \bibinfo {author} {\bibfnamefont {J.-C.}\ \bibnamefont {Besse}},
  \bibinfo {author} {\bibfnamefont {L.}~\bibnamefont {Wernli}}, \bibinfo
  {author} {\bibfnamefont {P.}~\bibnamefont {Magnard}}, \bibinfo {author}
  {\bibfnamefont {P.}~\bibnamefont {Kurpiers}}, \bibinfo {author}
  {\bibfnamefont {G.~J.}\ \bibnamefont {Norris}}, \bibinfo {author}
  {\bibfnamefont {A.}~\bibnamefont {Wallraff}},\ and\ \bibinfo {author}
  {\bibfnamefont {C.}~\bibnamefont {Eichler}},\ }\bibfield  {title} {\bibinfo
  {title} {Realization of a universal quantum gate set for itinerant microwave
  photons},\ }\href@noop {} {\bibfield  {journal} {\bibinfo  {journal} {Phys.
  Rev. X}\ }\textbf {\bibinfo {volume} {12}},\ \bibinfo {pages} {011008}
  (\bibinfo {year} {2022})}\BibitemShut {NoStop}%
\bibitem [{\citenamefont {Megrant}\ \emph {et~al.}(2012)\citenamefont
  {Megrant}, \citenamefont {Neill}, \citenamefont {Barends}, \citenamefont
  {Chiaro}, \citenamefont {Chen}, \citenamefont {Feigl}, \citenamefont {Kelly},
  \citenamefont {Lucero}, \citenamefont {Mariantoni}, \citenamefont
  {O’Malley} \emph {et~al.}}]{megrant2012planar}%
  \BibitemOpen
  \bibfield  {author} {\bibinfo {author} {\bibfnamefont {A.}~\bibnamefont
  {Megrant}}, \bibinfo {author} {\bibfnamefont {C.}~\bibnamefont {Neill}},
  \bibinfo {author} {\bibfnamefont {R.}~\bibnamefont {Barends}}, \bibinfo
  {author} {\bibfnamefont {B.}~\bibnamefont {Chiaro}}, \bibinfo {author}
  {\bibfnamefont {Y.}~\bibnamefont {Chen}}, \bibinfo {author} {\bibfnamefont
  {L.}~\bibnamefont {Feigl}}, \bibinfo {author} {\bibfnamefont
  {J.}~\bibnamefont {Kelly}}, \bibinfo {author} {\bibfnamefont
  {E.}~\bibnamefont {Lucero}}, \bibinfo {author} {\bibfnamefont
  {M.}~\bibnamefont {Mariantoni}}, \bibinfo {author} {\bibfnamefont {P.~J.}\
  \bibnamefont {O’Malley}}, \emph {et~al.},\ }\bibfield  {title} {\bibinfo
  {title} {Planar superconducting resonators with internal quality factors
  above one million},\ }\href@noop {} {\bibfield  {journal} {\bibinfo
  {journal} {Appl. Phys. Lett.}\ }\textbf {\bibinfo {volume} {100}},\ \bibinfo
  {pages} {113510} (\bibinfo {year} {2012})}\BibitemShut {NoStop}%
\bibitem [{\citenamefont {Calusine}\ \emph {et~al.}(2018)\citenamefont
  {Calusine}, \citenamefont {Melville}, \citenamefont {Woods}, \citenamefont
  {Das}, \citenamefont {Stull}, \citenamefont {Bolkhovsky}, \citenamefont
  {Braje}, \citenamefont {Hover}, \citenamefont {Kim}, \citenamefont {Miloshi}
  \emph {et~al.}}]{calusine2018analysis}%
  \BibitemOpen
  \bibfield  {author} {\bibinfo {author} {\bibfnamefont {G.}~\bibnamefont
  {Calusine}}, \bibinfo {author} {\bibfnamefont {A.}~\bibnamefont {Melville}},
  \bibinfo {author} {\bibfnamefont {W.}~\bibnamefont {Woods}}, \bibinfo
  {author} {\bibfnamefont {R.}~\bibnamefont {Das}}, \bibinfo {author}
  {\bibfnamefont {C.}~\bibnamefont {Stull}}, \bibinfo {author} {\bibfnamefont
  {V.}~\bibnamefont {Bolkhovsky}}, \bibinfo {author} {\bibfnamefont
  {D.}~\bibnamefont {Braje}}, \bibinfo {author} {\bibfnamefont
  {D.}~\bibnamefont {Hover}}, \bibinfo {author} {\bibfnamefont {D.~K.}\
  \bibnamefont {Kim}}, \bibinfo {author} {\bibfnamefont {X.}~\bibnamefont
  {Miloshi}}, \emph {et~al.},\ }\bibfield  {title} {\bibinfo {title} {Analysis
  and mitigation of interface losses in trenched superconducting coplanar
  waveguide resonators},\ }\href@noop {} {\bibfield  {journal} {\bibinfo
  {journal} {Appl. Phys. Lett.}\ }\textbf {\bibinfo {volume} {112}},\ \bibinfo
  {pages} {062601} (\bibinfo {year} {2018})}\BibitemShut {NoStop}%
\bibitem [{\citenamefont {Woods}\ \emph {et~al.}(2019)\citenamefont {Woods},
  \citenamefont {Calusine}, \citenamefont {Melville}, \citenamefont {Sevi},
  \citenamefont {Golden}, \citenamefont {Kim}, \citenamefont {Rosenberg},
  \citenamefont {Yoder},\ and\ \citenamefont {Oliver}}]{woods2019determining}%
  \BibitemOpen
  \bibfield  {author} {\bibinfo {author} {\bibfnamefont {W.}~\bibnamefont
  {Woods}}, \bibinfo {author} {\bibfnamefont {G.}~\bibnamefont {Calusine}},
  \bibinfo {author} {\bibfnamefont {A.}~\bibnamefont {Melville}}, \bibinfo
  {author} {\bibfnamefont {A.}~\bibnamefont {Sevi}}, \bibinfo {author}
  {\bibfnamefont {E.}~\bibnamefont {Golden}}, \bibinfo {author} {\bibfnamefont
  {D.~K.}\ \bibnamefont {Kim}}, \bibinfo {author} {\bibfnamefont
  {D.}~\bibnamefont {Rosenberg}}, \bibinfo {author} {\bibfnamefont {J.~L.}\
  \bibnamefont {Yoder}},\ and\ \bibinfo {author} {\bibfnamefont {W.~D.}\
  \bibnamefont {Oliver}},\ }\bibfield  {title} {\bibinfo {title} {Determining
  interface dielectric losses in superconducting coplanar-waveguide
  resonators},\ }\href@noop {} {\bibfield  {journal} {\bibinfo  {journal}
  {Phys. Rev. Appl.}\ }\textbf {\bibinfo {volume} {12}},\ \bibinfo {pages}
  {014012} (\bibinfo {year} {2019})}\BibitemShut {NoStop}%
\bibitem [{\citenamefont {Nguyen}\ \emph {et~al.}(2019)\citenamefont {Nguyen},
  \citenamefont {Lin}, \citenamefont {Somoroff}, \citenamefont {Mencia},
  \citenamefont {Grabon},\ and\ \citenamefont {Manucharyan}}]{nguyen2019high}%
  \BibitemOpen
  \bibfield  {author} {\bibinfo {author} {\bibfnamefont {L.~B.}\ \bibnamefont
  {Nguyen}}, \bibinfo {author} {\bibfnamefont {Y.-H.}\ \bibnamefont {Lin}},
  \bibinfo {author} {\bibfnamefont {A.}~\bibnamefont {Somoroff}}, \bibinfo
  {author} {\bibfnamefont {R.}~\bibnamefont {Mencia}}, \bibinfo {author}
  {\bibfnamefont {N.}~\bibnamefont {Grabon}},\ and\ \bibinfo {author}
  {\bibfnamefont {V.~E.}\ \bibnamefont {Manucharyan}},\ }\bibfield  {title}
  {\bibinfo {title} {High-coherence fluxonium qubit},\ }\href@noop {}
  {\bibfield  {journal} {\bibinfo  {journal} {Phys. Rev. X}\ }\textbf {\bibinfo
  {volume} {9}},\ \bibinfo {pages} {041041} (\bibinfo {year}
  {2019})}\BibitemShut {NoStop}%
\bibitem [{\citenamefont {Yurtalan}\ \emph {et~al.}(2021)\citenamefont
  {Yurtalan}, \citenamefont {Shi}, \citenamefont {Flatt},\ and\ \citenamefont
  {Lupascu}}]{yurtalan2021characterization}%
  \BibitemOpen
  \bibfield  {author} {\bibinfo {author} {\bibfnamefont {M.}~\bibnamefont
  {Yurtalan}}, \bibinfo {author} {\bibfnamefont {J.}~\bibnamefont {Shi}},
  \bibinfo {author} {\bibfnamefont {G.}~\bibnamefont {Flatt}},\ and\ \bibinfo
  {author} {\bibfnamefont {A.}~\bibnamefont {Lupascu}},\ }\bibfield  {title}
  {\bibinfo {title} {Characterization of multilevel dynamics and decoherence in
  a high-anharmonicity capacitively shunted flux circuit},\ }\href@noop {}
  {\bibfield  {journal} {\bibinfo  {journal} {Phys. Rev. Appl.}\ }\textbf
  {\bibinfo {volume} {16}},\ \bibinfo {pages} {054051} (\bibinfo {year}
  {2021})}\BibitemShut {NoStop}%
\bibitem [{\citenamefont {Yan}\ \emph {et~al.}(2020)\citenamefont {Yan},
  \citenamefont {Sung}, \citenamefont {Krantz}, \citenamefont {Kamal},
  \citenamefont {Kim}, \citenamefont {Yoder}, \citenamefont {Orlando},
  \citenamefont {Gustavsson},\ and\ \citenamefont
  {Oliver}}]{yan2020engineering}%
  \BibitemOpen
  \bibfield  {author} {\bibinfo {author} {\bibfnamefont {F.}~\bibnamefont
  {Yan}}, \bibinfo {author} {\bibfnamefont {Y.}~\bibnamefont {Sung}}, \bibinfo
  {author} {\bibfnamefont {P.}~\bibnamefont {Krantz}}, \bibinfo {author}
  {\bibfnamefont {A.}~\bibnamefont {Kamal}}, \bibinfo {author} {\bibfnamefont
  {D.~K.}\ \bibnamefont {Kim}}, \bibinfo {author} {\bibfnamefont {J.~L.}\
  \bibnamefont {Yoder}}, \bibinfo {author} {\bibfnamefont {T.~P.}\ \bibnamefont
  {Orlando}}, \bibinfo {author} {\bibfnamefont {S.}~\bibnamefont
  {Gustavsson}},\ and\ \bibinfo {author} {\bibfnamefont {W.~D.}\ \bibnamefont
  {Oliver}},\ }\bibfield  {title} {\bibinfo {title} {Engineering framework for
  optimizing superconducting qubit designs},\ }\href@noop {} {\bibfield
  {journal} {\bibinfo  {journal} {arXiv:2006.04130}\ } (\bibinfo {year}
  {2020})}\BibitemShut {NoStop}%
\bibitem [{\citenamefont {Shearrow}\ \emph {et~al.}(2018)\citenamefont
  {Shearrow}, \citenamefont {Koolstra}, \citenamefont {Whiteley}, \citenamefont
  {Earnest}, \citenamefont {Barry}, \citenamefont {Heremans}, \citenamefont
  {Awschalom}, \citenamefont {Shirokoff},\ and\ \citenamefont
  {Schuster}}]{shearrow2018atomic}%
  \BibitemOpen
  \bibfield  {author} {\bibinfo {author} {\bibfnamefont {A.}~\bibnamefont
  {Shearrow}}, \bibinfo {author} {\bibfnamefont {G.}~\bibnamefont {Koolstra}},
  \bibinfo {author} {\bibfnamefont {S.~J.}\ \bibnamefont {Whiteley}}, \bibinfo
  {author} {\bibfnamefont {N.}~\bibnamefont {Earnest}}, \bibinfo {author}
  {\bibfnamefont {P.~S.}\ \bibnamefont {Barry}}, \bibinfo {author}
  {\bibfnamefont {F.~J.}\ \bibnamefont {Heremans}}, \bibinfo {author}
  {\bibfnamefont {D.~D.}\ \bibnamefont {Awschalom}}, \bibinfo {author}
  {\bibfnamefont {E.}~\bibnamefont {Shirokoff}},\ and\ \bibinfo {author}
  {\bibfnamefont {D.~I.}\ \bibnamefont {Schuster}},\ }\bibfield  {title}
  {\bibinfo {title} {Atomic layer deposition of titanium nitride for quantum
  circuits},\ }\href@noop {} {\bibfield  {journal} {\bibinfo  {journal} {Appl.
  Phys. Lett.}\ }\textbf {\bibinfo {volume} {113}},\ \bibinfo {pages} {212601}
  (\bibinfo {year} {2018})}\BibitemShut {NoStop}%
\bibitem [{\citenamefont {Gr{\"u}nhaupt}\ \emph {et~al.}(2018)\citenamefont
  {Gr{\"u}nhaupt}, \citenamefont {Maleeva}, \citenamefont {Skacel},
  \citenamefont {Calvo}, \citenamefont {Levy-Bertrand}, \citenamefont
  {Ustinov}, \citenamefont {Rotzinger}, \citenamefont {Monfardini},
  \citenamefont {Catelani},\ and\ \citenamefont {Pop}}]{grunhaupt2018loss}%
  \BibitemOpen
  \bibfield  {author} {\bibinfo {author} {\bibfnamefont {L.}~\bibnamefont
  {Gr{\"u}nhaupt}}, \bibinfo {author} {\bibfnamefont {N.}~\bibnamefont
  {Maleeva}}, \bibinfo {author} {\bibfnamefont {S.~T.}\ \bibnamefont {Skacel}},
  \bibinfo {author} {\bibfnamefont {M.}~\bibnamefont {Calvo}}, \bibinfo
  {author} {\bibfnamefont {F.}~\bibnamefont {Levy-Bertrand}}, \bibinfo {author}
  {\bibfnamefont {A.~V.}\ \bibnamefont {Ustinov}}, \bibinfo {author}
  {\bibfnamefont {H.}~\bibnamefont {Rotzinger}}, \bibinfo {author}
  {\bibfnamefont {A.}~\bibnamefont {Monfardini}}, \bibinfo {author}
  {\bibfnamefont {G.}~\bibnamefont {Catelani}},\ and\ \bibinfo {author}
  {\bibfnamefont {I.~M.}\ \bibnamefont {Pop}},\ }\bibfield  {title} {\bibinfo
  {title} {Loss mechanisms and quasiparticle dynamics in superconducting
  microwave resonators made of thin-film granular aluminum},\ }\href@noop {}
  {\bibfield  {journal} {\bibinfo  {journal} {Phys. Rev. Lett.}\ }\textbf
  {\bibinfo {volume} {121}},\ \bibinfo {pages} {117001} (\bibinfo {year}
  {2018})}\BibitemShut {NoStop}%
\bibitem [{\citenamefont {Bienfait}\ \emph {et~al.}(2019)\citenamefont
  {Bienfait}, \citenamefont {Satzinger}, \citenamefont {Zhong}, \citenamefont
  {Chang}, \citenamefont {Chou}, \citenamefont {Conner}, \citenamefont {Dumur},
  \citenamefont {Grebel}, \citenamefont {Peairs}, \citenamefont {Povey} \emph
  {et~al.}}]{bienfait2019phonon}%
  \BibitemOpen
  \bibfield  {author} {\bibinfo {author} {\bibfnamefont {A.}~\bibnamefont
  {Bienfait}}, \bibinfo {author} {\bibfnamefont {K.~J.}\ \bibnamefont
  {Satzinger}}, \bibinfo {author} {\bibfnamefont {Y.}~\bibnamefont {Zhong}},
  \bibinfo {author} {\bibfnamefont {H.-S.}\ \bibnamefont {Chang}}, \bibinfo
  {author} {\bibfnamefont {M.-H.}\ \bibnamefont {Chou}}, \bibinfo {author}
  {\bibfnamefont {C.~R.}\ \bibnamefont {Conner}}, \bibinfo {author}
  {\bibfnamefont {{\'E}.}~\bibnamefont {Dumur}}, \bibinfo {author}
  {\bibfnamefont {J.}~\bibnamefont {Grebel}}, \bibinfo {author} {\bibfnamefont
  {G.~A.}\ \bibnamefont {Peairs}}, \bibinfo {author} {\bibfnamefont {R.~G.}\
  \bibnamefont {Povey}}, \emph {et~al.},\ }\bibfield  {title} {\bibinfo {title}
  {Phonon-mediated quantum state transfer and remote qubit entanglement},\
  }\href@noop {} {\bibfield  {journal} {\bibinfo  {journal} {Science}\ }\textbf
  {\bibinfo {volume} {364}},\ \bibinfo {pages} {368} (\bibinfo {year}
  {2019})}\BibitemShut {NoStop}%
\bibitem [{\citenamefont {Andersson}\ \emph {et~al.}(2019)\citenamefont
  {Andersson}, \citenamefont {Suri}, \citenamefont {Guo}, \citenamefont
  {Aref},\ and\ \citenamefont {Delsing}}]{andersson2019non}%
  \BibitemOpen
  \bibfield  {author} {\bibinfo {author} {\bibfnamefont {G.}~\bibnamefont
  {Andersson}}, \bibinfo {author} {\bibfnamefont {B.}~\bibnamefont {Suri}},
  \bibinfo {author} {\bibfnamefont {L.}~\bibnamefont {Guo}}, \bibinfo {author}
  {\bibfnamefont {T.}~\bibnamefont {Aref}},\ and\ \bibinfo {author}
  {\bibfnamefont {P.}~\bibnamefont {Delsing}},\ }\bibfield  {title} {\bibinfo
  {title} {Non-exponential decay of a giant artificial atom},\ }\href@noop {}
  {\bibfield  {journal} {\bibinfo  {journal} {Nat. Phys.}\ }\textbf {\bibinfo
  {volume} {15}},\ \bibinfo {pages} {1123} (\bibinfo {year}
  {2019})}\BibitemShut {NoStop}%
\bibitem [{\citenamefont {Dumur}\ \emph {et~al.}(2021)\citenamefont {Dumur},
  \citenamefont {Satzinger}, \citenamefont {Peairs}, \citenamefont {Chou},
  \citenamefont {Bienfait}, \citenamefont {Chang}, \citenamefont {Conner},
  \citenamefont {Grebel}, \citenamefont {Povey}, \citenamefont {Zhong} \emph
  {et~al.}}]{dumur2021quantum}%
  \BibitemOpen
  \bibfield  {author} {\bibinfo {author} {\bibfnamefont {{\'E}.}~\bibnamefont
  {Dumur}}, \bibinfo {author} {\bibfnamefont {K.}~\bibnamefont {Satzinger}},
  \bibinfo {author} {\bibfnamefont {G.}~\bibnamefont {Peairs}}, \bibinfo
  {author} {\bibfnamefont {M.-H.}\ \bibnamefont {Chou}}, \bibinfo {author}
  {\bibfnamefont {A.}~\bibnamefont {Bienfait}}, \bibinfo {author}
  {\bibfnamefont {H.-S.}\ \bibnamefont {Chang}}, \bibinfo {author}
  {\bibfnamefont {C.}~\bibnamefont {Conner}}, \bibinfo {author} {\bibfnamefont
  {J.}~\bibnamefont {Grebel}}, \bibinfo {author} {\bibfnamefont
  {R.}~\bibnamefont {Povey}}, \bibinfo {author} {\bibfnamefont
  {Y.}~\bibnamefont {Zhong}}, \emph {et~al.},\ }\bibfield  {title} {\bibinfo
  {title} {Quantum communication with itinerant surface acoustic wave
  phonons},\ }\href@noop {} {\bibfield  {journal} {\bibinfo  {journal} {npj
  Quantum Inf.}\ }\textbf {\bibinfo {volume} {7}},\ \bibinfo {pages} {1}
  (\bibinfo {year} {2021})}\BibitemShut {NoStop}%
\bibitem [{\citenamefont {Keller}\ \emph {et~al.}(2017)\citenamefont {Keller},
  \citenamefont {Dieterle}, \citenamefont {Fang}, \citenamefont {Berger},
  \citenamefont {Fink},\ and\ \citenamefont {Painter}}]{keller2017transmon}%
  \BibitemOpen
  \bibfield  {author} {\bibinfo {author} {\bibfnamefont {A.~J.}\ \bibnamefont
  {Keller}}, \bibinfo {author} {\bibfnamefont {P.~B.}\ \bibnamefont
  {Dieterle}}, \bibinfo {author} {\bibfnamefont {M.}~\bibnamefont {Fang}},
  \bibinfo {author} {\bibfnamefont {B.}~\bibnamefont {Berger}}, \bibinfo
  {author} {\bibfnamefont {J.~M.}\ \bibnamefont {Fink}},\ and\ \bibinfo
  {author} {\bibfnamefont {O.}~\bibnamefont {Painter}},\ }\bibfield  {title}
  {\bibinfo {title} {Al transmon qubits on silicon-on-insulator for quantum
  device integration},\ }\href@noop {} {\bibfield  {journal} {\bibinfo
  {journal} {Appl. Phys. Lett.}\ }\textbf {\bibinfo {volume} {111}},\ \bibinfo
  {pages} {042603} (\bibinfo {year} {2017})}\BibitemShut {NoStop}%
\bibitem [{\citenamefont {Mirhosseini}\ \emph {et~al.}(2019)\citenamefont
  {Mirhosseini}, \citenamefont {Kim}, \citenamefont {Zhang}, \citenamefont
  {Sipahigil}, \citenamefont {Dieterle}, \citenamefont {Keller}, \citenamefont
  {Asenjo-Garcia}, \citenamefont {Chang},\ and\ \citenamefont
  {Painter}}]{mirhosseini2019cavity}%
  \BibitemOpen
  \bibfield  {author} {\bibinfo {author} {\bibfnamefont {M.}~\bibnamefont
  {Mirhosseini}}, \bibinfo {author} {\bibfnamefont {E.}~\bibnamefont {Kim}},
  \bibinfo {author} {\bibfnamefont {X.}~\bibnamefont {Zhang}}, \bibinfo
  {author} {\bibfnamefont {A.}~\bibnamefont {Sipahigil}}, \bibinfo {author}
  {\bibfnamefont {P.~B.}\ \bibnamefont {Dieterle}}, \bibinfo {author}
  {\bibfnamefont {A.~J.}\ \bibnamefont {Keller}}, \bibinfo {author}
  {\bibfnamefont {A.}~\bibnamefont {Asenjo-Garcia}}, \bibinfo {author}
  {\bibfnamefont {D.~E.}\ \bibnamefont {Chang}},\ and\ \bibinfo {author}
  {\bibfnamefont {O.}~\bibnamefont {Painter}},\ }\bibfield  {title} {\bibinfo
  {title} {Cavity quantum electrodynamics with atom-like mirrors},\ }\href@noop
  {} {\bibfield  {journal} {\bibinfo  {journal} {Nature}\ }\textbf {\bibinfo
  {volume} {569}},\ \bibinfo {pages} {692} (\bibinfo {year}
  {2019})}\BibitemShut {NoStop}%
\bibitem [{\citenamefont {Krinner}\ \emph {et~al.}(2019)\citenamefont
  {Krinner}, \citenamefont {Storz}, \citenamefont {Kurpiers}, \citenamefont
  {Magnard}, \citenamefont {Heinsoo}, \citenamefont {Keller}, \citenamefont
  {Luetolf}, \citenamefont {Eichler},\ and\ \citenamefont
  {Wallraff}}]{krinner2019engineering}%
  \BibitemOpen
  \bibfield  {author} {\bibinfo {author} {\bibfnamefont {S.}~\bibnamefont
  {Krinner}}, \bibinfo {author} {\bibfnamefont {S.}~\bibnamefont {Storz}},
  \bibinfo {author} {\bibfnamefont {P.}~\bibnamefont {Kurpiers}}, \bibinfo
  {author} {\bibfnamefont {P.}~\bibnamefont {Magnard}}, \bibinfo {author}
  {\bibfnamefont {J.}~\bibnamefont {Heinsoo}}, \bibinfo {author} {\bibfnamefont
  {R.}~\bibnamefont {Keller}}, \bibinfo {author} {\bibfnamefont
  {J.}~\bibnamefont {Luetolf}}, \bibinfo {author} {\bibfnamefont
  {C.}~\bibnamefont {Eichler}},\ and\ \bibinfo {author} {\bibfnamefont
  {A.}~\bibnamefont {Wallraff}},\ }\bibfield  {title} {\bibinfo {title}
  {Engineering cryogenic setups for 100-qubit scale superconducting circuit
  systems},\ }\href@noop {} {\bibfield  {journal} {\bibinfo  {journal} {EPJ
  Quant. Tech.}\ }\textbf {\bibinfo {volume} {6}},\ \bibinfo {pages} {2}
  (\bibinfo {year} {2019})}\BibitemShut {NoStop}%
\bibitem [{\citenamefont {Macklin}\ \emph {et~al.}(2015)\citenamefont
  {Macklin}, \citenamefont {O’Brien}, \citenamefont {Hover}, \citenamefont
  {Schwartz}, \citenamefont {Bolkhovsky}, \citenamefont {Zhang}, \citenamefont
  {Oliver},\ and\ \citenamefont {Siddiqi}}]{Macklin2015}%
  \BibitemOpen
  \bibfield  {author} {\bibinfo {author} {\bibfnamefont {C.}~\bibnamefont
  {Macklin}}, \bibinfo {author} {\bibfnamefont {K.}~\bibnamefont {O’Brien}},
  \bibinfo {author} {\bibfnamefont {D.}~\bibnamefont {Hover}}, \bibinfo
  {author} {\bibfnamefont {M.~E.}\ \bibnamefont {Schwartz}}, \bibinfo {author}
  {\bibfnamefont {V.}~\bibnamefont {Bolkhovsky}}, \bibinfo {author}
  {\bibfnamefont {X.}~\bibnamefont {Zhang}}, \bibinfo {author} {\bibfnamefont
  {W.~D.}\ \bibnamefont {Oliver}},\ and\ \bibinfo {author} {\bibfnamefont
  {I.}~\bibnamefont {Siddiqi}},\ }\bibfield  {title} {\bibinfo {title} {A
  near–quantum-limited josephson traveling-wave parametric amplifier},\
  }\href {https://doi.org/10.1126/science.aaa8525} {\bibfield  {journal}
  {\bibinfo  {journal} {Science}\ }\textbf {\bibinfo {volume} {350}},\ \bibinfo
  {pages} {307} (\bibinfo {year} {2015})}\BibitemShut {NoStop}%
\bibitem [{\citenamefont {Engelen}\ \emph {et~al.}(2006)\citenamefont
  {Engelen}, \citenamefont {Sugimoto}, \citenamefont {Watanabe}, \citenamefont
  {Korterik}, \citenamefont {Ikeda}, \citenamefont {van Hulst}, \citenamefont
  {Asakawa},\ and\ \citenamefont {Kuipers}}]{engelen2006theeffect}%
  \BibitemOpen
  \bibfield  {author} {\bibinfo {author} {\bibfnamefont {R.}~\bibnamefont
  {Engelen}}, \bibinfo {author} {\bibfnamefont {Y.}~\bibnamefont {Sugimoto}},
  \bibinfo {author} {\bibfnamefont {Y.}~\bibnamefont {Watanabe}}, \bibinfo
  {author} {\bibfnamefont {J.}~\bibnamefont {Korterik}}, \bibinfo {author}
  {\bibfnamefont {N.}~\bibnamefont {Ikeda}}, \bibinfo {author} {\bibfnamefont
  {N.}~\bibnamefont {van Hulst}}, \bibinfo {author} {\bibfnamefont
  {K.}~\bibnamefont {Asakawa}},\ and\ \bibinfo {author} {\bibfnamefont
  {L.}~\bibnamefont {Kuipers}},\ }\bibfield  {title} {\bibinfo {title} {The
  effect of higher-order dispersion on slow light propagation in photonic
  crystal waveguides},\ }\href {https://doi.org/10.1364/OE.14.001658}
  {\bibfield  {journal} {\bibinfo  {journal} {Opt. Express}\ }\textbf {\bibinfo
  {volume} {14}},\ \bibinfo {pages} {1658} (\bibinfo {year}
  {2006})}\BibitemShut {NoStop}%
\bibitem [{\citenamefont {Calaj{\'o}}\ \emph {et~al.}(2016)\citenamefont
  {Calaj{\'o}}, \citenamefont {Ciccarello}, \citenamefont {Chang},\ and\
  \citenamefont {Rabl}}]{calajo2016atom}%
  \BibitemOpen
  \bibfield  {author} {\bibinfo {author} {\bibfnamefont {G.}~\bibnamefont
  {Calaj{\'o}}}, \bibinfo {author} {\bibfnamefont {F.}~\bibnamefont
  {Ciccarello}}, \bibinfo {author} {\bibfnamefont {D.}~\bibnamefont {Chang}},\
  and\ \bibinfo {author} {\bibfnamefont {P.}~\bibnamefont {Rabl}},\ }\bibfield
  {title} {\bibinfo {title} {Atom-field dressed states in slow-light waveguide
  qed},\ }\href@noop {} {\bibfield  {journal} {\bibinfo  {journal} {Phys. Rev.
  A}\ }\textbf {\bibinfo {volume} {93}},\ \bibinfo {pages} {033833} (\bibinfo
  {year} {2016})}\BibitemShut {NoStop}%
\bibitem [{\citenamefont {Dirac}(1927)}]{dirac1927quantum}%
  \BibitemOpen
  \bibfield  {author} {\bibinfo {author} {\bibfnamefont {P.~A.~M.}\
  \bibnamefont {Dirac}},\ }\bibfield  {title} {\bibinfo {title} {The quantum
  theory of the emission and absorption of radiation},\ }\href@noop {}
  {\bibfield  {journal} {\bibinfo  {journal} {Proceedings of the Royal Society
  of London. Series A, Containing Papers of a Mathematical and Physical
  Character}\ }\textbf {\bibinfo {volume} {114}},\ \bibinfo {pages} {243}
  (\bibinfo {year} {1927})}\BibitemShut {NoStop}%
\bibitem [{\citenamefont {Gonz{\'a}lez-Tudela}\ and\ \citenamefont
  {Cirac}(2017)}]{gonzalez2017markovian}%
  \BibitemOpen
  \bibfield  {author} {\bibinfo {author} {\bibfnamefont {A.}~\bibnamefont
  {Gonz{\'a}lez-Tudela}}\ and\ \bibinfo {author} {\bibfnamefont {J.~I.}\
  \bibnamefont {Cirac}},\ }\bibfield  {title} {\bibinfo {title} {Markovian and
  non-markovian dynamics of quantum emitters coupled to two-dimensional
  structured reservoirs},\ }\href@noop {} {\bibfield  {journal} {\bibinfo
  {journal} {Phys. Rev. A}\ }\textbf {\bibinfo {volume} {96}},\ \bibinfo
  {pages} {043811} (\bibinfo {year} {2017})}\BibitemShut {NoStop}%
\bibitem [{\citenamefont {Barends}\ \emph {et~al.}(2013)\citenamefont
  {Barends}, \citenamefont {Kelly}, \citenamefont {Megrant}, \citenamefont
  {Sank}, \citenamefont {Jeffrey}, \citenamefont {Chen}, \citenamefont {Yin},
  \citenamefont {Chiaro}, \citenamefont {Mutus}, \citenamefont {Neill} \emph
  {et~al.}}]{barends2013coherent}%
  \BibitemOpen
  \bibfield  {author} {\bibinfo {author} {\bibfnamefont {R.}~\bibnamefont
  {Barends}}, \bibinfo {author} {\bibfnamefont {J.}~\bibnamefont {Kelly}},
  \bibinfo {author} {\bibfnamefont {A.}~\bibnamefont {Megrant}}, \bibinfo
  {author} {\bibfnamefont {D.}~\bibnamefont {Sank}}, \bibinfo {author}
  {\bibfnamefont {E.}~\bibnamefont {Jeffrey}}, \bibinfo {author} {\bibfnamefont
  {Y.}~\bibnamefont {Chen}}, \bibinfo {author} {\bibfnamefont {Y.}~\bibnamefont
  {Yin}}, \bibinfo {author} {\bibfnamefont {B.}~\bibnamefont {Chiaro}},
  \bibinfo {author} {\bibfnamefont {J.}~\bibnamefont {Mutus}}, \bibinfo
  {author} {\bibfnamefont {C.}~\bibnamefont {Neill}}, \emph {et~al.},\
  }\bibfield  {title} {\bibinfo {title} {Coherent josephson qubit suitable for
  scalable quantum integrated circuits},\ }\href@noop {} {\bibfield  {journal}
  {\bibinfo  {journal} {Phys. Rev. Lett.}\ }\textbf {\bibinfo {volume} {111}},\
  \bibinfo {pages} {080502} (\bibinfo {year} {2013})}\BibitemShut {NoStop}%
\bibitem [{\citenamefont {Jeffrey}\ \emph {et~al.}(2014)\citenamefont
  {Jeffrey}, \citenamefont {Sank}, \citenamefont {Mutus}, \citenamefont
  {White}, \citenamefont {Kelly}, \citenamefont {Barends}, \citenamefont
  {Chen}, \citenamefont {Chen}, \citenamefont {Chiaro}, \citenamefont
  {Dunsworth}, \citenamefont {Megrant}, \citenamefont {O'Malley}, \citenamefont
  {Neill}, \citenamefont {Roushan}, \citenamefont {Vainsencher}, \citenamefont
  {Wenner}, \citenamefont {Cleland},\ and\ \citenamefont
  {Martinis}}]{jeffrey2014fast}%
  \BibitemOpen
  \bibfield  {author} {\bibinfo {author} {\bibfnamefont {E.}~\bibnamefont
  {Jeffrey}}, \bibinfo {author} {\bibfnamefont {D.}~\bibnamefont {Sank}},
  \bibinfo {author} {\bibfnamefont {J.~Y.}\ \bibnamefont {Mutus}}, \bibinfo
  {author} {\bibfnamefont {T.~C.}\ \bibnamefont {White}}, \bibinfo {author}
  {\bibfnamefont {J.}~\bibnamefont {Kelly}}, \bibinfo {author} {\bibfnamefont
  {R.}~\bibnamefont {Barends}}, \bibinfo {author} {\bibfnamefont
  {Y.}~\bibnamefont {Chen}}, \bibinfo {author} {\bibfnamefont {Z.}~\bibnamefont
  {Chen}}, \bibinfo {author} {\bibfnamefont {B.}~\bibnamefont {Chiaro}},
  \bibinfo {author} {\bibfnamefont {A.}~\bibnamefont {Dunsworth}}, \bibinfo
  {author} {\bibfnamefont {A.}~\bibnamefont {Megrant}}, \bibinfo {author}
  {\bibfnamefont {P.~J.~J.}\ \bibnamefont {O'Malley}}, \bibinfo {author}
  {\bibfnamefont {C.}~\bibnamefont {Neill}}, \bibinfo {author} {\bibfnamefont
  {P.}~\bibnamefont {Roushan}}, \bibinfo {author} {\bibfnamefont
  {A.}~\bibnamefont {Vainsencher}}, \bibinfo {author} {\bibfnamefont
  {J.}~\bibnamefont {Wenner}}, \bibinfo {author} {\bibfnamefont {A.~N.}\
  \bibnamefont {Cleland}},\ and\ \bibinfo {author} {\bibfnamefont {J.~M.}\
  \bibnamefont {Martinis}},\ }\bibfield  {title} {\bibinfo {title} {Fast
  accurate state measurement with superconducting qubits},\ }\href
  {https://doi.org/10.1103/PhysRevLett.112.190504} {\bibfield  {journal}
  {\bibinfo  {journal} {Phys. Rev. Lett.}\ }\textbf {\bibinfo {volume} {112}},\
  \bibinfo {pages} {190504} (\bibinfo {year} {2014})}\BibitemShut {NoStop}%
\bibitem [{\citenamefont {Bronn}\ \emph {et~al.}(2018)\citenamefont {Bronn},
  \citenamefont {Adiga}, \citenamefont {Olivadese}, \citenamefont {Wu},
  \citenamefont {Chow},\ and\ \citenamefont {Pappas}}]{bronn2018high}%
  \BibitemOpen
  \bibfield  {author} {\bibinfo {author} {\bibfnamefont {N.~T.}\ \bibnamefont
  {Bronn}}, \bibinfo {author} {\bibfnamefont {V.~P.}\ \bibnamefont {Adiga}},
  \bibinfo {author} {\bibfnamefont {S.~B.}\ \bibnamefont {Olivadese}}, \bibinfo
  {author} {\bibfnamefont {X.}~\bibnamefont {Wu}}, \bibinfo {author}
  {\bibfnamefont {J.~M.}\ \bibnamefont {Chow}},\ and\ \bibinfo {author}
  {\bibfnamefont {D.~P.}\ \bibnamefont {Pappas}},\ }\bibfield  {title}
  {\bibinfo {title} {High coherence plane breaking packaging for
  superconducting qubits},\ }\href@noop {} {\bibfield  {journal} {\bibinfo
  {journal} {Quantum Sci. Technol.}\ }\textbf {\bibinfo {volume} {3}},\
  \bibinfo {pages} {024007} (\bibinfo {year} {2018})}\BibitemShut {NoStop}%
\bibitem [{\citenamefont {Sete}\ \emph {et~al.}(2015)\citenamefont {Sete},
  \citenamefont {Martinis},\ and\ \citenamefont {Korotkov}}]{sete2015quantum}%
  \BibitemOpen
  \bibfield  {author} {\bibinfo {author} {\bibfnamefont {E.~A.}\ \bibnamefont
  {Sete}}, \bibinfo {author} {\bibfnamefont {J.~M.}\ \bibnamefont {Martinis}},\
  and\ \bibinfo {author} {\bibfnamefont {A.~N.}\ \bibnamefont {Korotkov}},\
  }\bibfield  {title} {\bibinfo {title} {Quantum theory of a bandpass purcell
  filter for qubit readout},\ }\href
  {https://doi.org/10.1103/PhysRevA.92.012325} {\bibfield  {journal} {\bibinfo
  {journal} {Phys. Rev. A}\ }\textbf {\bibinfo {volume} {92}},\ \bibinfo
  {pages} {012325} (\bibinfo {year} {2015})}\BibitemShut {NoStop}%
\bibitem [{\citenamefont {Cleland}\ \emph {et~al.}(2019)\citenamefont
  {Cleland}, \citenamefont {Pechal}, \citenamefont {Stas}, \citenamefont
  {Sarabalis}, \citenamefont {Wollack},\ and\ \citenamefont
  {Safavi-Naeini}}]{cleland2019mechanical}%
  \BibitemOpen
  \bibfield  {author} {\bibinfo {author} {\bibfnamefont {A.~Y.}\ \bibnamefont
  {Cleland}}, \bibinfo {author} {\bibfnamefont {M.}~\bibnamefont {Pechal}},
  \bibinfo {author} {\bibfnamefont {P.-J.~C.}\ \bibnamefont {Stas}}, \bibinfo
  {author} {\bibfnamefont {C.~J.}\ \bibnamefont {Sarabalis}}, \bibinfo {author}
  {\bibfnamefont {E.~A.}\ \bibnamefont {Wollack}},\ and\ \bibinfo {author}
  {\bibfnamefont {A.~H.}\ \bibnamefont {Safavi-Naeini}},\ }\bibfield  {title}
  {\bibinfo {title} {Mechanical purcell filters for microwave quantum
  machines},\ }\href {https://doi.org/10.1063/1.5111151} {\bibfield  {journal}
  {\bibinfo  {journal} {Applied Physics Letters}\ }\textbf {\bibinfo {volume}
  {115}},\ \bibinfo {pages} {263504} (\bibinfo {year} {2019})}\BibitemShut
  {NoStop}%
\bibitem [{\citenamefont {Pozar}(2005)}]{pozar2005microwave}%
  \BibitemOpen
  \bibfield  {author} {\bibinfo {author} {\bibfnamefont {D.~M.}\ \bibnamefont
  {Pozar}},\ }\href {https://cds.cern.ch/record/882338} {\emph {\bibinfo
  {title} {{Microwave engineering; 3rd ed.}}}}\ (\bibinfo  {publisher}
  {Wiley},\ \bibinfo {address} {Hoboken, NJ},\ \bibinfo {year}
  {2005})\BibitemShut {NoStop}%
\bibitem [{\citenamefont {Rol}\ \emph {et~al.}(2020)\citenamefont {Rol},
  \citenamefont {Ciorciaro}, \citenamefont {Malinowski}, \citenamefont
  {Tarasinski}, \citenamefont {Sagastizabal}, \citenamefont {Bultink},
  \citenamefont {Salathe}, \citenamefont {Haandb{\ae}k}, \citenamefont
  {Sedivy},\ and\ \citenamefont {DiCarlo}}]{rol2020time}%
  \BibitemOpen
  \bibfield  {author} {\bibinfo {author} {\bibfnamefont {M.~A.}\ \bibnamefont
  {Rol}}, \bibinfo {author} {\bibfnamefont {L.}~\bibnamefont {Ciorciaro}},
  \bibinfo {author} {\bibfnamefont {F.~K.}\ \bibnamefont {Malinowski}},
  \bibinfo {author} {\bibfnamefont {B.~M.}\ \bibnamefont {Tarasinski}},
  \bibinfo {author} {\bibfnamefont {R.~E.}\ \bibnamefont {Sagastizabal}},
  \bibinfo {author} {\bibfnamefont {C.~C.}\ \bibnamefont {Bultink}}, \bibinfo
  {author} {\bibfnamefont {Y.}~\bibnamefont {Salathe}}, \bibinfo {author}
  {\bibfnamefont {N.}~\bibnamefont {Haandb{\ae}k}}, \bibinfo {author}
  {\bibfnamefont {J.}~\bibnamefont {Sedivy}},\ and\ \bibinfo {author}
  {\bibfnamefont {L.}~\bibnamefont {DiCarlo}},\ }\bibfield  {title} {\bibinfo
  {title} {Time-domain characterization and correction of on-chip distortion of
  control pulses in a quantum processor},\ }\href@noop {} {\bibfield  {journal}
  {\bibinfo  {journal} {Appl. Phys. Lett.}\ }\textbf {\bibinfo {volume}
  {116}},\ \bibinfo {pages} {054001} (\bibinfo {year} {2020})}\BibitemShut
  {NoStop}%
\bibitem [{\citenamefont {Johnson}(2011)}]{johnson2011controlling}%
  \BibitemOpen
  \bibfield  {author} {\bibinfo {author} {\bibfnamefont {B.~R.}\ \bibnamefont
  {Johnson}},\ }\href@noop {} {\emph {\bibinfo {title} {Controlling photons in
  superconducting electrical circuits}}}\ (\bibinfo  {publisher} {Yale
  University},\ \bibinfo {year} {2011})\BibitemShut {NoStop}%
\bibitem [{\citenamefont {Didier}\ \emph {et~al.}(2018)\citenamefont {Didier},
  \citenamefont {Sete}, \citenamefont {da~Silva},\ and\ \citenamefont
  {Rigetti}}]{didier2018analytical}%
  \BibitemOpen
  \bibfield  {author} {\bibinfo {author} {\bibfnamefont {N.}~\bibnamefont
  {Didier}}, \bibinfo {author} {\bibfnamefont {E.~A.}\ \bibnamefont {Sete}},
  \bibinfo {author} {\bibfnamefont {M.~P.}\ \bibnamefont {da~Silva}},\ and\
  \bibinfo {author} {\bibfnamefont {C.}~\bibnamefont {Rigetti}},\ }\bibfield
  {title} {\bibinfo {title} {Analytical modeling of parametrically modulated
  transmon qubits},\ }\href {https://doi.org/10.1103/PhysRevA.97.022330}
  {\bibfield  {journal} {\bibinfo  {journal} {Phys. Rev. A}\ }\textbf {\bibinfo
  {volume} {97}},\ \bibinfo {pages} {022330} (\bibinfo {year}
  {2018})}\BibitemShut {NoStop}%
\bibitem [{\citenamefont {Eichler}(2013)}]{eichler2013thesis}%
  \BibitemOpen
  \bibfield  {author} {\bibinfo {author} {\bibfnamefont {C.}~\bibnamefont
  {Eichler}},\ }\emph {\bibinfo {title} {Experimental characterization of
  quantum microwave radiation and its entanglement with a superconducting
  qubit}},\ \href@noop {} {Ph.D. thesis},\ \bibinfo  {school} {ETH Zurich}
  (\bibinfo {year} {2013})\BibitemShut {NoStop}%
\bibitem [{\citenamefont {Koch}\ \emph {et~al.}(2007)\citenamefont {Koch},
  \citenamefont {Yu}, \citenamefont {Gambetta}, \citenamefont {Houck},
  \citenamefont {Schuster}, \citenamefont {Majer}, \citenamefont {Blais},
  \citenamefont {Devoret}, \citenamefont {Girvin},\ and\ \citenamefont
  {Schoelkopf}}]{Koch2007}%
  \BibitemOpen
  \bibfield  {author} {\bibinfo {author} {\bibfnamefont {J.}~\bibnamefont
  {Koch}}, \bibinfo {author} {\bibfnamefont {T.~M.}\ \bibnamefont {Yu}},
  \bibinfo {author} {\bibfnamefont {J.}~\bibnamefont {Gambetta}}, \bibinfo
  {author} {\bibfnamefont {A.~A.}\ \bibnamefont {Houck}}, \bibinfo {author}
  {\bibfnamefont {D.~I.}\ \bibnamefont {Schuster}}, \bibinfo {author}
  {\bibfnamefont {J.}~\bibnamefont {Majer}}, \bibinfo {author} {\bibfnamefont
  {A.}~\bibnamefont {Blais}}, \bibinfo {author} {\bibfnamefont {M.~H.}\
  \bibnamefont {Devoret}}, \bibinfo {author} {\bibfnamefont {S.~M.}\
  \bibnamefont {Girvin}},\ and\ \bibinfo {author} {\bibfnamefont {R.~J.}\
  \bibnamefont {Schoelkopf}},\ }\bibfield  {title} {\bibinfo {title}
  {Charge-insensitive qubit design derived from the cooper pair box},\ }\href
  {https://doi.org/10.1103/PhysRevA.76.042319} {\bibfield  {journal} {\bibinfo
  {journal} {Phys. Rev. A}\ }\textbf {\bibinfo {volume} {76}},\ \bibinfo
  {pages} {042319} (\bibinfo {year} {2007})}\BibitemShut {NoStop}%
\bibitem [{\citenamefont {Hoi}(2013)}]{hoi2013quantum}%
  \BibitemOpen
  \bibfield  {author} {\bibinfo {author} {\bibfnamefont {I.-C.}\ \bibnamefont
  {Hoi}},\ }\href@noop {} {\emph {\bibinfo {title} {Quantum optics with
  propagating microwaves in superconducting circuits}}}\ (\bibinfo  {publisher}
  {Chalmers University of Technology},\ \bibinfo {year} {2013})\BibitemShut
  {NoStop}%
\bibitem [{\citenamefont {Chow}\ \emph {et~al.}(2012)\citenamefont {Chow},
  \citenamefont {Gambetta}, \citenamefont {C{\'o}rcoles}, \citenamefont
  {Merkel}, \citenamefont {Smolin}, \citenamefont {Rigetti}, \citenamefont
  {Poletto}, \citenamefont {Keefe}, \citenamefont {Rothwell}, \citenamefont
  {Rozen}, \citenamefont {Ketchen},\ and\ \citenamefont
  {Steffen}}]{chow2012universal}%
  \BibitemOpen
  \bibfield  {author} {\bibinfo {author} {\bibfnamefont {J.~M.}\ \bibnamefont
  {Chow}}, \bibinfo {author} {\bibfnamefont {J.~M.}\ \bibnamefont {Gambetta}},
  \bibinfo {author} {\bibfnamefont {A.~D.}\ \bibnamefont {C{\'o}rcoles}},
  \bibinfo {author} {\bibfnamefont {S.~T.}\ \bibnamefont {Merkel}}, \bibinfo
  {author} {\bibfnamefont {J.~A.}\ \bibnamefont {Smolin}}, \bibinfo {author}
  {\bibfnamefont {C.}~\bibnamefont {Rigetti}}, \bibinfo {author} {\bibfnamefont
  {S.}~\bibnamefont {Poletto}}, \bibinfo {author} {\bibfnamefont {G.~A.}\
  \bibnamefont {Keefe}}, \bibinfo {author} {\bibfnamefont {M.~B.}\ \bibnamefont
  {Rothwell}}, \bibinfo {author} {\bibfnamefont {J.~R.}\ \bibnamefont {Rozen}},
  \bibinfo {author} {\bibfnamefont {M.~B.}\ \bibnamefont {Ketchen}},\ and\
  \bibinfo {author} {\bibfnamefont {M.}~\bibnamefont {Steffen}},\ }\bibfield
  {title} {\bibinfo {title} {Universal quantum gate set approaching
  fault-tolerant thresholds with superconducting qubits},\ }\href@noop {}
  {\bibfield  {journal} {\bibinfo  {journal} {Phys. Rev. Lett.}\ }\textbf
  {\bibinfo {volume} {109}},\ \bibinfo {pages} {060501} (\bibinfo {year}
  {2012})}\BibitemShut {NoStop}%
\bibitem [{\citenamefont {Eichler}\ \emph {et~al.}(2012)\citenamefont
  {Eichler}, \citenamefont {Bozyigit},\ and\ \citenamefont
  {Wallraff}}]{eichler2012characterizing}%
  \BibitemOpen
  \bibfield  {author} {\bibinfo {author} {\bibfnamefont {C.}~\bibnamefont
  {Eichler}}, \bibinfo {author} {\bibfnamefont {D.}~\bibnamefont {Bozyigit}},\
  and\ \bibinfo {author} {\bibfnamefont {A.}~\bibnamefont {Wallraff}},\
  }\bibfield  {title} {\bibinfo {title} {Characterizing quantum microwave
  radiation and its entanglement with superconducting qubits using linear
  detectors},\ }\href@noop {} {\bibfield  {journal} {\bibinfo  {journal} {Phys.
  Rev. A}\ }\textbf {\bibinfo {volume} {86}},\ \bibinfo {pages} {032106}
  (\bibinfo {year} {2012})}\BibitemShut {NoStop}%
\bibitem [{\citenamefont {Diamond}\ and\ \citenamefont
  {Boyd}(2016)}]{diamond2016cvxpy}%
  \BibitemOpen
  \bibfield  {author} {\bibinfo {author} {\bibfnamefont {S.}~\bibnamefont
  {Diamond}}\ and\ \bibinfo {author} {\bibfnamefont {S.}~\bibnamefont {Boyd}},\
  }\bibfield  {title} {\bibinfo {title} {{CVXPY}: {A} {P}ython-embedded
  modeling language for convex optimization},\ }\href@noop {} {\bibfield
  {journal} {\bibinfo  {journal} {Journal of Machine Learning Research}\
  }\textbf {\bibinfo {volume} {17}},\ \bibinfo {pages} {1} (\bibinfo {year}
  {2016})}\BibitemShut {NoStop}%
\bibitem [{\citenamefont {McKay}\ \emph {et~al.}(2017)\citenamefont {McKay},
  \citenamefont {Wood}, \citenamefont {Sheldon}, \citenamefont {Chow},\ and\
  \citenamefont {Gambetta}}]{mckay2017efficient}%
  \BibitemOpen
  \bibfield  {author} {\bibinfo {author} {\bibfnamefont {D.~C.}\ \bibnamefont
  {McKay}}, \bibinfo {author} {\bibfnamefont {C.~J.}\ \bibnamefont {Wood}},
  \bibinfo {author} {\bibfnamefont {S.}~\bibnamefont {Sheldon}}, \bibinfo
  {author} {\bibfnamefont {J.~M.}\ \bibnamefont {Chow}},\ and\ \bibinfo
  {author} {\bibfnamefont {J.~M.}\ \bibnamefont {Gambetta}},\ }\bibfield
  {title} {\bibinfo {title} {Efficient z gates for quantum computing},\
  }\href@noop {} {\bibfield  {journal} {\bibinfo  {journal} {Phys. Rev. A}\
  }\textbf {\bibinfo {volume} {96}},\ \bibinfo {pages} {022330} (\bibinfo
  {year} {2017})}\BibitemShut {NoStop}%
\bibitem [{\citenamefont {da~Silva}\ \emph {et~al.}(2010)\citenamefont
  {da~Silva}, \citenamefont {Bozyigit}, \citenamefont {Wallraff},\ and\
  \citenamefont {Blais}}]{da2010schemes}%
  \BibitemOpen
  \bibfield  {author} {\bibinfo {author} {\bibfnamefont {M.~P.}\ \bibnamefont
  {da~Silva}}, \bibinfo {author} {\bibfnamefont {D.}~\bibnamefont {Bozyigit}},
  \bibinfo {author} {\bibfnamefont {A.}~\bibnamefont {Wallraff}},\ and\
  \bibinfo {author} {\bibfnamefont {A.}~\bibnamefont {Blais}},\ }\bibfield
  {title} {\bibinfo {title} {Schemes for the observation of photon correlation
  functions in circuit qed with linear detectors},\ }\href@noop {} {\bibfield
  {journal} {\bibinfo  {journal} {Phys.Rev. A}\ }\textbf {\bibinfo {volume}
  {82}},\ \bibinfo {pages} {043804} (\bibinfo {year} {2010})}\BibitemShut
  {NoStop}%
\bibitem [{\citenamefont {Nielsen}\ and\ \citenamefont
  {Chuang}(2000)}]{nielsenchuang}%
  \BibitemOpen
  \bibfield  {author} {\bibinfo {author} {\bibfnamefont {M.}~\bibnamefont
  {Nielsen}}\ and\ \bibinfo {author} {\bibfnamefont {I.~L.}\ \bibnamefont
  {Chuang}},\ }\href@noop {} {\emph {\bibinfo {title} {Quantum Computation and
  Quantum Information}}}\ (\bibinfo  {publisher} {Cambridge University Press},\
  \bibinfo {year} {2000})\BibitemShut {NoStop}%
\bibitem [{\citenamefont {Nielsen}\ \emph {et~al.}(2021)\citenamefont
  {Nielsen}, \citenamefont {Gamble}, \citenamefont {Rudin}, \citenamefont
  {Scholten}, \citenamefont {Young},\ and\ \citenamefont
  {Blume-Kohout}}]{nielsen2021gate}%
  \BibitemOpen
  \bibfield  {author} {\bibinfo {author} {\bibfnamefont {E.}~\bibnamefont
  {Nielsen}}, \bibinfo {author} {\bibfnamefont {J.~K.}\ \bibnamefont {Gamble}},
  \bibinfo {author} {\bibfnamefont {K.}~\bibnamefont {Rudin}}, \bibinfo
  {author} {\bibfnamefont {T.}~\bibnamefont {Scholten}}, \bibinfo {author}
  {\bibfnamefont {K.}~\bibnamefont {Young}},\ and\ \bibinfo {author}
  {\bibfnamefont {R.}~\bibnamefont {Blume-Kohout}},\ }\bibfield  {title}
  {\bibinfo {title} {Gate set tomography},\ }\href@noop {} {\bibfield
  {journal} {\bibinfo  {journal} {Quantum}\ }\textbf {\bibinfo {volume} {5}},\
  \bibinfo {pages} {557} (\bibinfo {year} {2021})}\BibitemShut {NoStop}%
\bibitem [{\citenamefont {Nielsen}\ \emph {et~al.}(2020)\citenamefont
  {Nielsen}, \citenamefont {Blume-Kohout}, \citenamefont {Saldyt},
  \citenamefont {Gross}, \citenamefont {Scholten}, \citenamefont {Rudinger},
  \citenamefont {Proctor}, \citenamefont {Gamble},\ and\ \citenamefont
  {Russo}}]{nielsen2020pyGSTi}%
  \BibitemOpen
  \bibfield  {author} {\bibinfo {author} {\bibfnamefont {E.}~\bibnamefont
  {Nielsen}}, \bibinfo {author} {\bibfnamefont {R.}~\bibnamefont
  {Blume-Kohout}}, \bibinfo {author} {\bibfnamefont {L.}~\bibnamefont
  {Saldyt}}, \bibinfo {author} {\bibfnamefont {J.}~\bibnamefont {Gross}},
  \bibinfo {author} {\bibfnamefont {T.}~\bibnamefont {Scholten}}, \bibinfo
  {author} {\bibfnamefont {K.}~\bibnamefont {Rudinger}}, \bibinfo {author}
  {\bibfnamefont {T.}~\bibnamefont {Proctor}}, \bibinfo {author} {\bibfnamefont
  {J.~K.}\ \bibnamefont {Gamble}},\ and\ \bibinfo {author} {\bibfnamefont
  {A.}~\bibnamefont {Russo}},\ }\href {https://doi.org/10.5281/zenodo.3675466}
  {\emph {\bibinfo {title} {PyGSTi version 0.9.9.1}}}\ (\bibinfo {year}
  {2020})\BibitemShut {NoStop}%
\bibitem [{\citenamefont {Keith}\ \emph {et~al.}(2018)\citenamefont {Keith},
  \citenamefont {Baldwin}, \citenamefont {Glancy},\ and\ \citenamefont
  {Knill}}]{keith2018joint}%
  \BibitemOpen
  \bibfield  {author} {\bibinfo {author} {\bibfnamefont {A.~C.}\ \bibnamefont
  {Keith}}, \bibinfo {author} {\bibfnamefont {C.~H.}\ \bibnamefont {Baldwin}},
  \bibinfo {author} {\bibfnamefont {S.}~\bibnamefont {Glancy}},\ and\ \bibinfo
  {author} {\bibfnamefont {E.}~\bibnamefont {Knill}},\ }\bibfield  {title}
  {\bibinfo {title} {Joint quantum-state and measurement tomography with
  incomplete measurements},\ }\href
  {https://doi.org/10.1103/PhysRevA.98.042318} {\bibfield  {journal} {\bibinfo
  {journal} {Phys. Rev. A}\ }\textbf {\bibinfo {volume} {98}},\ \bibinfo
  {pages} {042318} (\bibinfo {year} {2018})}\BibitemShut {NoStop}%
\bibitem [{\citenamefont {Bylander}\ \emph {et~al.}(2011)\citenamefont
  {Bylander}, \citenamefont {Gustavsson}, \citenamefont {Yan}, \citenamefont
  {Yoshihara}, \citenamefont {Harrabi}, \citenamefont {Fitch}, \citenamefont
  {Cory}, \citenamefont {Nakamura}, \citenamefont {Tsai},\ and\ \citenamefont
  {Oliver}}]{bylander2011noise}%
  \BibitemOpen
  \bibfield  {author} {\bibinfo {author} {\bibfnamefont {J.}~\bibnamefont
  {Bylander}}, \bibinfo {author} {\bibfnamefont {S.}~\bibnamefont
  {Gustavsson}}, \bibinfo {author} {\bibfnamefont {F.}~\bibnamefont {Yan}},
  \bibinfo {author} {\bibfnamefont {F.}~\bibnamefont {Yoshihara}}, \bibinfo
  {author} {\bibfnamefont {K.}~\bibnamefont {Harrabi}}, \bibinfo {author}
  {\bibfnamefont {G.}~\bibnamefont {Fitch}}, \bibinfo {author} {\bibfnamefont
  {D.~G.}\ \bibnamefont {Cory}}, \bibinfo {author} {\bibfnamefont
  {Y.}~\bibnamefont {Nakamura}}, \bibinfo {author} {\bibfnamefont {J.-S.}\
  \bibnamefont {Tsai}},\ and\ \bibinfo {author} {\bibfnamefont {W.~D.}\
  \bibnamefont {Oliver}},\ }\bibfield  {title} {\bibinfo {title} {Noise
  spectroscopy through dynamical decoupling with a superconducting flux
  qubit},\ }\href@noop {} {\bibfield  {journal} {\bibinfo  {journal} {Nat.
  Phys.}\ }\textbf {\bibinfo {volume} {7}},\ \bibinfo {pages} {565} (\bibinfo
  {year} {2011})}\BibitemShut {NoStop}%
\bibitem [{\citenamefont {Krantz}\ \emph {et~al.}(2019)\citenamefont {Krantz},
  \citenamefont {Kjaergaard}, \citenamefont {Yan}, \citenamefont {Orlando},
  \citenamefont {Gustavsson},\ and\ \citenamefont
  {Oliver}}]{krantz2019quantum}%
  \BibitemOpen
  \bibfield  {author} {\bibinfo {author} {\bibfnamefont {P.}~\bibnamefont
  {Krantz}}, \bibinfo {author} {\bibfnamefont {M.}~\bibnamefont {Kjaergaard}},
  \bibinfo {author} {\bibfnamefont {F.}~\bibnamefont {Yan}}, \bibinfo {author}
  {\bibfnamefont {T.~P.}\ \bibnamefont {Orlando}}, \bibinfo {author}
  {\bibfnamefont {S.}~\bibnamefont {Gustavsson}},\ and\ \bibinfo {author}
  {\bibfnamefont {W.~D.}\ \bibnamefont {Oliver}},\ }\bibfield  {title}
  {\bibinfo {title} {A quantum engineer's guide to superconducting qubits},\
  }\href@noop {} {\bibfield  {journal} {\bibinfo  {journal} {Appl. Phys. Rev.}\
  }\textbf {\bibinfo {volume} {6}},\ \bibinfo {pages} {021318} (\bibinfo {year}
  {2019})}\BibitemShut {NoStop}%
\bibitem [{\citenamefont {Hutchings}\ \emph {et~al.}(2017)\citenamefont
  {Hutchings}, \citenamefont {Hertzberg}, \citenamefont {Liu}, \citenamefont
  {Bronn}, \citenamefont {Keefe}, \citenamefont {Brink}, \citenamefont {Chow},\
  and\ \citenamefont {Plourde}}]{hutchings2017tunable}%
  \BibitemOpen
  \bibfield  {author} {\bibinfo {author} {\bibfnamefont {M.}~\bibnamefont
  {Hutchings}}, \bibinfo {author} {\bibfnamefont {J.~B.}\ \bibnamefont
  {Hertzberg}}, \bibinfo {author} {\bibfnamefont {Y.}~\bibnamefont {Liu}},
  \bibinfo {author} {\bibfnamefont {N.~T.}\ \bibnamefont {Bronn}}, \bibinfo
  {author} {\bibfnamefont {G.~A.}\ \bibnamefont {Keefe}}, \bibinfo {author}
  {\bibfnamefont {M.}~\bibnamefont {Brink}}, \bibinfo {author} {\bibfnamefont
  {J.~M.}\ \bibnamefont {Chow}},\ and\ \bibinfo {author} {\bibfnamefont
  {B.}~\bibnamefont {Plourde}},\ }\bibfield  {title} {\bibinfo {title} {Tunable
  superconducting qubits with flux-independent coherence},\ }\href@noop {}
  {\bibfield  {journal} {\bibinfo  {journal} {Phys. Rev. Appl.}\ }\textbf
  {\bibinfo {volume} {8}},\ \bibinfo {pages} {044003} (\bibinfo {year}
  {2017})}\BibitemShut {NoStop}%
\bibitem [{\citenamefont {Chen}\ \emph {et~al.}(2014)\citenamefont {Chen},
  \citenamefont {Neill}, \citenamefont {Roushan}, \citenamefont {Leung},
  \citenamefont {Fang}, \citenamefont {Barends}, \citenamefont {Kelly},
  \citenamefont {Campbell}, \citenamefont {Chen}, \citenamefont {Chiaro} \emph
  {et~al.}}]{chen2014qubit}%
  \BibitemOpen
  \bibfield  {author} {\bibinfo {author} {\bibfnamefont {Y.}~\bibnamefont
  {Chen}}, \bibinfo {author} {\bibfnamefont {C.}~\bibnamefont {Neill}},
  \bibinfo {author} {\bibfnamefont {P.}~\bibnamefont {Roushan}}, \bibinfo
  {author} {\bibfnamefont {N.}~\bibnamefont {Leung}}, \bibinfo {author}
  {\bibfnamefont {M.}~\bibnamefont {Fang}}, \bibinfo {author} {\bibfnamefont
  {R.}~\bibnamefont {Barends}}, \bibinfo {author} {\bibfnamefont
  {J.}~\bibnamefont {Kelly}}, \bibinfo {author} {\bibfnamefont
  {B.}~\bibnamefont {Campbell}}, \bibinfo {author} {\bibfnamefont
  {Z.}~\bibnamefont {Chen}}, \bibinfo {author} {\bibfnamefont {B.}~\bibnamefont
  {Chiaro}}, \emph {et~al.},\ }\bibfield  {title} {\bibinfo {title} {Qubit
  architecture with high coherence and fast tunable coupling},\ }\href@noop {}
  {\bibfield  {journal} {\bibinfo  {journal} {Phys. Rev. Lett.}\ }\textbf
  {\bibinfo {volume} {113}},\ \bibinfo {pages} {220502} (\bibinfo {year}
  {2014})}\BibitemShut {NoStop}%
\bibitem [{\citenamefont {Yan}\ \emph {et~al.}(2018)\citenamefont {Yan},
  \citenamefont {Krantz}, \citenamefont {Sung}, \citenamefont {Kjaergaard},
  \citenamefont {Campbell}, \citenamefont {Orlando}, \citenamefont
  {Gustavsson},\ and\ \citenamefont {Oliver}}]{yan2018tunable}%
  \BibitemOpen
  \bibfield  {author} {\bibinfo {author} {\bibfnamefont {F.}~\bibnamefont
  {Yan}}, \bibinfo {author} {\bibfnamefont {P.}~\bibnamefont {Krantz}},
  \bibinfo {author} {\bibfnamefont {Y.}~\bibnamefont {Sung}}, \bibinfo {author}
  {\bibfnamefont {M.}~\bibnamefont {Kjaergaard}}, \bibinfo {author}
  {\bibfnamefont {D.~L.}\ \bibnamefont {Campbell}}, \bibinfo {author}
  {\bibfnamefont {T.~P.}\ \bibnamefont {Orlando}}, \bibinfo {author}
  {\bibfnamefont {S.}~\bibnamefont {Gustavsson}},\ and\ \bibinfo {author}
  {\bibfnamefont {W.~D.}\ \bibnamefont {Oliver}},\ }\bibfield  {title}
  {\bibinfo {title} {Tunable coupling scheme for implementing high-fidelity
  two-qubit gates},\ }\href@noop {} {\bibfield  {journal} {\bibinfo  {journal}
  {Phys. Rev. Appl.}\ }\textbf {\bibinfo {volume} {10}},\ \bibinfo {pages}
  {054062} (\bibinfo {year} {2018})}\BibitemShut {NoStop}%
\bibitem [{\citenamefont {Killey}\ \emph {et~al.}(2006)\citenamefont {Killey},
  \citenamefont {Watts}, \citenamefont {Glick},\ and\ \citenamefont
  {Bayvel}}]{killey2006electronic}%
  \BibitemOpen
  \bibfield  {author} {\bibinfo {author} {\bibfnamefont {R.}~\bibnamefont
  {Killey}}, \bibinfo {author} {\bibfnamefont {P.}~\bibnamefont {Watts}},
  \bibinfo {author} {\bibfnamefont {M.}~\bibnamefont {Glick}},\ and\ \bibinfo
  {author} {\bibfnamefont {P.}~\bibnamefont {Bayvel}},\ }\bibfield  {title}
  {\bibinfo {title} {Electronic dispersion compensation by signal
  predistortion},\ }in\ \href@noop {} {\emph {\bibinfo {booktitle} {2006
  Optical Fiber Communication Conference and the National Fiber Optic Engineers
  Conference}}}\ (\bibinfo {organization} {IEEE},\ \bibinfo {year} {2006})\
  pp.\ \bibinfo {pages} {3--pp}\BibitemShut {NoStop}%
\bibitem [{\citenamefont {Ramachandran}(2007)}]{ramachandran2007fiber}%
  \BibitemOpen
  \bibfield  {author} {\bibinfo {author} {\bibfnamefont {S.}~\bibnamefont
  {Ramachandran}},\ }\href@noop {} {\emph {\bibinfo {title} {Fiber based
  dispersion compensation}}},\ Vol.~\bibinfo {volume} {5}\ (\bibinfo
  {publisher} {Springer Science \& Business Media},\ \bibinfo {year}
  {2007})\BibitemShut {NoStop}%
\bibitem [{\citenamefont {Huang}\ \emph
  {et~al.}(2021{\natexlab{a}})\citenamefont {Huang}, \citenamefont {Clerk},\
  and\ \citenamefont {Martin}}]{huang2021nondispersing}%
  \BibitemOpen
  \bibfield  {author} {\bibinfo {author} {\bibfnamefont {Z.}~\bibnamefont
  {Huang}}, \bibinfo {author} {\bibfnamefont {A.}~\bibnamefont {Clerk}},\ and\
  \bibinfo {author} {\bibfnamefont {I.}~\bibnamefont {Martin}},\ }\bibfield
  {title} {\bibinfo {title} {Nondispersing wave packets in lattice floquet
  systems},\ }\href@noop {} {\bibfield  {journal} {\bibinfo  {journal}
  {Physical Review Letters}\ }\textbf {\bibinfo {volume} {126}},\ \bibinfo
  {pages} {100601} (\bibinfo {year} {2021}{\natexlab{a}})}\BibitemShut
  {NoStop}%
\bibitem [{\citenamefont {Huang}\ \emph
  {et~al.}(2021{\natexlab{b}})\citenamefont {Huang}, \citenamefont {Lienhard},
  \citenamefont {Calusine}, \citenamefont {Veps{\"a}l{\"a}inen}, \citenamefont
  {Braum{\"u}ller}, \citenamefont {Kim}, \citenamefont {Melville},
  \citenamefont {Niedzielski}, \citenamefont {Yoder}, \citenamefont {Kannan}
  \emph {et~al.}}]{huang2021microwave}%
  \BibitemOpen
  \bibfield  {author} {\bibinfo {author} {\bibfnamefont {S.}~\bibnamefont
  {Huang}}, \bibinfo {author} {\bibfnamefont {B.}~\bibnamefont {Lienhard}},
  \bibinfo {author} {\bibfnamefont {G.}~\bibnamefont {Calusine}}, \bibinfo
  {author} {\bibfnamefont {A.}~\bibnamefont {Veps{\"a}l{\"a}inen}}, \bibinfo
  {author} {\bibfnamefont {J.}~\bibnamefont {Braum{\"u}ller}}, \bibinfo
  {author} {\bibfnamefont {D.~K.}\ \bibnamefont {Kim}}, \bibinfo {author}
  {\bibfnamefont {A.~J.}\ \bibnamefont {Melville}}, \bibinfo {author}
  {\bibfnamefont {B.~M.}\ \bibnamefont {Niedzielski}}, \bibinfo {author}
  {\bibfnamefont {J.~L.}\ \bibnamefont {Yoder}}, \bibinfo {author}
  {\bibfnamefont {B.}~\bibnamefont {Kannan}}, \emph {et~al.},\ }\bibfield
  {title} {\bibinfo {title} {Microwave package design for superconducting
  quantum processors},\ }\href@noop {} {\bibfield  {journal} {\bibinfo
  {journal} {PRX Quant.}\ }\textbf {\bibinfo {volume} {2}},\ \bibinfo {pages}
  {020306} (\bibinfo {year} {2021}{\natexlab{b}})}\BibitemShut {NoStop}%
\bibitem [{\citenamefont {Bartolucci}\ \emph {et~al.}(2021)\citenamefont
  {Bartolucci}, \citenamefont {Birchall}, \citenamefont {Bombin}, \citenamefont
  {Cable}, \citenamefont {Dawson}, \citenamefont {Gimeno-Segovia},
  \citenamefont {Johnston}, \citenamefont {Kieling}, \citenamefont {Nickerson},
  \citenamefont {Pant} \emph {et~al.}}]{bartolucci2021fusion}%
  \BibitemOpen
  \bibfield  {author} {\bibinfo {author} {\bibfnamefont {S.}~\bibnamefont
  {Bartolucci}}, \bibinfo {author} {\bibfnamefont {P.}~\bibnamefont
  {Birchall}}, \bibinfo {author} {\bibfnamefont {H.}~\bibnamefont {Bombin}},
  \bibinfo {author} {\bibfnamefont {H.}~\bibnamefont {Cable}}, \bibinfo
  {author} {\bibfnamefont {C.}~\bibnamefont {Dawson}}, \bibinfo {author}
  {\bibfnamefont {M.}~\bibnamefont {Gimeno-Segovia}}, \bibinfo {author}
  {\bibfnamefont {E.}~\bibnamefont {Johnston}}, \bibinfo {author}
  {\bibfnamefont {K.}~\bibnamefont {Kieling}}, \bibinfo {author} {\bibfnamefont
  {N.}~\bibnamefont {Nickerson}}, \bibinfo {author} {\bibfnamefont
  {M.}~\bibnamefont {Pant}}, \emph {et~al.},\ }\bibfield  {title} {\bibinfo
  {title} {Fusion-based quantum computation},\ }\href@noop {} {\bibfield
  {journal} {\bibinfo  {journal} {arXiv:2101.09310}\ } (\bibinfo {year}
  {2021})}\BibitemShut {NoStop}%
\bibitem [{\citenamefont {Place}\ \emph {et~al.}(2021)\citenamefont {Place},
  \citenamefont {Rodgers}, \citenamefont {Mundada}, \citenamefont {Smitham},
  \citenamefont {Fitzpatrick}, \citenamefont {Leng}, \citenamefont {Premkumar},
  \citenamefont {Bryon}, \citenamefont {Vrajitoarea}, \citenamefont {Sussman}
  \emph {et~al.}}]{place2021new}%
  \BibitemOpen
  \bibfield  {author} {\bibinfo {author} {\bibfnamefont {A.~P.}\ \bibnamefont
  {Place}}, \bibinfo {author} {\bibfnamefont {L.~V.}\ \bibnamefont {Rodgers}},
  \bibinfo {author} {\bibfnamefont {P.}~\bibnamefont {Mundada}}, \bibinfo
  {author} {\bibfnamefont {B.~M.}\ \bibnamefont {Smitham}}, \bibinfo {author}
  {\bibfnamefont {M.}~\bibnamefont {Fitzpatrick}}, \bibinfo {author}
  {\bibfnamefont {Z.}~\bibnamefont {Leng}}, \bibinfo {author} {\bibfnamefont
  {A.}~\bibnamefont {Premkumar}}, \bibinfo {author} {\bibfnamefont
  {J.}~\bibnamefont {Bryon}}, \bibinfo {author} {\bibfnamefont
  {A.}~\bibnamefont {Vrajitoarea}}, \bibinfo {author} {\bibfnamefont
  {S.}~\bibnamefont {Sussman}}, \emph {et~al.},\ }\bibfield  {title} {\bibinfo
  {title} {New material platform for superconducting transmon qubits with
  coherence times exceeding 0.3 milliseconds},\ }\href@noop {} {\bibfield
  {journal} {\bibinfo  {journal} {Nature Comm.}\ }\textbf {\bibinfo {volume}
  {12}},\ \bibinfo {pages} {1} (\bibinfo {year} {2021})}\BibitemShut {NoStop}%
\bibitem [{\citenamefont {Wenner}\ \emph {et~al.}(2014)\citenamefont {Wenner},
  \citenamefont {Yin}, \citenamefont {Chen}, \citenamefont {Barends},
  \citenamefont {Chiaro}, \citenamefont {Jeffrey}, \citenamefont {Kelly},
  \citenamefont {Megrant}, \citenamefont {Mutus}, \citenamefont {Neill} \emph
  {et~al.}}]{wenner2014catching}%
  \BibitemOpen
  \bibfield  {author} {\bibinfo {author} {\bibfnamefont {J.}~\bibnamefont
  {Wenner}}, \bibinfo {author} {\bibfnamefont {Y.}~\bibnamefont {Yin}},
  \bibinfo {author} {\bibfnamefont {Y.}~\bibnamefont {Chen}}, \bibinfo {author}
  {\bibfnamefont {R.}~\bibnamefont {Barends}}, \bibinfo {author} {\bibfnamefont
  {B.}~\bibnamefont {Chiaro}}, \bibinfo {author} {\bibfnamefont
  {E.}~\bibnamefont {Jeffrey}}, \bibinfo {author} {\bibfnamefont
  {J.}~\bibnamefont {Kelly}}, \bibinfo {author} {\bibfnamefont
  {A.}~\bibnamefont {Megrant}}, \bibinfo {author} {\bibfnamefont
  {J.}~\bibnamefont {Mutus}}, \bibinfo {author} {\bibfnamefont
  {C.}~\bibnamefont {Neill}}, \emph {et~al.},\ }\bibfield  {title} {\bibinfo
  {title} {Catching time-reversed microwave coherent state photons with 99.4\%
  absorption efficiency},\ }\href@noop {} {\bibfield  {journal} {\bibinfo
  {journal} {Phy. Rev. Lett.}\ }\textbf {\bibinfo {volume} {112}},\ \bibinfo
  {pages} {210501} (\bibinfo {year} {2014})}\BibitemShut {NoStop}%
\bibitem [{\citenamefont {Kurpiers}\ \emph {et~al.}(2018)\citenamefont
  {Kurpiers}, \citenamefont {Magnard}, \citenamefont {Walter}, \citenamefont
  {Royer}, \citenamefont {Pechal}, \citenamefont {Heinsoo}, \citenamefont
  {Salath{\'e}}, \citenamefont {Akin}, \citenamefont {Storz}, \citenamefont
  {Besse} \emph {et~al.}}]{kurpiers2018deterministic}%
  \BibitemOpen
  \bibfield  {author} {\bibinfo {author} {\bibfnamefont {P.}~\bibnamefont
  {Kurpiers}}, \bibinfo {author} {\bibfnamefont {P.}~\bibnamefont {Magnard}},
  \bibinfo {author} {\bibfnamefont {T.}~\bibnamefont {Walter}}, \bibinfo
  {author} {\bibfnamefont {B.}~\bibnamefont {Royer}}, \bibinfo {author}
  {\bibfnamefont {M.}~\bibnamefont {Pechal}}, \bibinfo {author} {\bibfnamefont
  {J.}~\bibnamefont {Heinsoo}}, \bibinfo {author} {\bibfnamefont
  {Y.}~\bibnamefont {Salath{\'e}}}, \bibinfo {author} {\bibfnamefont
  {A.}~\bibnamefont {Akin}}, \bibinfo {author} {\bibfnamefont {S.}~\bibnamefont
  {Storz}}, \bibinfo {author} {\bibfnamefont {J.-C.}\ \bibnamefont {Besse}},
  \emph {et~al.},\ }\bibfield  {title} {\bibinfo {title} {Deterministic quantum
  state transfer and remote entanglement using microwave photons},\ }\href@noop
  {} {\bibfield  {journal} {\bibinfo  {journal} {Nature}\ }\textbf {\bibinfo
  {volume} {558}},\ \bibinfo {pages} {264} (\bibinfo {year}
  {2018})}\BibitemShut {NoStop}%
\end{thebibliography}

%

\end{document}